\documentclass[trackchanges,twocolumn]{aastex7}
\usepackage[utf8]{inputenc}
\usepackage{newunicodechar}
\newunicodechar{ʻ}{\textquotesingle}
\usepackage{hyperref}
\usepackage{longtable,booktabs,xcolor}
\usepackage{placeins}
\usepackage{caption}
\usepackage{subcaption}

\begin{document}



\title{The JWST Sub-Jupiters Survey: Observed Sensitivities to the Wide-Separation, Cold, Sub-Jupiter Exoplanet Population}

\newcommand{\IfA}{Institute for Astronomy, University of Edinburgh, Royal Observatory, Blackford Hill, Edinburgh, EH9 3HJ, UK}
\newcommand{\STScI}{Space Telescope Science Institute, 3700 San Martin Dr, Baltimore, MD 21218, USA}
\newcommand{\CfES}{Centre for Exoplanet Science, University of Edinburgh, Edinburgh EH9 3HJ, UK}
\newcommand{\UTA}{Department of Astronomy, University of Texas at Austin, Austin, TX 78712, USA}
\newcommand{\UCSC}{Department of Astronomy \& Astrophysics, University of California, Santa Cruz, 1156 High Street, Santa Cruz, CA 95064}
\newcommand{\CIERA}{Center for Interdisciplinary Exploration and Research in Astrophysics (CIERA), Northwestern University, 1800 Sherman Ave, Evanston, IL 60201, USA}
\newcommand{\NU}{Department of Physics and Astronomy, Northwestern University, 2145 Sheridan Rd, Evanston, IL 60208, USA}
\newcommand{\CSST}{Center for Space Sciences and Technology, University of Maryland, Baltimore County, 1000 Hilltop Circle, Baltimore, MD 21250, USA}
\newcommand{\ASD}{Astrophysics Science Division, NASA-GSFC, 8800 Greenbelt Rd, Greenbelt, MD 20771, USA}
\newcommand{\CRESST}{Center for Research and Exploration in Space Science and Technology, NASA-GSFC, 8800 Greenbelt Rd, Greenbelt, MD 20771, USA}
\newcommand{\UKATC}{UK Astronomy Technology Centre, Royal Observatory Edinburgh, Blackford Hill, Edinburgh EH9 3HJ, UK}
\newcommand{\ESI}{Eureka Scientific Inc., 2542 Delmar Ave., Suite 100, Oakland, CA 94602, USA}
\newcommand{\UoE}{Department of Physics and Astronomy, University of Exeter, Stocker Road, Exeter EX4 4QL, UK}
\newcommand{\ESA}{European Space Agency (ESA), ESA Office, Space Telescope Science Institute, 3700 San Martin Drive, Baltimore, MD 21218, USA}
\newcommand{\LAM}{Laboratoire d'Astrophysique de Marseille, 38 Rue Frédéric Joliot Curie, 13013 Marseille, France}
\newcommand{\UVic}{Department of Physics and Astronomy University of Victoria 3800 Finnerty Road Elliott Building, Room 207. Victoria BC V8P 5C2, Canada}
\newcommand{\NRC}{Herzberg Astronomy and Astrophysics Research Centre 5071 W Saanich Rd, Victoria, BC, V9E 2E7, Canada}
\newcommand{\JHU}{Department of Physics $\&$ Astronomy, Johns Hopkins University, 3400 N. Charles Street, Baltimore, MD 21218, USA}
\newcommand{\UoA}{Department of Astronomy / Steward Observatory, University of Arizona, Tucson, AZ 85721, USA}
\newcommand{\UoM}{Department of Astronomy, University of Michigan, Ann Arbor, MI 48109, USA}
\newcommand{\UH}{Institute for Astronomy, University of Hawaiʻi, 2680 Woodlawn Drive, Honolulu HI 96822}

\author[orcid=0000-0002-1652-420X,sname='Giovanni M. Strampelli']{Giovanni M. Strampelli}
\altaffiliation{Newton International Fellow}
\affiliation{\IfA}
\email[show]{gstrampelli@roe.ac.uk}  

\author[orcid=0000-0001-5365-4815,sname='Aarynn L. Carter']{Aarynn L. Carter}
\affiliation{\STScI}
\email{aacarter@stsci.edu}  

\author[orcid=0000-0003-4614-7035,sname='Beth Biller']{Beth Biller}
\affiliation{\IfA}
\affiliation{\CfES}
\email{bbiller@ed.ac.uk}  

\author[orcid=0000-0002-2428-9932,sname='Clémence Fontanive']{Clémence Fontanive}
\affiliation{\IfA}
\email{clemence.fontanive@roe.ac.uk}  

\author[orcid=0009-0004-4349-006X,sname='Seongbong Han']{Seongbong Han}
\affiliation{\UoM}
\email{sbonghan@umich.edu}  

\author[orcid=0000-0001-5864-9599,sname='James Mang']{James Mang}
\altaffiliation{NSF Graduate Research Fellow}
\affiliation{\UTA}
\email{j_mang@utexas.edu}  

\author[orcid=0000-0002-2683-2396,sname='Rodrigo Ferrer-Chavez']{Rodrigo Ferrer-Chavez}
\affiliation{\CIERA}
\affiliation{\NU}
\email{rodrigoferrer-chavez2029@u.northwestern.edu}  

\author[orcid=0000-0002-3122-6809,sname='Vito Squicciarini']{Vito Squicciarini}
\affiliation{\UoE}
\email{v.squicciarini@exeter.ac.uk} 

\author[orcid=0000-0001-5831-9530,sname='Rachel Bowens-Rubin']{Rachel Bowens-Rubin}
\affiliation{\ESI}
\email{rbowens-rubin@mit.edu}

\author[orcid=0000-0002-9962-132X,sname='Ben J. Sutlieff']{Ben J. Sutlieff}
\affiliation{\IfA}
\affiliation{\CfES}
\email{ben.sutlieff@roe.ac.uk}

\author[orcid=0000-0003-4909-256X,sname='Katie A. Crotts']{Katie A. Crotts}
\affiliation{\UoA}
\email{ktcrotts@arizona.edu}

\author[orcid=0000-0001-6396-8439, sname='William O. Balmer']{William O. Balmer}
\altaffiliation{51 Pegasi b Fellow}
\affiliation{\CIERA}
\email{wbalmer1@jhu.edu}

\author[orcid=0009-0000-0303-2145, sname='Rapha\"el Bendahan-West']{Rapha\"el Bendahan-West}
\affiliation{Department of Physics and Astronomy, University of Exeter, Stocker Road, Exeter EX4 4QL, UK}
\email{rb941@exeter.ac.uk}

\author[orcid=0000-0002-7325-5990,sname='Ellis Bogat']{Ellis Bogat}
\affiliation{\LAM}
\email{ellis.bogat@lam.fr}

\author[orcid=0009-0001-0275-7811,sname='Evelyn L. Bruinsma']{Evelyn L. Bruinsma}
\affiliation{\JHU}
\affiliation{\STScI}
\email{ebruins1@jhu.edu}

\author[orcid=0009-0005-9339-2369,sname='Xueqing Chen']{Xueqing Chen}
\affiliation{\IfA}
\email{xueqing.chen@ed.ac.uk}

\author[orcid=0000-0003-4557-414X, sname='Kyle Franson']{Kyle Franson}
\affiliation{\UCSC}
\email{kfranson@ucsc.edu}

\author[orcid=0000-0001-8627-0404,sname='Julien H. Girard']{Julien H. Girard}
\affiliation{\STScI}
\email{jgirard@stsci.edu}

\author[orcid=0000-0001-8074-2562, sname='Sasha Hinkley']{Sasha Hinkley}
\affiliation{University of Exeter, Physics Building, Stocker Road, Exeter EX4 4QL, UK}
\email{S.Hinkley@exeter.ac.uk}

\author[orcid=0009-0005-6943-6819, sname='Andrew D. James']{Andrew D. James}
\affiliation{Department of Physics and Astronomy, University of Exeter, Stocker Road, Exeter EX4 4QL, UK}
\email{adj207@exeter.ac.uk}

\author[orcid=0000-0003-2769-0438, sname='Jens Kammerer']{Jens Kammerer}
\affiliation{European Southern Observatory, Karl-Schwarzschild-Straße 2, 85748 Garching, Germany}
\email{jkammere@eso.org}

\author[orcid=0009-0009-1074-3696, sname='Rohan Kane']{Rohan Kane}
\affiliation{Space Telescope Science Institute, 3700 San Martin Dr, Baltimore, MD 21218, USA}
\email{rohan.kane22@gmail.com}

\author[orcid=0000-0002-6964-8732,sname='Kellen Lawson']{Kellen Lawson}
\affiliation{\CSST}
\affiliation{\ASD}
\affiliation{\CRESST}
\email{kellen.d.lawson@nasa.gov}

\author[orcid=0000-0002-0834-6140, sname='Jarron M. Leisenring']{Jarron M. Leisenring}
\affiliation{Department of Astronomy / Steward Observatory, University of Arizona, Tucson, AZ 85721, USA}
\email{jarronl@arizona.edu}

\author[orcid=0000-0003-1227-3084,sname='Michael Meyer']{Michael Meyer}
\affiliation{\UoM}
\email{mrmeyer@umich.edu}

\author[orcid=0000-0001-6205-9233, sname='Maxwell A. Millar-Blanchaer']{Maxwell A. Millar-Blanchaer}
\affiliation{Department of Physics, University of California, Santa Barbara, CA 93106, USA}
\email{maxmb@ucsb.edu}

\author[orcid=0000-0001-5053-2660, sname='Skyler Palatnick']{Skyler Palatnick}
\affiliation{Department of Physics, University of California, Santa Barbara, CA 93106, USA}
\email{skylerpalatnick@ucsb.edu}

\author[orcid=0009-0006-3044-9011, sname='Mia Belle Parkinson']{Mia Belle Parkinson}
\affiliation{UK Centre for Astrobiology, School of Physics and Astronomy, University of Edinburgh, King's Buildings, Edinburgh EH9 3FD}
\email{m.b.parkinson@sms.ed.ac.uk}

\author[orcid=0000-0003-3818-408X, sname='Laurent Pueyo']{Laurent Pueyo}
\affiliation{Space Telescope Science Institute, 3700 San Martin Dr, Baltimore, MD 21218, USA}
\email{pueyo@stsci.edu}

\author[orcid=0000-0003-4203-9715,sname='Emily Rickman']{Emily Rickman}
\affiliation{\ESA}
\email{erickman@stsci.edu}

\author[orcid=0000-0002-9372-5056,sname='Nour Skaf']{Nour Skaf}
\affiliation{\UH}
\email{nskaf@hawaii.edu}

\author[orcid=0000-0001-6098-3924,sname='Andy Skemer']{Andy Skemer}
\affiliation{\UCSC}
\email{askemer@ucsc.edu}

\author[orcid=0009-0008-2252-7969,sname='Klaus Subbotina Stephenson']{Klaus Subbotina Stephenson}
\affiliation{\UVic}
\affiliation{\NRC}
\email{klausss@uvic.ca}

\author[orcid=0000-0002-4388-6417, sname='Isabel Rebollido']{Isabel Rebollido}
\affiliation{Centro de Astrobiolog\'ia (CAB, CSIC-INTA), Camino Bajo del Castillo s/n, 28692 Villanueva de la Ca\~nada, Madrid, Spain}
\email{irebollido@cab.inta-csic.es}

\author[orcid=0000-0003-0774-6502, sname='Jason Wang']{Jason Wang}
\affiliation{Center for Interdisciplinary Exploration and Research in Astrophysics (CIERA) and Department of Physics and Astronomy, Northwestern University, Evanston, IL, 60208, USA}
\email{jason.wang@northwestern.edu}
 
\begin{abstract}
We present a sensitivity analysis of three recent JWST direct imaging programs (GO-4050, GO-5835, and SURVEY-6005) designed to build a demographic study of sub-Jupiter mass ($0.1 - 1$ M$_{\mathrm{J}}$) exoplanets in nearby young associations and moving groups. 
We find that these programs are sensitive to sub-Saturn to Jupiter mass exoplanets with effective temperature (T$_\mathrm{eff}$) as low as $\lesssim 250$ K, already at $\sim  2$ arcsec.
A comparison with the SHINE survey confirms that the improvements in mass sensitivity provided by JWST high-contrast imaging surveys outpace current ground observation 
for masses $\lesssim 1$ M$_{\mathrm{J}}$ and separations between $\sim20-500$ au, opening up a novel parameter space to explore.
Overall, this analysis provides a robust framework for interpreting future exoplanet detections. By correlating photometrically derived properties (e.g. mass, $T_{\mathrm{eff}}$, and separation) with theoretical model predictions, we show how the access to potentially low detectable masses (sub-Saturn to Jupiter) strongly depends on the expected atmospheric chemistry of the exoplanet population.

\end{abstract}



\section{Intro} 
Despite the continuous effort from ground-based surveys \cite[e.g.;][]{Bowler2016,Niel19,Fern19,Vigan2021,Desidera2021,Fult21,Squicciarini2025_2} and the known exoplanet population rapidly growing every year, empirical constraints on the properties and occurrence rates of low-mass ($\leq 1 M_\mathrm{Jup}$) wide separation ($\gtrsim 10$ au) giant exoplanets do not exist. Transiting surveys probe the exoplanet population from rocky super-Earths to gaseous super-Jupiters, but are mostly limited to within 1~au separations. Observations through radial velocity (RV) and microlensing techniques push the known population of exoplanets out to $\lesssim$10~au separations, but cannot perform follow-up atmospheric characterization. Astrometry can also provides evidence for the presence of exoplanets at closer separation ($\lesssim$10~au), but solutions at  wider separation are still mostly limited to Brown Dwarfs or more massive companions \citep{Thompson2026}.

While Gaia DR4 will improve these results,
to be able to detect \textit{and} characterize exoplanets beyond $\sim$10~au, we still need to turn to direct imaging techniques. 
Near- to mid-infrared wavelengths serve as the sweet spot for direct imaging observations.
Since young planets retain much of the heat of their primordial formation, and their spectral energy distributions (SEDs) reach their peak, making them relatively warm and luminous at these wavelengths. At the same time, the corresponding SEDs of stellar hosts gradually diminish, and the star-planet contrasts are at a minimum \citep{Pathak2021,Hinkley2022}. However, the vast majority of exoplanet imaging observations have so far been performed from the ground where the absorption of Earth's atmosphere limit our exploration of infrared wavelengths, and those wavelengths that are accessible are strongly affected by the thermal background. 
In the recent years $\mathrm{JWST}$ has changed this paradigm, providing access with direct imaging to a dramatically expanded observable parameter space, as new discoveries of both old and young, cold, exoplanets  recently demonstrated 
($\sim275$ K, $\sim3.5$ Gyr, $\sim6.3$ $M_\mathrm{Jup}$ \citealt{Matthews2024};
$\sim300$ K, $\sim4.5$ Gyr, $\sim6$ $M_\mathrm{Jup}$ \citealt{Bardalez2025}; $\sim275$ K, $\sim10$ Myr, $\sim0.2-1$ $M_\mathrm{Sat}$ \citealt{Lagrange2025,Crotts2025}), thanks to its outstanding performance in both infrared sensitivity and wavelength coverage.

Microlensing surveys \citep{Suzu16} show evidence  for an increase of occurrence rates of 0.05$-$1.6~$M_\mathrm{Jup}$ exoplanets out to 10~au. This is also supported by RV surveys \citep[e.g.;][]{Witt20} that show no drop-off in the occurrence of $>0.3$ $M_\mathrm{Jup}$ exoplanets out to 10~au. In addition, ALMA observations of substructure in protoplanetary disks seems to suggest occurrence rates of planets between $0.1 - 1$ $M_\mathrm{Jup}$ consistent with results from microlensing surveys \cite{Zhang2023}. 
High-precision RV measurements of FGKM stars by \citet{Fult21} seems to indicate a peak at $2-5$ au, and a drop off in the occurrence rate of Jupiter-like exoplanets beyond that. However, if we separate the population of exoplanets into their sub-Jupiter and super-Jupiter components ($\sim 1 - 20$ M$_{\mathrm{J}}$), the picture becomes less clear. When looking at the super-Jupiter population in isolation, the significance of a drop-off at 10~au is much less apparent, and the population seems to plateau. 
Conversely, the sub-Jupiter population is only measured out to 10 au. Thus it is not clear if the frequency of sub-Jupiter exoplanets plateau or rise past 10 au, or if sub-Jupiter exoplanets are more or less common than super-Jupiter beyond 10 au.

To facilitate answering these questions, 
the JWST Sub-Jupiters Survey Collaboration was established.
Three JWST programs have been performed in the last few years (GO-4050 and GO-5835: \citealt{2023jwst.prop.4050C,2024jwst.prop.5835C}; and SURVEY-6005: \citealt{2025jwst.prop.6005B}) to build a demographic study of sub-Jupiter mass ($0.1 - 1$ M$_{\mathrm{J}}$) exoplanets in nearby young associations and moving groups (loose, low-density groups of coeval stars). These coeval structures represent one of the best possible environments to search for planetary companions, since being part of an association helps to constrain both the age and the distance of their constituent members \citep{Sode14,Gagne2024}. Additionally, closer distances provides better access to smaller physical separations, and younger ages provide access to lower mass companions. These programs have already yielded their first results, with the very recent detection of RX J0534.0-0221 b \cite[a $\sim 2.8$ M$_\mathrm{J}$ planet at $\sim 14$~au;][]{Chavez2026}, and the scheduling of 35 new exoplanet candidate follow-ups for Cycle 5 \citep[GO-10764;][]{2026jwst.prop.10764}.

In this paper we discuss the results from the sensitivity analysis performed on the three JWST programs (GO-4050, GO-5835 and SURVEY-6005).
Section \ref{sec: Programs design} motivates target selection and observation strategy for the 3 JWST programs.
In Section \ref{sec: Analysis}, we describe the data reduction and analysis techniques used to identify point source candidates and measure sensitivity curves. 
In Section \ref{subsec: Contrast curves} we discuss how we derive the calibrated contrast curves, while in Section \ref{subsec: Contrast curves} we translate these contrast curves into magnitude limits. 
Finally, in Section \ref{sec: Discussion} we discuss the survey mass/T$_{\mathrm{eff}}$ sensitivity limits, the detection probability maps, and the expected yield from our three JWST programs combined, before presenting our conclusions in Section \ref{sec: Conclusion}.

\begin{deluxetable*}{cccccccccc}[!th]
\tablecaption{
Survey summary showing the respective GO/SURVEY ID, number of targets, dedicated reference stars, misaligned targets, filter employed, the average charged time per visit, and the beginning and end of the observations.
\label{tab: Surveys summary}}
\tablehead{\colhead{ID} &\colhead{Targets} &\colhead{References} & \colhead{Misaligned} & \colhead{Filters} & \colhead{Hours} & \colhead{Start} & \colhead{End}}
\startdata
4050 & 23 & 7 & 0 & F200W, F444W & 1.58 & Jun 6, 2024 & Jun 16, 2024\\
5835 & 56 & 0 & 0 & F200W, F210M, F444W & 1.86 & Sep 17, 2024 & Sep 20, 2025\\
6005 & 86 & 0 & 10 & F200W, F444W & 1.87 & Aug 20, 2024 & Jul 5, 2025\\
\hline
TOTAL & 165 & 7 & 10 & 3 & 331.08 & & & \\
\enddata
\end{deluxetable*}


\begin{deluxetable}{lcccccc}[!t]
\tablecaption{Targets summary showing for each GO/SURVEY ID, the number of isolated targets, binaries, isolated targets with disks, or both observed per program, and their respective totals.\label{tab: Targets types}}
\setlength{\tabcolsep}{4pt}          
\renewcommand{\arraystretch}{0.95}   
\tablewidth{\columnwidth} 
\tablehead{\colhead{ID} & \colhead{Singles} & \colhead{Binaries} & \colhead{Disks} & \colhead{Binaries/Disks} & \colhead{Total}}
\startdata
4050 & 15 & 3 & 4 & 1 & 23 \\
5835 & 32 & 16 & 6 & 2 & 56 \\
6005 & 54 & 26 & 5 & 1 & 86 \\
\hline
TOTAL & 101 & 45 & 15 & 4 & 165 \\
\enddata
\end{deluxetable}

\section{Program design} 
\label{sec: Programs design}
Between June 2024 and September 2025, three JWST NIRCam programs have been completed, targeting a total of 182 targets among nearby young moving groups and associations. 
A complete summary of their respective observation strategies is presented in Table \ref{tab: Surveys summary}. 
GO-4050 observed a total of 23 targets plus 7 dedicated reference stars using a coronagraphic setup in filters F200W+MASK335R and F444W+MASK335R.
GO-5835 had a total of 45 coronagraphic targets observed in filters F200W+MASK335R and F444W+MASK335R and 11 imaging targets observed in filters F210M and F444W. SURVEY-6005 had a total of 96 coronagraphic targets observed in filters F200W+MASK335R and F444W+MASK335R, of which 10 where later rejected due to coronagraphic misalignment following acquisition. 
We include all of these targets in our analysis, with the exception of the 7 reference targets from GO-4050, and the 10 misaligned targets from SURVEY-6005.
This brings the total number of available targets for the analysis to 165.

Table \ref{tab: Targets types} shows a breakdown of our targets by their type: if they are either flagged as isolated, hosting a binary or a disk, or both.

\begin{deluxetable}{lccccccccc}[!t]
\tablecaption{Target summary showing for each assigned membership, the number of targets per spectral type, the total number of targets observed in the group, the age and uncertainties, and the median distance of our targets in that group, with uncertainties represented by the \textit{16th/84th} percentiles when more than one target is in the group. Memberships, spectral types, ages and distances for each target are obtained from \texttt{MOCAdb}.
\label{tab: Targets summary}}
\tabletypesize{\scriptsize}          
\setlength{\tabcolsep}{4pt}          
\renewcommand{\arraystretch}{0.95}   
\tablewidth{\columnwidth}            
\tablehead{\colhead{Membership} &\colhead{A} & \colhead{F} & \colhead{G} & \colhead{K} & \colhead{M} & \colhead{L} & \colhead{Total} & \colhead{Age} & \colhead{Distance}\\
\colhead{} &\colhead{} & \colhead{} & \colhead{} & \colhead{} & \colhead{} & \colhead{} & \colhead{} & \colhead{\textit{Myr}} & \colhead{\textit{pc}}}
\startdata
ABDMG & 0 & 4 & 3 & 8 & 0 & 1 & 16 & 133$^{+15}_{-10}$ & 37$^{+17}_{-13}$ \\
AUSC & 0 & 0 & 0 & 1 & 2 & 1 & 4 & 342$\pm318$ & 48$^{+7}_{-3}$ \\
BPMG & 0 & 2 & 1 & 4 & 36 & 0 & 43 & 26$\pm3$ & 37$^{+14}_{-9}$ \\
CAR & 0 & 0 & 1 & 3 & 0 & 0 & 4 & 33$^{+2}_{-1}$ & 57$^{+4}_{-4}$ \\
CARN & 0 & 0 & 1 & 2 & 0 & 1 & 4 & 200$\pm50$ & 31$^{+1}_{-6}$ \\
COL & 0 & 3 & 3 & 0 & 4 & 0 & 10 & 33$^{+2}_{-1}$ & 53$^{+9}_{-3}$ \\
FCCN & 0 & 0 & 0 & 1 & 0 & 0 & 1 & 29$\pm4$ & 46 \\
FIELD & 0 & 0 & 1 & 2 & 4 & 0 & 7 & 11200$\pm750$ & 34$^{+18}_{-11}$ \\
MEL4 & 0 & 0 & 0 & 1 & 0 & 0 & 1 & 24$\pm1$ & 109 \\
OCTN & 0 & 0 & 0 & 0 & 1 & 0 & 1 & 55$^{+45}_{-25}$ & 52 \\
SCOCEN & 0 & 0 & 0 & 1 & 0 & 0 & 1 & 15$^{+18}_{-15}$ & 136 \\
THA & 0 & 6 & 8 & 11 & 3 & 0 & 28 & 40$\pm3$ & 46$^{+9}_{-2}$ \\
TWA & 2 & 0 & 0 & 3 & 19 & 0 & 24 & 10$\pm3$ & 58$^{+16}_{-11}$ \\
UMA & 0 & 5 & 9 & 6 & 0 & 0 & 20 & 414$\pm23$ & 24$^{+7}_{-7}$ \\
VUL & 0 & 1 & 0 & 0 & 0 & 0 & 1 & 162$\pm30$ & 32 \\
\hline
TOTAL & 2 & 21 & 27 & 43 & 69 & 3 & 165 &  &  \\
\enddata
\end{deluxetable}

\begin{deluxetable}{lccccccc}[!t]
\tablecaption{Target summary showing for each GO/SURVEY ID, the number of targets observed per spectral type and the total number of targets in the program.\label{tab: GO summary}}
\tablewidth{\columnwidth}            
\tablehead{\colhead{ID} &\colhead{A} & \colhead{F} & \colhead{G} & \colhead{K} & \colhead{M} & \colhead{L} & \colhead{Total}}
\startdata
4050 & 2 & 0 & 0 & 2 & 19 & 0 & 23 \\
5835 & 0 & 0 & 1 & 3 & 49 & 3 & 56 \\
6005 & 0 & 21 & 26 & 38 & 1 & 0 & 86 \\
\hline
TOTAL & 2 & 21 & 27 & 43 & 69 & 3 & 165 \\
\enddata
\end{deluxetable}


\subsection{Target selection}
\label{subsec: Target selection}
We targeted a range of different associations and moving groups (henceforth ``group") as part of our three programs.  
Table \ref{tab: Targets summary} shows a breakdown of the targets by spectral type and group, with the age,  the median distance, and relative uncertainties, for our targets in that group, as reported by \texttt{MOCAdb}\footnote{\url{https://mocadb.ca/}} \citep{Gagne18a,Gagne2024}: the Beta Pictoris moving group (BPMG: 43 targets, or $\sim 26 \%$ of the total sources), the Tucana-Horologium association (THA: 28 targets, $\sim 17 \%$), TW Hya association (TWA: 24 targets, $\sim 14.5 \%$), Ursa Major association (UMA: 20 targets, $\sim 12 \%$), and the AB Doradus moving group (ABDMG: 16 targets, $\sim 10 \%$), plus a few additional groups with very few targets each: Austral complex (AUSC: 4 targets, $\sim 2.4 \%$), Carina association (CAR: 4 targets, $\sim 2.4 \%$), Carina-Near Moving Group (CARN: 4 targets, $\sim 2.4 \%$), Columba association (COL: 10 targets, $\sim 6 \%$), FORH-Carina-Columba node of the Austral Complex (FCCN: 1 targets, $\lesssim 1 \%$), Melange-4 moving group (MEL4: 1 targets, $\lesssim 1 \%$), Octans association (OCTN: 1 targets, $\lesssim 1 \%$), Scorpio-Centaurus region (SCOCEN: 1 targets, $\lesssim 1 \%$), Vulpecula system (VUL: 1 targets, $\lesssim 1 \%$), and field stars (FIELD: 7 targets, $\sim 4.2 \%$).
GO-4050 targeted primarily M stars in TWA association, while GO-5835 targeted primarily M stars in BPMG moving group.  In contrast, SURVEY-6005 was a survey-mode program, designed to fill gaps in the JWST observing schedule, and hence observed FGK stars across several groups that were well-spread across the sky.  SURVEY-6005 originally targeted members of BPMG, ABDMG, THA, COL, CAR, CARN, and UMA, however, updated group memberships have shifted SURVEY-6005 targets into a wider range of groups. 

A complete breakdown of how many targets have been investigated for each program divided by spectral type is shown in Table \ref{tab: GO summary}.
When divided by spectral types, L stars constitute roughly $\sim 2 \%$ of all the targets, M $\sim 42 \%$, K $\sim 26 \%$,  G and F stars comprising about $\sim 16 \%$ and $\sim13\%$ respectively, and the rest are A type stars ($\sim1\%$).

Finally, a complete description of each target is also presented in Appendix \ref{apx: The target catalog}.

\subsection{Filter selection}
\label{subsec: Filter selection}

NIRCam and MIRI coronagraphic simulations were employed using \texttt{PanCAKE} \citep[an extension of the official JWST exposure time calculator;][]{Carter2021} to identify the most sensitive filters to sub-Jupiter mass companions \citep[assuming clear and near solar metallicity atmospheric models from][]{linder19}. NIRCam filter F444W has been selected as the primary filter for this purpose, supported by the NIRCam F200W (for coronagraphy) and the F210M filter (for direct imaging; only for GO-5835), which provides the best mass sensitivity between the available short wavelength filters. One very important advantage coming from this set-up is that the combination of F200W+F444W (or F210M+F444W for direct imaging) filters enables a preliminary screening of our candidates based on their colors, since exoplanets are fainter in the F200W/F210M compared to F444W, whereas background stars will be brighter.

\subsection{Observing strategy}
\label{subsec: Observing strategy}

We have performed NIRCam dual-band coronagraphic and imaging observations of a total of 165 targets (excluding references and misaligned). 
All coronagraphic observations have been taken using the \texttt{MASK335R} coronagraph mask (inner working angle: $0.82$"\footnote{\url{https://jwst-docs.stsci.edu/jwst-near-infrared-camera/nircam-instrumentation/nircam-filters-for-coronagraphy\#gsc.tab=0}}), in the \texttt{SUB320A335R} sub-array, while for the imaging observations we have used the \texttt{SUBARRAY\_DITHER} as the primary dither pattern with $4$ Primary Dithers and the \texttt{SUB160P} sub-array. Each target exposure was set to approximately 30 minutes, and the specific readout parameters for any given observation were obtained and optimized using \texttt{PanCAKE} simulations \citep{Carter2021}. 

While for GO-4050 we performed dedicated reference target observations, with 7 references taken, one for each imaging sequence, for GO-5835 and SURVEY-6005, we employed a self-referencing strategy to facilitate point spread function (PSF) subtraction through reference differential imaging \citep[RDI -][]{Lafreniere2009,Ruane2019}. 
The goal behind the similar setup for all three surveys is to produce a homogeneous PSF library for all data, as will be explained in more detail in Section \ref{subsec: PSF-Library}.

\section{Analysis}
\label{sec: Analysis}

\subsection{Data reduction}
\label{subsec: Data reduction}

All data were reduced using the \texttt{spaceKLIP}\footnote{\url{https://github.com/spacetelescope/spaceKLIP}} pipeline \citep{Girard2022, Kammerer2022,Carter2023}. \texttt{spaceKLIP} is a Python-based data reduction and analysis toolkit for JWST high-contrast imaging and now also non-coronagraphic imaging data, designed to apply a specific sets of tools to transform ``raw'' uncalibrated data, to the final product suitable for science. 
\begin{figure*}[!t]
    \centering
\includegraphics[width=\textwidth]{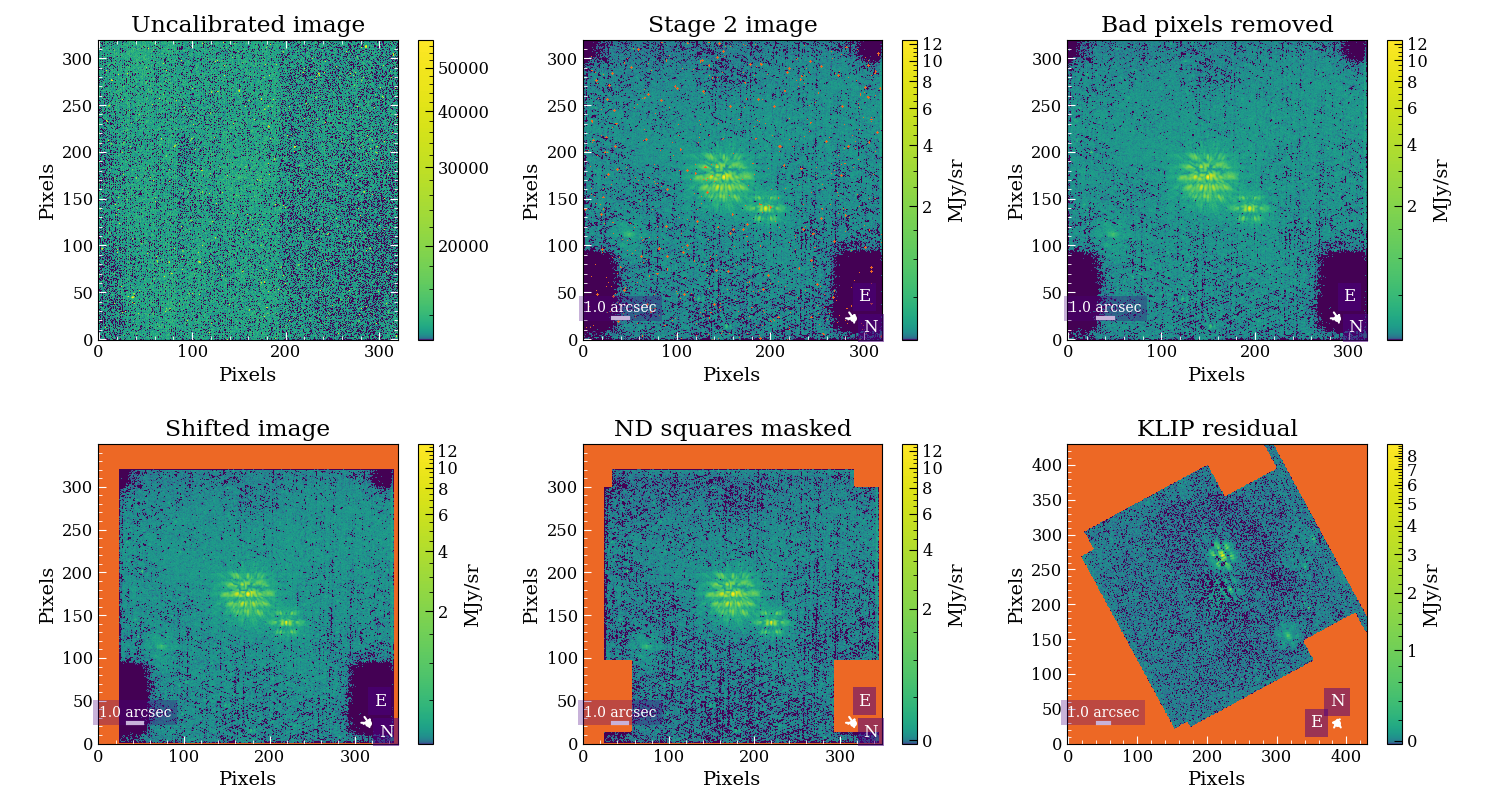}
    \caption{Example of results from the data reduction steps for the coronagraphic dataset. Top row, from left to right: the original \textit{uncal} data, as downloaded by MAST, \textit{detector calibrated calints} data, and \textit{instrument calibrated} data (after \textit{bad-pixels} have been identified and removed by the pipeline). 
    Bottom row, from left to right: \textit{pre-PSF subtraction shifted frame}, additional padding has been added to the frame to allow for the shifting of the images; \textit{ND squares masked frame}, to remove their impact on the sensitivity on overlapping areas; \textit{post-PSF subtraction residual}. Now the sources are clearly visible in the derotated residuals, comprising one close-in (background) source not entirely visible before.
    }
    \label{fig: data reduction}
\end{figure*}
Figure \ref{fig: data reduction} summarize the major steps of the pipeline and their outputs for the case of a coronagraphic target. The full description for this process is presented in Ferrer-Chavez et al, submitted.  The same steps have been followed for the imaging dataset as well excluding the ND squares masking, since they are not present in the imaging data.

The raw JWST data (exposure raw data products designated by a file name suffix of “uncal” that usually contain only the raw detector pixel values from an exposure - left panel, top row, in Figure \ref{fig: data reduction}) are read into \texttt{spaceKLIP} and processed using the  stages 1 and 2 from the \texttt{jwst} pipeline.
During this phase, 
the pipeline account for dark subtraction, photometric calibration, saturation, non-linearity, ramp-fitting, background subtraction and flat-fielding, transforming the raw uncalibrated data into flux-calibrated \textit{cal/calints} images (central panel, top row, in Figure \ref{fig: data reduction}).
Once the images are calibrated, tools from the \texttt{ImageTools} module can be used to further process the \textit{cal/calints} images and prepare them for PSF subtraction 
: i.e. to identify and clean bad pixels and apply  median-subtraction - right panel, top row, in Figure \ref{fig: data reduction};  properly recenter and align all the images to a common center - left panel, bottom row, in Figure \ref{fig: data reduction};  mask the four Neutral Density (ND) squares present in NIRCam coronagraphic images - central panel, bottom row, in Figure \ref{fig: data reduction}. ND squares are used to perform target acquisition for NIRCam coronagraphy  without saturating the detector. 
Because they heavily attenuate the flux in these regions, they must be masked before PSF subtraction to prevent artificial suppression during multi-image dither or rotation combinations.

\subsection{RDI PSF subtraction}
\label{subsec: PSF subtraction}
We performed PSF subtraction using the \texttt{spaceKLIP} implementation of \texttt{pyKLIP}\footnote{\url{https://pyklip.readthedocs.io/en/latest/}} \citep{Wang2015}, a python library for direct imaging of exoplanets and disks. \texttt{pyKLIP} uses an implementation of Karhunen-Loeve Image Processing \citep[KLIP; ][]{Soummer2012} algorithms based off of principal component analysis to model and subtract off the stellar PSF, and KLIP-Forward Modeling \citep[FM -][]{Pueyo2016} to perform the calibration of the contrast curves and companion fitting. 
Since \texttt{pyKLIP} requires a reference PSF library to perform RDI based PSF subtraction,
for this project, we created two curated PSF libraries: one for the coronagraphic dataset and one for the imaging dataset that will be discussed in more detail in the next section \ref{subsec: PSF-Library}. 
During this process, the science image can also be separated into a number of annuli and position-angle subsections. This allows us to perform the PSF subtraction independently in multiple subsections of the image, helping avoiding biases affecting the PSF subtraction quality.

An additional parameter to fine tune is the number of KL modes (i.e., principal components) used by the final PSF model for subtraction. One can either select a specific KL mode or allow \texttt{spaceKLIP} to use multiple modes, providing a cube of residual images after PSF subtraction, where each layer corresponds to removal of a different KL mode. This second approach has two major advantages. First it  allows us to pick the best residual for each astronomical source. Since removing more modes (up to higher KL modes) is more aggressive at subtracting the PSF closer to the center of the star, having access to subtractions removing differing numbers of KL modes allows us to pick the best residual for each candidate, depending on its separation. Second, it allows to detect false astronomical source that can be injected by reference stars in the science residual image. This usually happen when some of the reference stars used to build the PSF library have astrophysical sources in their field-of-view. These sources can end up injected in residual images after subtracting out KL modes that have inherited these sources, and usually appear and disappear between subtracting different KL modes, as the spurious sources are present in specific KL mode images.  In contrast, a true astrophysical source will remain present as more and more KL modes are subtracted. 
These injected sources are usually first detected in these residuals and then tracked down and masked in their original frames when the final PSF-library is assembled for subtraction. For a more in depth analysis of the KLIP parameters that were adopted to reduce our dataset, we refer the reader to Ferrer-Chavez et al, submitted.

The right panel, bottom row, of Figure \ref{fig: data reduction} shows the typical outcome of this subtraction step, where candidate sources clearly appear in the residuals post KLIP subtraction. 

\subsection{Building A PSF Library for Robust RDI}
\label{subsec: PSF-Library}
For this project we assembled two curated PSF libraries built from sources drawn from the entire programs sample (i.e. self-referencing strategy): one for NIRCam coronagraphy, where we combined the F200W and F444W coronagraphic data from the three programs in a unique RDI PSF Library; and one for NIRCam imaging, where we only utilized F210M and F444W data from the GO-5835 program, since GO-4050 and SURVEY-6005 do not include any imaging targets.

\subsubsection{Coronagraphic PSF library}

All coronagraphic observations from 
programs GO-4050, GO-5835 and SURVEY-6005 are compiled into a single, large reference PSF library. 
Since the coronagraphic PSF library data reduction and performance characterization are laid out in more detail in a separate publication (Ferrer-Chavez et al. submitted), here we will provide the reader with a summary of the process.

A large PSF library provides useful PSF diversity (from variations in stellar spectral type, wavefront error realizations and precise coronagraph alignment) that can enhance the contrast achieved following PSF subtraction. 
However, non-PSF astrophysical signals present in the observations of the library (such as background stars or galaxies, close stellar binaries or bright debris disks) can contaminate the PSF subtracted image, in some cases even introducing false positives resembling planetary companions. For this reason, we visually inspect in detail all observations from the programs to remove from the
PSF library 
any source of contamination. To do so, we first excluded all the
observations with any contamination at close separations (within $\sim3$ arcseconds), and observations with exceedingly bright off-axis sources (such as bright foreground stars). These are contaminants that are bright enough to affect large portions of the image, and so they are difficult to properly mask.
The remaining 
observations are included in 
a preliminary PSF library 
used to perform a preliminary PSF subtraction.
The output of this process is then visually inspected to identify
any remaining faint contaminants (see next section \ref{subsec: Off-axis sources masking} for an example of this process).
We clean these contaminants by masking a circular region centered at the source, and replacing those pixels with pixel-wise median across a set of contaminant-free stars similar to them (weighted by the flux of the star, to preserve PSF structure as much as possible). 

\subsubsection{Imaging PSF library} 
Similar to the coronagraphic case, to ensure that each science image could also serve as a reference PSF in the imaging PSF library (i.e., free of contaminating off-axis sources), we performed the same additional step of cleaning each reference of any contaminants.

Additionally, some specific targets have been completely removed from the list of imaging PSF references due to the presence of a bright off-axis companions that could not be efficiently masked, specifically: 2MASS-J04231498-1533245, 2MASS-J02495639-0557352, and 2MASS-J03550477-1032415.
This left us with only 8 targets available as references in the imaging PSF library. 

Identification of astrophysical objects in the final residual images
\label{subsec: Off-axis sources masking}
Once the reduction and subtraction phases have been completed for all 165 scientific targets, the next step is to inspect each residual image for the presence of candidates sources in both the short and long wavelength images that may impact our sensitivity estimates. To be able to reliably inspect all the available data, we use an internal web application, built using \texttt{Flask} and utilizing js9 \citep[a Javascript implementation of SAOimage ds9,][]{SAOImage_DS9} that allows multiple users to visualize and interact with FITS images, and mark the positions of visually identified candidate sources.

As observations were taken we organized multiple campaigns where members of the team visually inspected and annotated sources present in the final residual images. To ensure sources are not duplicated based on the uncertainty in the by-hand marking process, any marked coordinates within 4 pixels of each other are median-combined to produce a single source position. This method has drawbacks for precise candidate identification, as it is certainly feasible for two sources to be within 4 pixels of each other, but is nevertheless sufficient for masking purposes. Due to the qualitative nature of visual inspection, it is possible that sources may have been missed by all annotators. However, with an average of $\sim 10$ annotators for each image, it is unlikely that any sources bright enough to impact our understanding of the sensitivity limits have been missed. With the final coordinates of source objects as annotated by team-members, we mask the object position in the residual image using a circular mask whit a size depending on the brightness of the source. This insures that these objects do not affect our estimation of the sensitivity limits of the star around which they are found.

\section{Results}
\subsection{Contrast curves}
\label{subsec: Contrast curves}
We constructed contrast curves to provide a preliminary measure of how well faint objects can be detected relative to the bright star under the coronagraph mask.
\texttt{spaceKLIP} provides two separate metrics as part of its \texttt{analysistools} package:  the \textit{raw} contrast curves and \textit{calibrated} contrast curves.

To compute a \textit{raw} contrast curve for each source, we first used \texttt{spaceKLIP} \texttt{starphot.get\_stellar\_magnitudes} tool that evaluates the source brightness and zero point flux for the relevant filter by fitting a scaled version of a stellar atmosphere model to available photometry\footnote{\url{http://drgmk.com/sdb/}} \citep{Yelverton2019}.
Then, the measured pixel brightness and the estimated source brightness are combined to determine the contrast at five times the standard deviation of the noise in a given annular region ($5\sigma$ contrast), accounting also for small sample statistics \citep{Mawet2014}. This threshold represents the faintest object that can be reliably detected above the noise. 

Coronagraphic dataset reduction requires also accounting for mask throughput \citep[i.e. the efficiency with which light passes through the coronagraph mask to the detector as a function of angular separation and wavelength ][]{Krist2009, Krist2010}. To correct for the spatial variations, \texttt{spaceKLIP} applies an instrumental throughput scaling factor to each location across the 2D image. At small separations, the mask significantly attenuates the host starlight but at the same time dims also any companion flux. This reduced throughput, paired with elevated residual stellar noise, steepens the contrast curve and degrades inner sensitivity limits. Conversely, at larger separations, the throughput approaches unity as mask suppression decreases, increasing sensitivity thresholds for faint companions. Because throughput also varies with wavelength, different filters produce distinct contrast curves depending on the coronagraph performance at those wavelengths.
\begin{figure}[!t]
    \centering
    \includegraphics[width=0.49\textwidth]{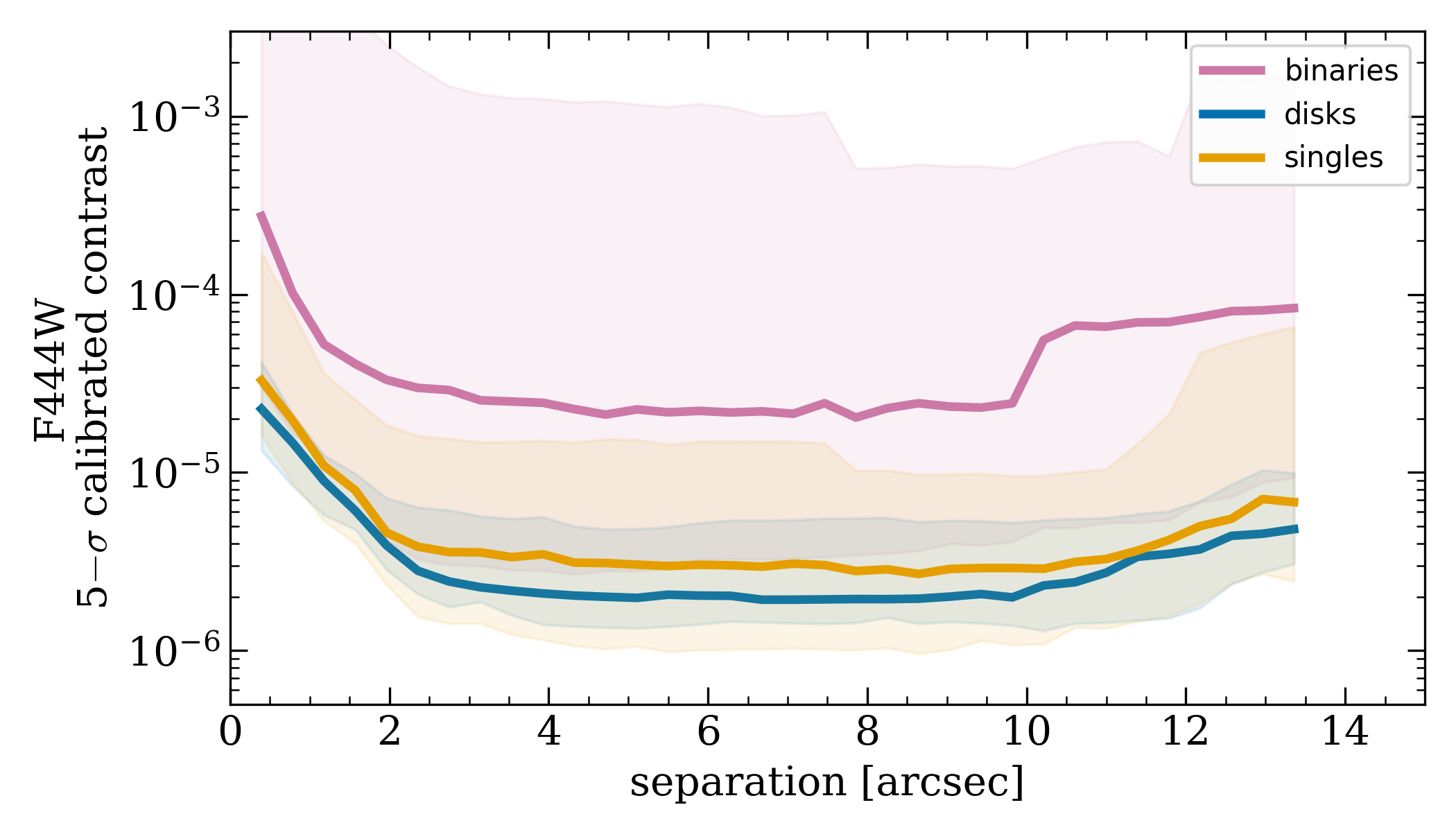}
    \caption{Calibrated contrast curve comparison between isolated sources (orange), single stars with disks (blue) and binaries without disks (magenta) in the GO-4050, GO-5835, and SURVEY-6005 surveys combined. To avoid confusion, the 4 targets with both binaries and disks were removed from this comparison}. The plot shows the impact that binaries have on the contrast sensitivity compared to isolated source, or isolated sources hosting disks.
    \label{fig: fkstn_calcon_binary_disk_F444W}
\end{figure}

\begin{figure*}[!t]
    \centering
    \includegraphics[width=0.45\textwidth]{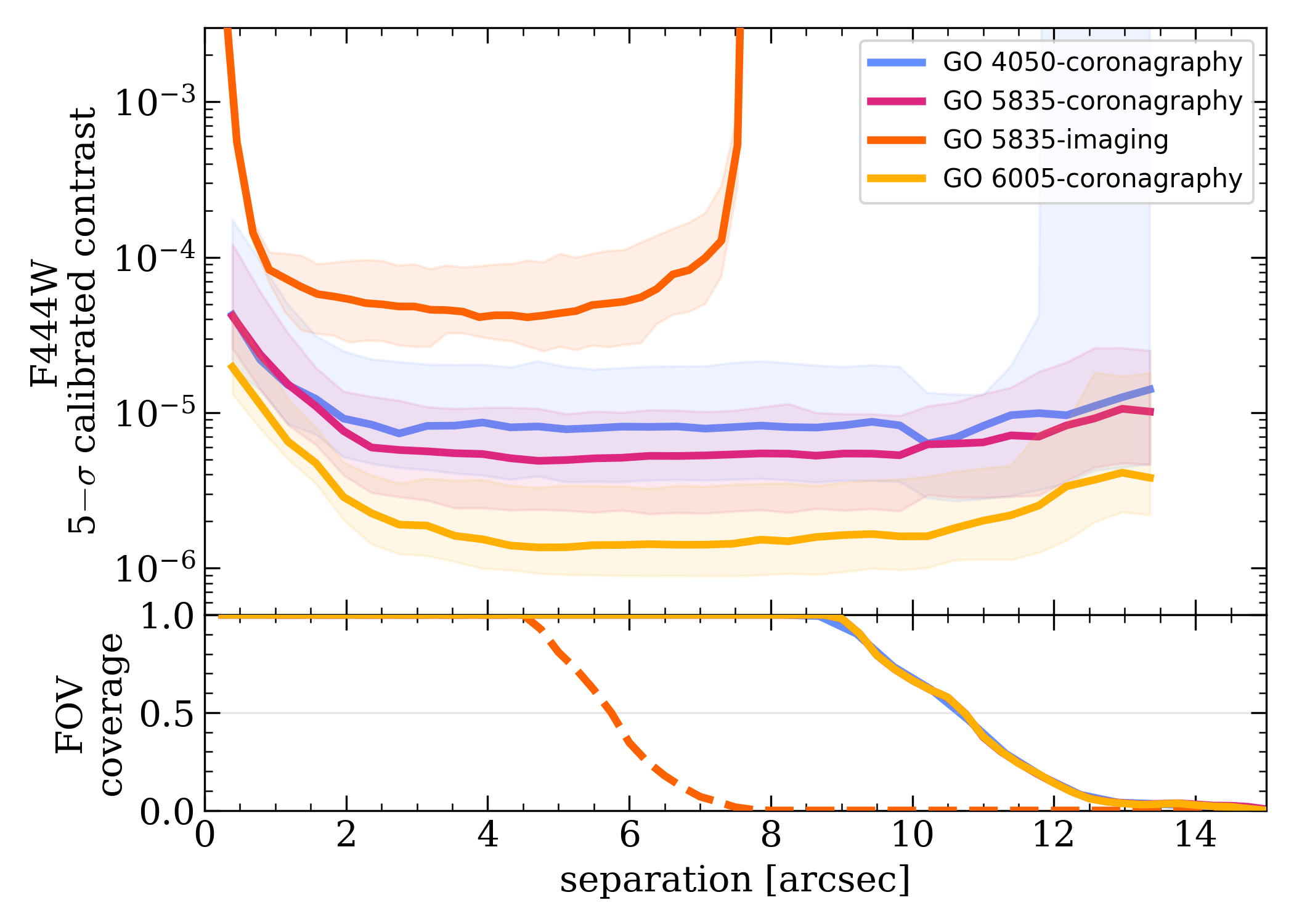}
    \includegraphics[width=0.45\textwidth]{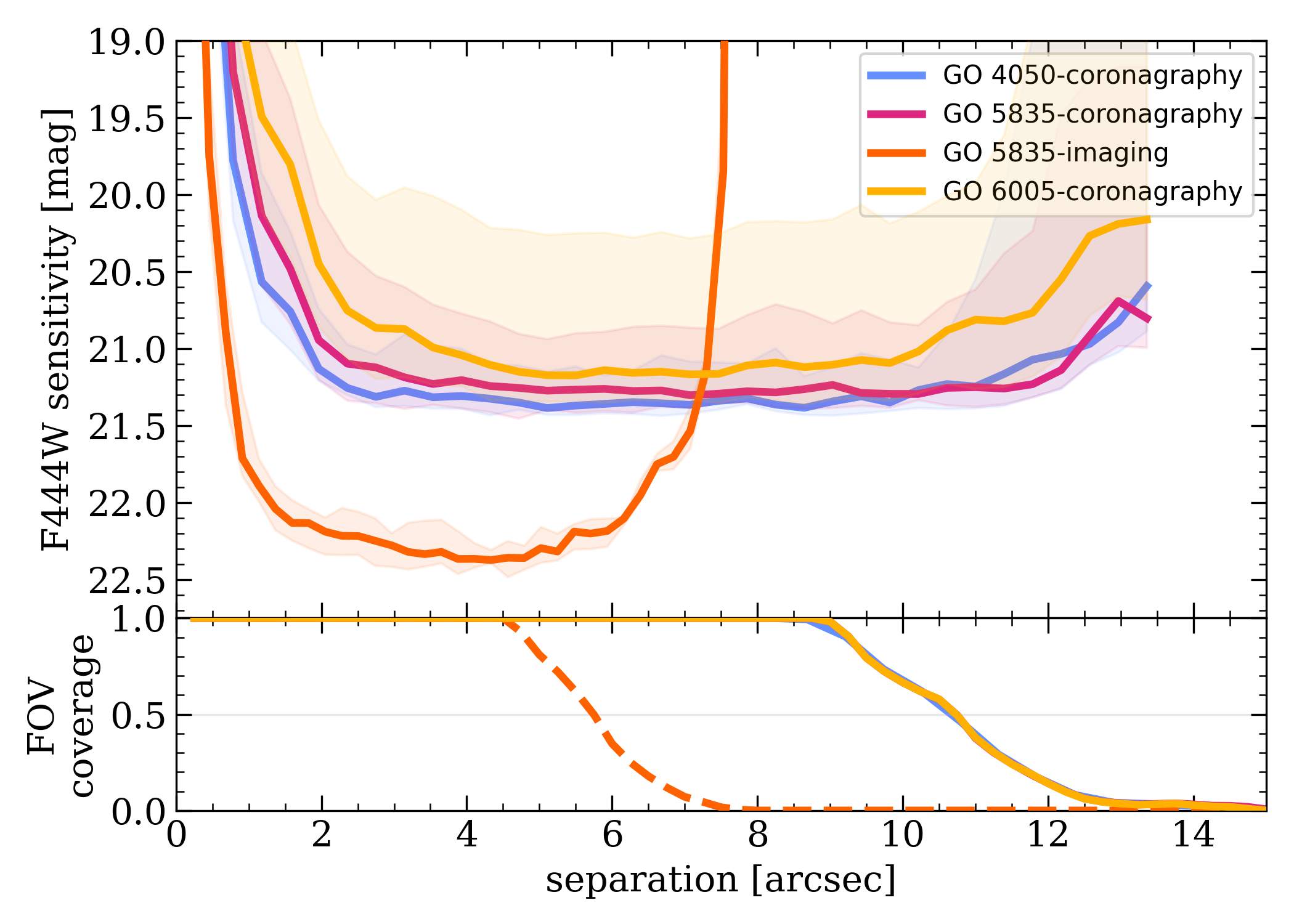}
    \caption{Left: Calibrated contrast curves for each program. Right: Magnitude limits for each program. 
    The thick continuous lines show the median of the distribution of curves, while the shaded areas cover the $16th/84th$ percentile for each distribution. The lower part of each plot shows the FOV coverage of the data (coronagraphic - solid line, imaging - dashed line) as a function of the distance from the center.
    The absolute magnitude sensitivity floor is governed by the instrument background noise rather than coronagraphic performance: GO-4050 and 5835 (mostly focused on M-dwarfs) exhibit poorer contrast  while converging to identical magnitudes limits as SURVEY-6005 (primarily FGK hosts).
    For the GO-5835 imaging component, the absence of a coronagraph increased the overall throughput and signal-to-noise ratio achieved for this faint M-dwarf sample.}
    \label{fig: Contrast/Magnitude}
\end{figure*}

While the \textit{raw} contrast curves provide a quick, intuitive estimate of sensitivity, they do not fully reproduce the real performance of an imaging system because they do not account for post-processing effects that may impact the flux of a potential companion. To obtain a more accurate picture of the sensitivity of a system then, we need to compute a \textit{calibrated} contrast curve that accounts for the effects of over-subtraction and self-subtraction of a companion during the PSF subtraction process. Throughput for high-contrast imaging algorithms like \texttt{KLIP} are typically low close to the star due to over-subtraction, where the algorithm aggressively removes residual starlight, and can unintentionally suppress a companion. At greater distances, the \texttt{KLIP} throughput improves, allowing the companion’s signal to remain largely unaffected by the subtraction process.

To compute the \textit{calibrated} contrast curves, \texttt{spaceKLIP} performs injection and recovery of simulated companions into the data at known separations and position angles. To simulate the companion, the tool takes advantage of \texttt{STPSF} to generate the PSF.
The algorithm then retrieves these injected signals and applies scale factors to correct for inaccuracies introduced during PSF subtraction. The peak flux of the injected companions is then measured in sigma relative to the $5\sigma$ contrast at the injected separation. The retrieved scale factors are used to adjust the \textit{raw} contrast curves, providing a more accurate representation of the true sensitivity achieved in a given observation. 

\subsubsection{Contrast Achieved on Isolated and Disk Targets}
When analyzing the residuals, two major source of pollution are observed: a bright binary companion, and a disk.
The heavy impact of binaries on F444W contrast sensitivity, is clearly shown in Figure \ref{fig: fkstn_calcon_binary_disk_F444W}, where the achieved calibrated contrast limit for binaries is almost an order of magnitude  worst compared to either isolated stars or disks, which reach very similar contrast limits instead. On the other hand, analysis shows 
This shows that for our purposes, the impact that disks have on our sensitivity limits is mostly negligible in F444W (where the disk is barely present).
Since our study is derived by the F444W sensitivity, the presence of the disk do not  impact negatively our results, and they can safely be included in the final sample. Even in the filter F200W (where disks are usually brighter), analysis shows that the impact of the disks on contrast limits in this filter is mostly negligible as well.


Binaries instead can not be included without an apriori careful treatment. In appendix \ref{apx: Accounting for Binarity in the Imaging Data}, we will present a new method developed for \texttt{spaceKLIP} to recenter and remove visually resolved binary companions tested on non-coronagraphic data, but since this will be the focus of an upcoming paper (e.g. Bruinsma et al. in prep.), they will they will be excluded by our current analysis.

For all further analysis then, we will focus only on the sensitivity limits of isolated targets and disks \citep[with studies of specific disks being the focus of other papers in the series, e.g.][]{Crotts2025, Crotts2026, Palatnick2025} in the observed moving groups and associations. This selection leaves 111 isolated targets and disks for the sensitivity analysis. 

Figure \ref{fig: Contrast/Magnitude} shows the \textit{calibrated} contrast curves for each program, where the thick line marks the median of the distribution, while the shaded area shows the $16th/84th$ percentile for each distribution. The available field of view (FOV) coverage in the NIRCam coronagraphic (solid line) and imaging (dashed line) results in a lower achieved contrast at wide angular
separations. For some specific datasets, small sample statistics have a major impact near the edge, 
causing the contrast to greatly deteriorate at wide separations.
While the GO-4050 data roughly reach a minimum median contrast level of $\sim 10^{-5}$ at $\gtrsim 2$ arcsec, the GO-5835 coronagraphic and SURVEY-6005 data are able to go much deeper in contrast ($\sim 5 \times 10^{-6}$ and $\sim  2 \times 10^{-6}$ respectively). Since the targets in SURVEY-6005 are the only FGK targets in the surveys (excluding 1 K star in GO-4050 - see Table \ref{tab: GO summary}), the contrast curves for these sources are able to explore a much larger parameter space in contrast since the sources in this surveys are generally brighter compared to the M stars observed by GO-4050 and GO-5835.
Because all targets in the GO-5835 imaging survey are faint M-dwarfs (hence the decision to not use the coronagraph), the performance is the lowest, with a median contrast of only $\sim 6 \times 10^{-5}$ at similar separations.

\subsection{Magnitude limits}
\label{subsec: Contrast curves}

The right-hand panel of Figure \ref{fig: Contrast/Magnitude} shows the magnitude sensitivity limits for each program. These curves are obtained by converting the contrast curve presented in Figure \ref{fig: Contrast/Magnitude} into a magnitude difference,
using the apparent magnitude of the corresponding target. 
To obtain the apparent magnitude of a target star, we used the \texttt{starphot.get\_stellar\_magnitudes} tool from \texttt{spaceKLIP} (see Section \ref{subsec: Contrast curves}). 

Despite a significant spread in contrast sensitivity across the various coronagraphic programs, the median magnitude sensitivities consistently plateau at comparable limits. Specifically, SURVEY-6005 achieves a median F444W magnitude of $\sim 21.3$ at separations $\gtrsim 4\text{ arcsec}$, whereas GO-4050 and GO-5835 plateau at $\sim 21.4$ at separations $\gtrsim 2\text{ arcsec}$. 

Ultimately, the absolute magnitude sensitivity floor is governed by the instrument background noise rather than coronagraphic performance. This background-limited regime explains why GO-4050 and 5835 (mostly focused on M-dwarfs) exhibit poorer contrast  while converging to identical magnitudes limits as SURVEY-6005. Moreover, SURVEY-6005’s superior contrast does not improve its absolute magnitude sensitivity at small separations, primarily because its FGK hosts are intrinsically brighter than the M dwarfs in GO-4050/5835. Additionally, SURVEY-6005 exhibits marginally lower background sensitivity, as shorter readout ramps are required to prevent saturation, causing higher read noise compared to the longer \(\texttt{DEEP8}\) ramps used in GO-4050/5835. The findings suggest that the necessary detector configuration for brighter targets limits the achievable sensitivity in the background-limited regime.

For the GO-5835 imaging component, the absence of a coronagraph increased the overall throughput and signal-to-noise ratio achieved for this faint M-dwarf sample. This yielded improved performance at closer separation, and a background magnitude limit in F444W that is $\sim 1$ magnitude deeper than the coronagraphic limits ($\sim 22.2$ at $\gtrsim2$ arcsec).

\section{Discussion}
\label{sec: Discussion}

\subsection{Survey sensitivity limits}
\label{sec: Surveys sensitivity limits}

To assess the sensitivity limits of the three JWST programs analyzed in this work, we converted the \textit{calibrated} contrast curves presented in section~\ref{subsec: Contrast curves}, corrected for the finite field of view, into minimum detectable companion mass and effective-temperature detection limits.

To this end, we employed the Python tool {\tt{MADYS}}\footnote{\url{https://madys.readthedocs.io/en/latest/}} \citep{Squicciarini22}, which derives statistical constraints on the detectability of self-luminous companions based on detection limits derived from high-contrast imaging observations. Suitable evolutionary models (described in section ~\ref{subsec: Models}) are used to translate photometric contrast into companion masses and/or effective temperatures as a function of age.

\begin{figure*}[!t]
    \centering
    \includegraphics[width=0.99\textwidth]{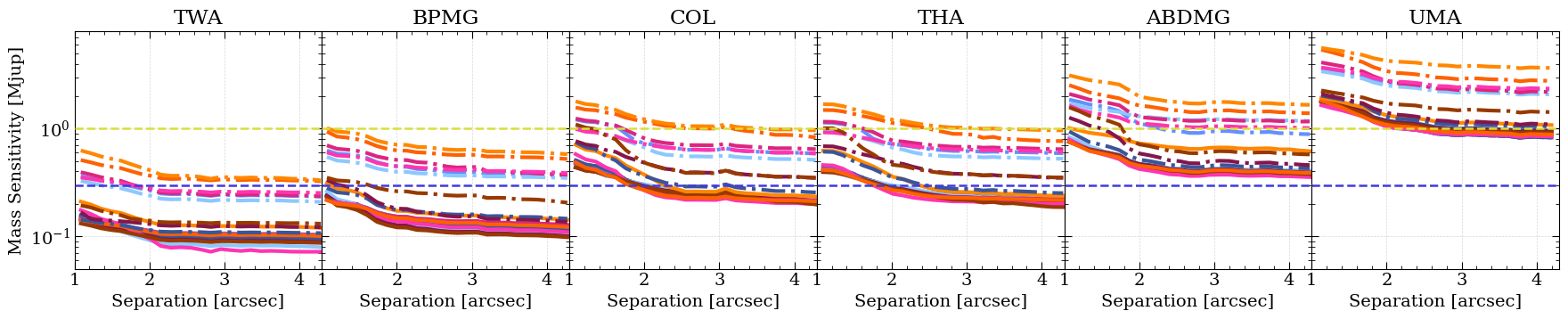}\\
    \includegraphics[width=0.99\textwidth]{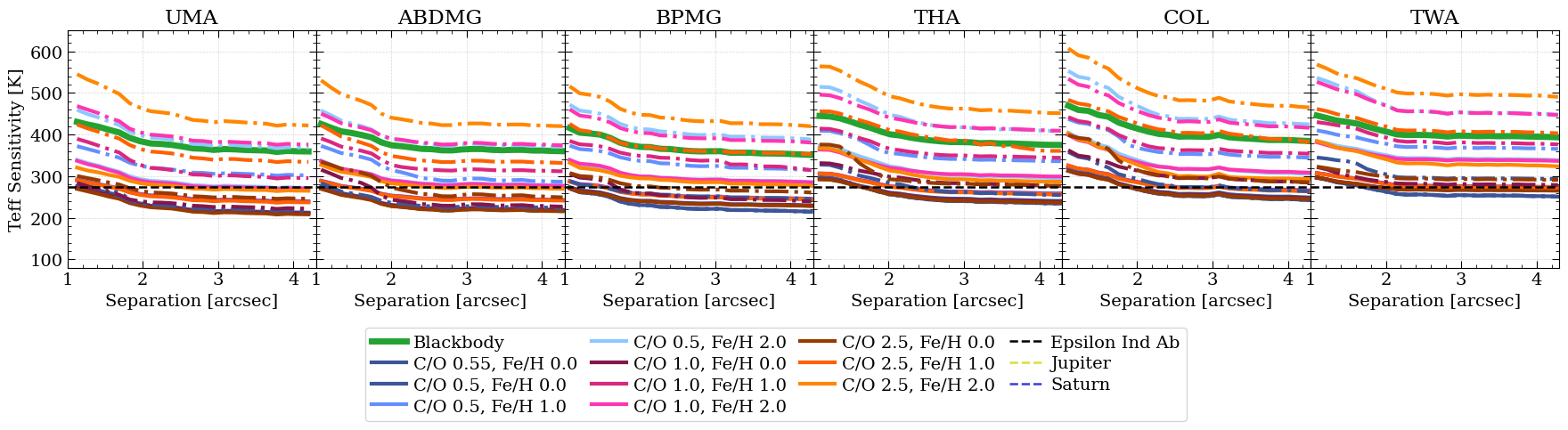}
    \caption{Mass/T$_{\mathrm{eff}}$ sensitivity comparison between different groups. Cloud-free \texttt{BEX-ATMO} and \texttt{Sonora Flame Skimmer} \texttt{CEQ} (continuous lines) and \texttt{NEQ} (dash-dotted lines), and a blackbody model (to act as a cloudy case scenario - green thick line) are shown for comparison. The yellow and blue horizontal dashed lines show the Jupiter and Saturn reference masses (top row), while the  black horizontal dashed line shows the Epsilon Ind Ab reference temperature \citep[bottom row -][]{Matthews2024}. Note that in the upper row, the moving groups have been ordered by their respective age (younger on the left). In the lower row instead, the moving groups have been ordered by their respective median distance (closer on the left), following the values reported in Table \ref{tab: Targets summary}.}
    \label{fig: Mass_Teff_sens_limits_zoomin}
\end{figure*}

\subsection{Model selection}
\label{subsec: Models}

\begin{table*}[t]
\centering
\tabletypesize{\scriptsize}          
\setlength{\tabcolsep}{2pt}          
\renewcommand{\arraystretch}{0.95}   
\tablewidth{\columnwidth} 
\caption{Sensitivity limits at 2.0 arcsec for Mass (M$_{\rm Jup}$) and T$_{\rm eff}$ (K). Note that the moving groups have been arranged following the same order shown in Figure \ref{fig: Mass_Teff_sens_limits_zoomin}.\label{Tab: Mass_Teff_sensitivity_at_2}}
\begin{tabular}{lc|cccccc|cccccc}
\hline
 &  & \multicolumn{6}{c|}{Mass} & \multicolumn{6}{c}{T$_{\rm eff}$} \\
Model & CO/FeH & TWA & BPMG & COL & THA & ABDMG & UMA & UMA & ABDMG & BPMG & THA & COL & TWA \\
\hline
blackbody & -- & -- & -- & -- & -- & -- & -- & 383 & 374 & 371 & 402 & 415 & 407 \\
\hline
bex-atmo2023-ceq & 0.55/0.0 & 0.11 & 0.14 & 0.29 & 0.28 & 0.46 & 1.07 & 233 & 229 & 230 & 257 & 268 & 263 \\
\hline
bex-atmo2023-neq-s & 0.55/0.0 & 0.11 & 0.14 & 0.29 & 0.28 & 0.46 & 1.06 & 233 & 229 & 230 & 257 & 268 & 263 \\
\hline
bex-atmo2023-neq-w & 0.55/0.0 & 0.11 & 0.14 & 0.29 & 0.28 & 0.46 & 1.07 & 234 & 229 & 230 & 257 & 268 & 263 \\
\hline
sonora-flame-skimmer-ceq & 0.5/0.0 & 0.10 & 0.13 & 0.30 & 0.29 & 0.47 & 1.20 & 232 & 231 & 241 & 261 & 272 & 272 \\
 & 0.5/1.0 & 0.11 & 0.15 & 0.28 & 0.28 & 0.46 & 1.11 & 257 & 254 & 258 & 277 & 287 & 283 \\
 & 0.5/2.0 & 0.09 & 0.13 & 0.30 & 0.30 & 0.46 & 1.13 & 292 & 290 & 300 & 325 & 339 & 351 \\
 & 1.0/0.0 & 0.10 & 0.13 & 0.29 & 0.28 & 0.47 & 1.17 & 230 & 230 & 240 & 259 & 270 & 272 \\
 & 1.0/1.0 & 0.11 & 0.15 & 0.27 & 0.27 & 0.45 & 1.10 & 255 & 253 & 257 & 276 & 286 & 282 \\
 & 1.0/2.0 & 0.10 & 0.14 & 0.27 & 0.25 & 0.42 & 1.05 & 288 & 289 & 299 & 321 & 334 & 347 \\
 & 2.5/0.0 & 0.10 & 0.12 & 0.29 & 0.28 & 0.46 & 1.15 & 228 & 228 & 240 & 258 & 269 & 271 \\
 & 2.5/1.0 & 0.11 & 0.15 & 0.27 & 0.27 & 0.45 & 1.09 & 254 & 252 & 256 & 275 & 285 & 282 \\
 & 2.5/2.0 & 0.14 & 0.17 & 0.37 & 0.36 & 0.73 & 1.34 & 282 & 283 & 295 & 315 & 328 & 341 \\
\hline
sonora-flame-skimmer-neq & 0.5/0.0 & 0.11 & 0.18 & 0.36 & 0.36 & 0.52 & 1.29 & 234 & 236 & 259 & 285 & 293 & 305 \\
 & 0.5/1.0 & 0.27 & 0.45 & 0.74 & 0.72 & 1.12 & 2.69 & 324 & 315 & 337 & 356 & 368 & 378 \\
 & 0.5/2.0 & 0.24 & 0.40 & 0.67 & 0.67 & 1.30 & 2.53 & 397 & 390 & 413 & 449 & 469 & 471 \\
 & 1.0/0.0 & 0.13 & 0.18 & 0.48 & 0.49 & 0.59 & 1.41 & 239 & 243 & 258 & 299 & 313 & 286 \\
 & 1.0/1.0 & 0.28 & 0.51 & 0.82 & 0.78 & 1.29 & 2.79 & 325 & 327 & 349 & 367 & 381 & 393 \\
 & 1.0/2.0 & 0.28 & 0.45 & 0.72 & 0.72 & 1.12 & 2.81 & 404 & 390 & 404 & 441 & 457 & 470 \\
 & 2.5/0.0 & 0.14 & 0.26 & 0.49 & 0.47 & 0.72 & 1.72 & 264 & 273 & 276 & 294 & 315 & 299 \\
 & 2.5/1.0 & 0.37 & 0.63 & 1.16 & 1.14 & 1.64 & 3.43 & 364 & 349 & 371 & 412 & 427 & 420 \\
 & 2.5/2.0 & 0.41 & 0.71 & 1.22 & 1.22 & 1.99 & 4.29 & 462 & 441 & 448 & 494 & 512 & 510 \\
\hline
\end{tabular}
\end{table*}

At the lowest temperatures expected to be probed by our programs (i.e. $\lesssim$ 350 K), exoplanet fluxes in the NIRCam bands will be at least partially affected by the presence of clouds \citep{Burrows2003,Sudarsky2003,Morley2014,Bowens2025}. However, existing exoplanet atmospheric models \citep[e.g.,][]{Saumon2008, Allard2012, Linder2019, Phillips2020, Marley2021, Lacy2023, Morley2024, Mukherjee2024, Radcliffe2026} do not adequately cover the low temperatures required to evaluate the sensitivity we expect to reach, do not include clouds, or are uncoupled from evolutionary models.
Additionally, recent JWST discoveries \citep[such as TWA7b - ][]{Lagrange2025,Crotts2025} highlight the potential need for disequilibrium chemistry and enhanced metallicity to better characterize the flux of low-mass sub-Jupiter exoplanets, that current planet evolutionary models appears to overestimate.
To compute sensitivity limits and detection probability maps, we then need to resort to 2 specific evolutionary model grids that cover sufficiently low temperatures for our purposes, combining the measured luminosity and the estimated age of each source to infer their mass and T$_{\mathrm{eff}}$. 

The first model, hereafter referred to as \texttt{BEX-ATMO} (or \texttt{bex-atmo2023-ceq/neq-s/neq-w}), combines the \texttt{ATMO} chemical-equilibrium atmospheric models \citep{phillips20} with the latest equation of state \citep{chabrier23} and the \texttt{BEX} models \citep{linder19}, following the coupling scheme described in \citet{carter21}. The second set of models is the \texttt{Sonora Flame Skimmer} models \citep[or \texttt{sonora-flame-skimmer-ceq/neq};][]{Mang2026,mang_2026_20030439}, the latest set of clear atmospheric and evolutionary models in the Sonora family \citep{Marley2021, Karalidi2021, Morley2024, Mukherjee2024, Wogan2025, Davis2025}. These models span $T_{\rm eff}$ = 50~K to 2400~K, log(g) = 2 - 5.5 (cgs), a wide range of metallicity ranging from 10x sub-solar ([M/H] = -1.0) to 100x super-solar ([M/H] = +2.0), and C/O ratios from 0.5x solar to 2.5x solar (solar = 0.458) for both equilibrium (\texttt{CEQ}) and disequilibrium (\texttt{NEQ}) chemistry. 

While all our models are currently cloud-free, the main difference between the \texttt{CEQ} models and the \texttt{NEQ} models is that the \texttt{CEQ} models allow all the chemical species to mix to reach a steady state with one another, while the \texttt{NEQ} models assume that vertical mixing dredges up different species into the upper layers of the atmosphere before they fully react with one another. Where these species mix is determined by when vertical mixing begins to dominate. This is parameterized by the eddy diffusion parameter, $K_{\rm zz}$, with various strengths of mixing ranging from $K_{\rm zz} = 10^2, 10^4, 10^7, 10^8,~\rm and~10^9$, where larger values of $K_{\rm zz}$ represent stronger mixing. 
The parameters $\mathrm{Fe/H}$ and $\mathrm{C/O}$ instead regulate the composition of the objects (e.g. Jupiter has enhanced metallicity relative to the sun). The C/O, for example, is typically dictated by where in the disk the exoplanet formed relative to various ice/snow lines \citep{Oberg2011}.  The T$_{\mathrm{eff}}$ of the object also affects the dominant molecular species C and O are found in. For colder objects (late-T to Y dwarfs; $\lesssim 700$ K), equilibrium chemistry favors $\mathrm{O}$ in $\mathrm{H_2O}$ and $\mathrm{C}$ in $\mathrm{CH_4}$, while for hotter objects (L to early-T dwarfs;  $\gtrsim 1000$ K) carbon is increasingly partitioned into $\mathrm{CO}$ and $\mathrm{CO_2}$. These temperature boundaries are approximate because the dominant reservoirs depend on the specific pressure–temperature profile of a given object \citep{MarleyRobinson2015,Lodders2002,Zahnle2014}.

\subsection{Mass/Temperature limits}
\label{subsec: Sensitivity limits}
To study the sensitivity limits of our programs, we investigated models belonging to the \texttt{BEX-ATMO} ($\mathrm{C/O}=0.55$, $\mathrm{Fe/H}=0.0$) and \texttt{Sonora Flame Skimmer} families ($\mathrm{C/O}$=[$0.5$,$1$,$2.5$], $\mathrm{Fe/H}$=[$0$,$1$,$2$]). This range of parameters has been chosen to have a common point between Sonora and BEX-ATMO ($\mathrm{C/O} = 0.5$ and $\mathrm{Fe/H}= 0$) families. The Sonora models also offer $\mathrm{FeH} = -1$, but since the candidate companions we expect to detect are likely to have solar or super-solar metallicity, we do not consider that case.

In Figure \ref{fig: Mass_Teff_sens_limits_zoomin},
we show a zoomed view of the mass/T$_{\mathrm{eff}}$ average sensitivity limits for each moving group.  We limit the separation range plotted to [1, 4] arcsec, since this is the region where the majority of changes happen to our sensitivity, before plateauing to a constant value.  We note that these results are dependent on the atmospheric model adopted and that we only plot moving groups with observations of $N\geq 10$ members.  Table \ref{Tab: Mass_Teff_sensitivity_at_2} provide the mass/T$_{\mathrm{eff}}$ sensitivities at a separation of $2$ arcsec for each moving group.

The top row of Figure \ref{fig: Mass_Teff_sens_limits_zoomin} and the left side of Table \ref{Tab: Mass_Teff_sensitivity_at_2} highlight the wide mass sensitivity range we will be able to reach with our programs. 
In TWA we can expect sensitivity to extend to exoplanets with sub-Saturn mass between $\sim 0.09 - 0.41$ M$_{\mathrm{J}}$. In UMA instead, we are more likely to only reach super-Jupiter masses between $\sim 1.06 - 4.29$ M$_{\mathrm{J}}$ for enhanced metallicity. In both cases, the spread is dictated by the selected $\mathrm{CEQ}/\mathrm{NEQ}$ models and the $\mathrm{Fe/H}$ and $\mathrm{C/O}$ parameters.

The bottom row of Figure \ref{fig: Mass_Teff_sens_limits_zoomin} and the right side of Table \ref{Tab: Mass_Teff_sensitivity_at_2} instead, show the T$_{\mathrm{eff}}$ versus separation space explored by our programs. As we stated at the beginning of section \ref{subsec: Filter selection}, at the temperatures probed by our programs we would expect clouds to be present in the atmosphere of the exoplanets. Similar to \cite{Bowens2025}, we investigate the impact of cloud opacity on our $T_{\mathrm{eff}}$ sensitivity limits by estimating a maximally cloudy exoplanet as a blackbody.
For each target, the F444W apparent magnitudes were computed as a function of the blackbody temperature and interpolated over the corresponding magnitude sensitivity curves (Figure \ref{fig: Contrast/Magnitude}). The resulting $T_{\mathrm{eff}}$–separation curves were then averaged within each group to produce the final blackbody sensitivity limit (green thick lines) for the group.
The bottom row of Figure \ref{fig: Mass_Teff_sens_limits_zoomin} shows that while the presence of clouds can heavily impact our $T_{\mathrm{eff}}$ sensitivity limits compared to $\mathrm{CEQ}$ models, the impact of the $\mathrm{Fe/H}$ and $\mathrm{C/O}$ chemistry in $\mathrm{NEQ}$ models also have a strong effect on the suppression of flux received from exoplanets, pushing upward our overall $T_{\mathrm{eff}}$ sensitivity limits.

With all this considered, in UMA we could be sensitive to T$_{\mathrm{eff}}$ between $\sim 228 - 462$ K for enhanced metallicity (the spreads again dictated by the amount of $\mathrm{C/O}$). For TWA instead, we could be sensitive to T$_{\mathrm{eff}}$ $\sim 263 - 510$ K for exoplanets with enhanced metallicity. 

\begin{figure*}[!th]
    \centering
    \includegraphics[width=0.98\textwidth]{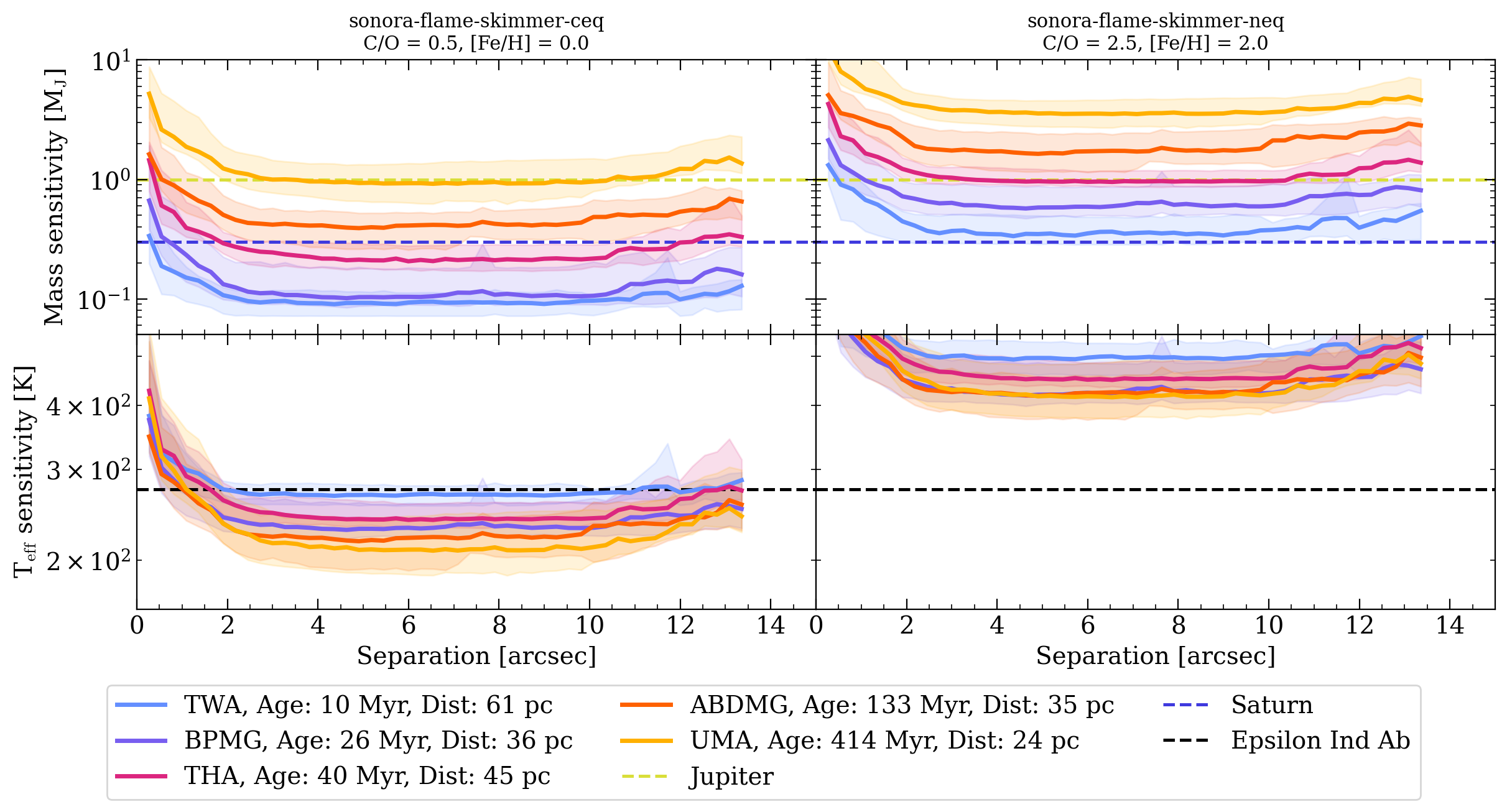}
    \caption{Mass (top)/T$_{\mathrm{eff}}$ (bottom) sensitivity limits for the \texttt{Sonora Flame Skimmer} family models with $\mathrm{C/O}=0.5$, $\mathrm{Fe/H}=0.0$, \texttt{CEQ} (left); and $\mathrm{C/O}=2.5$, $\mathrm{Fe/H}=2.0$, \texttt{NEQ} (right). The yellow and blue horizontal dashed lines show the Jupiter and Saturn reference masses, while the  black horizontal dashed line shows the Epsilon Ind Ab reference temperature \citep{Matthews2024}. The shaded area cover the $16th/84th$ percentile for each distribution.}
    \label{fig: Mass_Teff_sens_limits}
\end{figure*}

To conclude this section, we provide in  Figure \ref{fig: Mass_Teff_sens_limits}, an overview of the sensitivity limits of our programs for the full separation range [0, $\sim14$ arcsec]. This analysis, coupled with future atmospheric characterization, will be able to provide a better picture of what the true sensitivity limits of our programs are. Currently though, since there is not a real consensus on what the common atmospheric parameters for exoplanets in our programs should be, we 
decide to highlight 
what our worst/best-case constraints look like. 
For these specific reasons, we selected two models from the \texttt{Sonora Flame Skimmer} family, one for the \texttt{CEQ} case, with solar $\mathrm{C/O}$ and metallicity ($\mathrm{C/O}=0.5$ and $\mathrm{Fe/H}=0.0$) , and one for the \texttt{NEQ} case, with more extreme parameters ($\mathrm{C/O}=2.5$ and $\mathrm{Fe/H}=2.0$), since they allow for a greater portion of parameter space to be explored compare to all the other model parameters, and the \texttt{BEX-ATMO} model as well. 
Similar to Figure \ref{fig: Mass_Teff_sens_limits_zoomin}, we grouped together the different sources based on their groups, but in this case we overplotted one on to the other for a quick comparison. The assumed age and median distance of each group (as reported in Table \ref{tab: Targets summary}) are displayed for reference. 
Figure \ref{fig: Mass_Teff_sens_limits} clearly shows that past $\sim 2"$, Jupiter masses (or Saturn masses in the most optimistic case) and T$_{\mathrm{eff}} \lesssim 275$ are well within reach of our programs.

\subsection{Temperature sensitivity limits comparison between NIRCAM and MIRI}
\label{subsection: Teff sensitivity limits comparison between  NIRCAM and MIRI}

\begin{figure*}[!t]  
    \centering
    \includegraphics[width=0.98\textwidth]{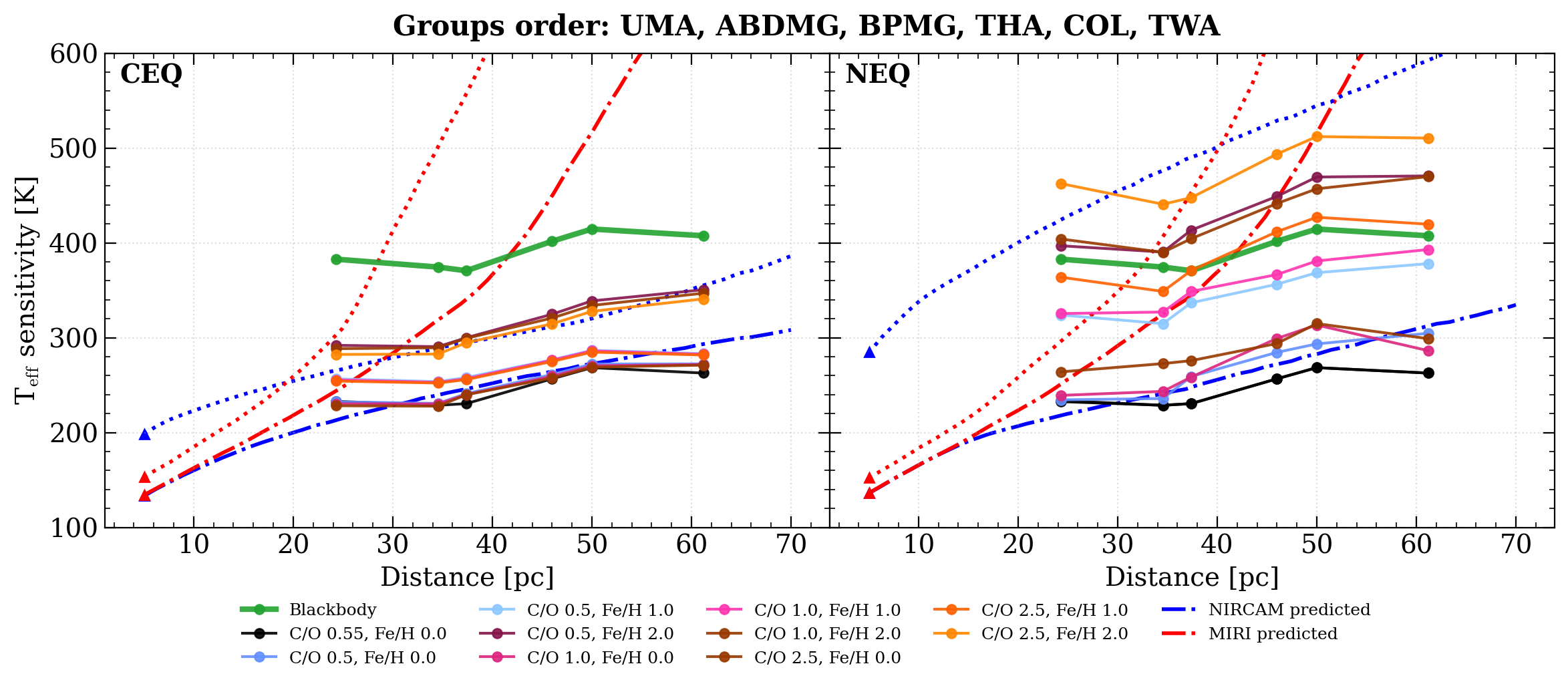}\\
    \caption{T$_{\mathrm{eff}}$ sensitivity at 2 arcsecond, for the \texttt{CEQ} (left) and \texttt{NEQ} models (right), showing an overall good agreement between our sensitivity limits and \cite{Bowens2025} predictions, even though the pace at which MIRI can potentially outperform NIRCam T$_{\mathrm{eff}}$ sensitivity depends on atmospheric chemistry of the expected population of exoplanets.
    Each dot represent the T$_{\mathrm{eff}}$ sensitivity for a different group: UMA, ABDMG, BPMG, THA, COL and TWA. Each continuous line instead shows either the \texttt{BEX-ATMO} (black), or the \texttt{Sonora Flame Skimmer} (color coded by their different parameters) models. A blackbody model is overplotted to act as a cloudy case comparison  (green thick line).    
    5-$\sigma$ sensitivity from \cite{Bowens2025} for F444W (blue) and F2100W (red), converted into our \texttt{Sonora Flame Skimmer} models, are overplotted for the [$\mathrm{C/O}$, $\mathrm{Fe/H}$] = [0.5, 0.0] (dashed line), and [$\mathrm{C/O}$, $\mathrm{Fe/H}$] = [2.5, 2.0] (dotted line) case. The triangles represent the original point from where the extrapolation to higher distances begun.}
    \label{fig: mass_teff_trends}
\end{figure*}

Similar to \cite{Bowens2025}, we perform a comparison between the expected T$_{\mathrm{eff}}$ sensitivities for both JWST MIRI F2100W imaging and NIRCam F444W coronagraphy, based on our ample sample. In their analysis, they compared equilibrium models for three  atmospheric cases: clear, cloudy (fsed = 8), and maximally cloudy, concluding that MIRI imaging offers the advantage for systems within 20 pc. 
In our analysis instead, we explored NIRCam performances for both equilibrium and non-equilibrium cloud-free models, for different values of $\mathrm{C/O}$ and $\mathrm{Fe/H}$, and then compared against their predictions. A blackbody model is also included to act as a cloudy case comparison scenario.
To facilitate a direct comparison to the their NIRCam/MIRI performance, we converted their 5-$\sigma$ magnitude limits in the background-limited regime for EV Lac (20.2 mag in F444W, 15.1 mag in F2100W, at 2.6 arcsec) to the expected T$_{\mathrm{eff}}$ in the \texttt{Sonora Flame Skimmer} models with $\mathrm{C/O}$=[0.5,2.5] and $\mathrm{Fe/H}$=[0,2], after rescaling them to different distances to trace how their estimated performances would look like in our reference system.

\begin{figure*}[!thp]
    \centering
    \includegraphics[width=0.95\textwidth]{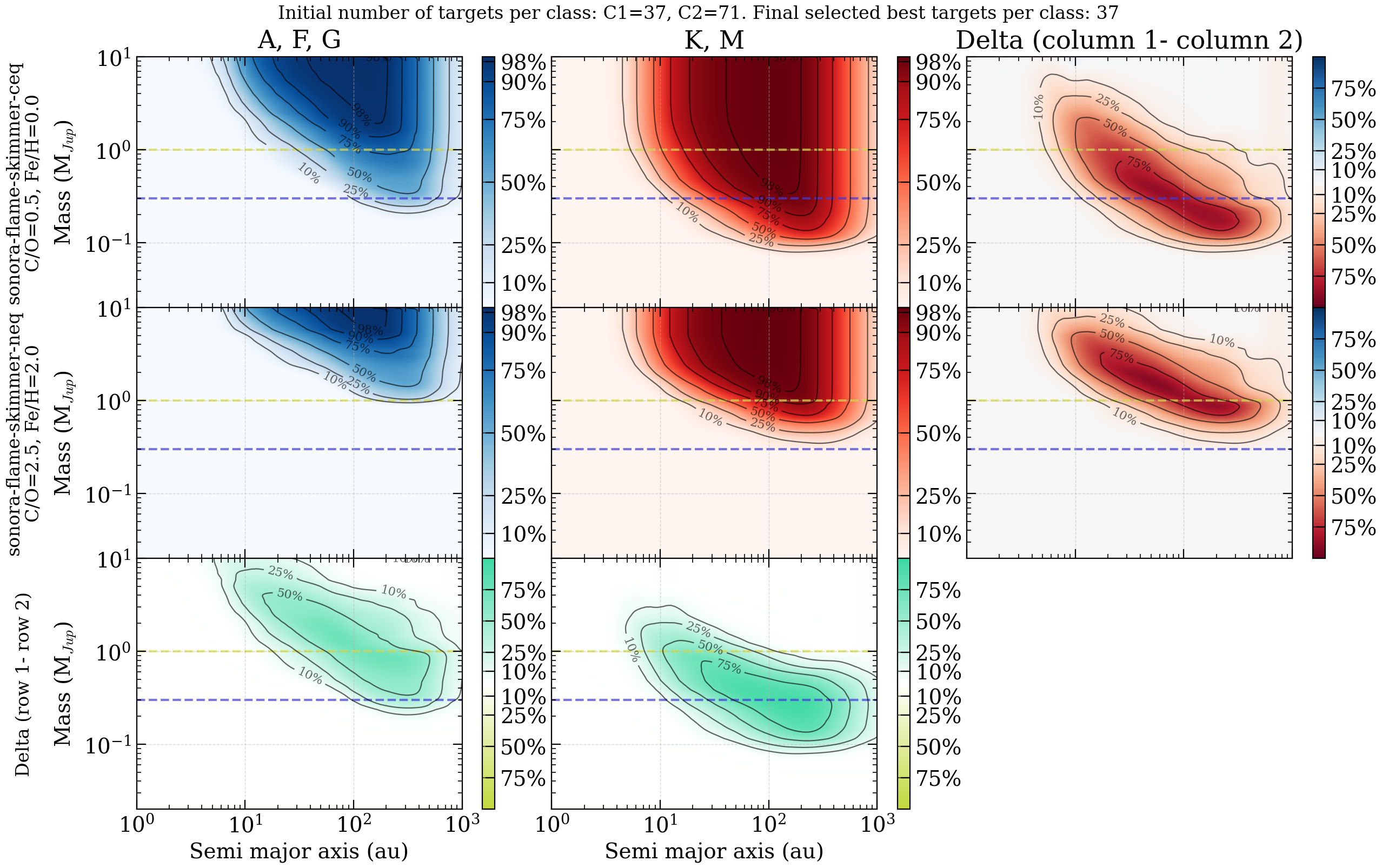}\\
    \includegraphics[width=0.95\textwidth]{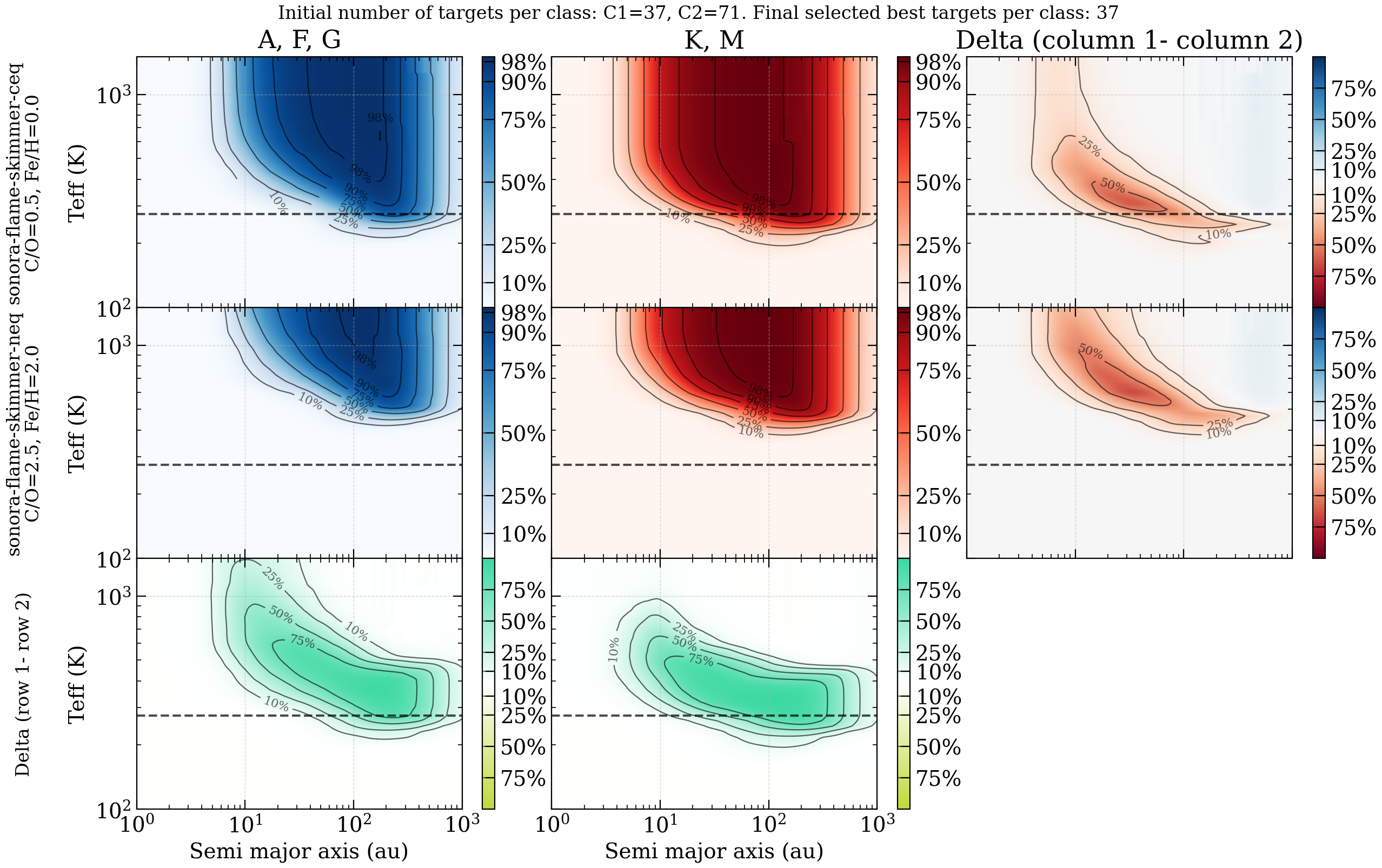}
    \caption{Mass/T$_{\mathrm{eff}}$ detection probability maps comparison for two \texttt{Sonora Flame Skimmer} \texttt{CEQ} and \texttt{NEQ} models, for the A, F, G (combined), and K and M (combined) spectral types. All maps are derived from the F444W sensitivity limits shown in Figure \ref{fig: Contrast/Magnitude}.
    The yellow and blue horizontal dashed lines show the Jupiter and Saturn reference masses, while the  black horizontal dashed line shows the Epsilon Ind AB reference temperature \citep{Matthews2024}. The last row in both panels clearly show not only the considered atmospheric chemistry, but also the spectral type of the targeted star has an heavy impact on the expected planet Mass/T$_{\mathrm{eff}}$ detection probabilities.}
    \label{fig: Sdmaps}
\end{figure*}

In Figure \ref{fig: mass_teff_trends} we present the predicted value of
T$_{\mathrm{eff}}$ sensitivity at 2 arcsec NIRCam and MIRI instruments.
Similar to Figure \ref{fig: Mass_Teff_sens_limits_zoomin}, a blackbody model (green thick line) is overplotted to both panels to act as a cloudy case comparison scenario.  Continuous color code lines shows the different model parameters. The red (F444W) or blue (F2100W) lines shows \cite{Bowens2025} converted 5-$\sigma$ magnitude limits in the background-limited regime for the $\mathrm{C/O}=0.5$, $\mathrm{Fe/H}=0$ (dashed line), and $\mathrm{C/O}=2.5$, $\mathrm{Fe/H}=2.0$ (dotted line) case. The triangles represent their converted 5-$\sigma$ background-limited estimation of EV Lac, from where the extrapolation to higher distances begun.
The Figure shows (as expected) how the minimum T$_{\mathrm{eff}}$ sensitivity decrease as farther the group is. Distant sources require higher T$_{\mathrm{eff}}$ to offset distance-related flux attenuation (F $\propto $ T$_{\mathrm{eff}}^{4}/d^2$) and remain above the instrumental detection threshold.
Also in this case we see an outlier in UMA ($\sim 25$ pc), that seems to perform much worse compared to the rest of the trend, or TWA compared to COL, that performs better. One possible reason is that while we used the age of the cluster as reference age for all the members of the same group, for the distance we used individual reported values for each target, leaving us with a wider distribution of distances in each group (sometimes overlapping with in 1$\sigma$), that could lead to much similar sensitivity differences when the median is taken. Additionally, the spectral type composition of each group plays a role in the sensitivity limits. While the UMA sample is mostly composed by F and G stars, the BPMG sample, the next group in distance, is mostly composed of M-dwarfs. The same reasoning goes for the comparison between COL and TWA samples, where COL has a more balanced spectral type composition, with slightly more F, G stars compared to K or M, while TWA sample is mostly composed of M-dwarfs. As we have seen from Figure \ref{fig: Sdmaps}, the spectral composition has an impact on the performances the T$_{\mathrm{eff}}$ sensitivities.
Overall, our sensitivities mostly match \cite{Bowens2025} predictions, with a particular good agreement in the \texttt{CEQ} plot.
When we compare to their MIRI results instead, we see that our NIRCam sensitivity limits start to out pace their MIRI predictions at $\sim 25-30$ pc (best/worse case scenario) for the \texttt{Sonora Flame Skimmer CEQ} models, and at $\sim 25-50$ pc (best/worse case scenario) for the \texttt{Sonora Flame Skimmer NEQ} models. 
\begin{figure*}[!th]
    \centering
    \includegraphics[width=0.95\textwidth]{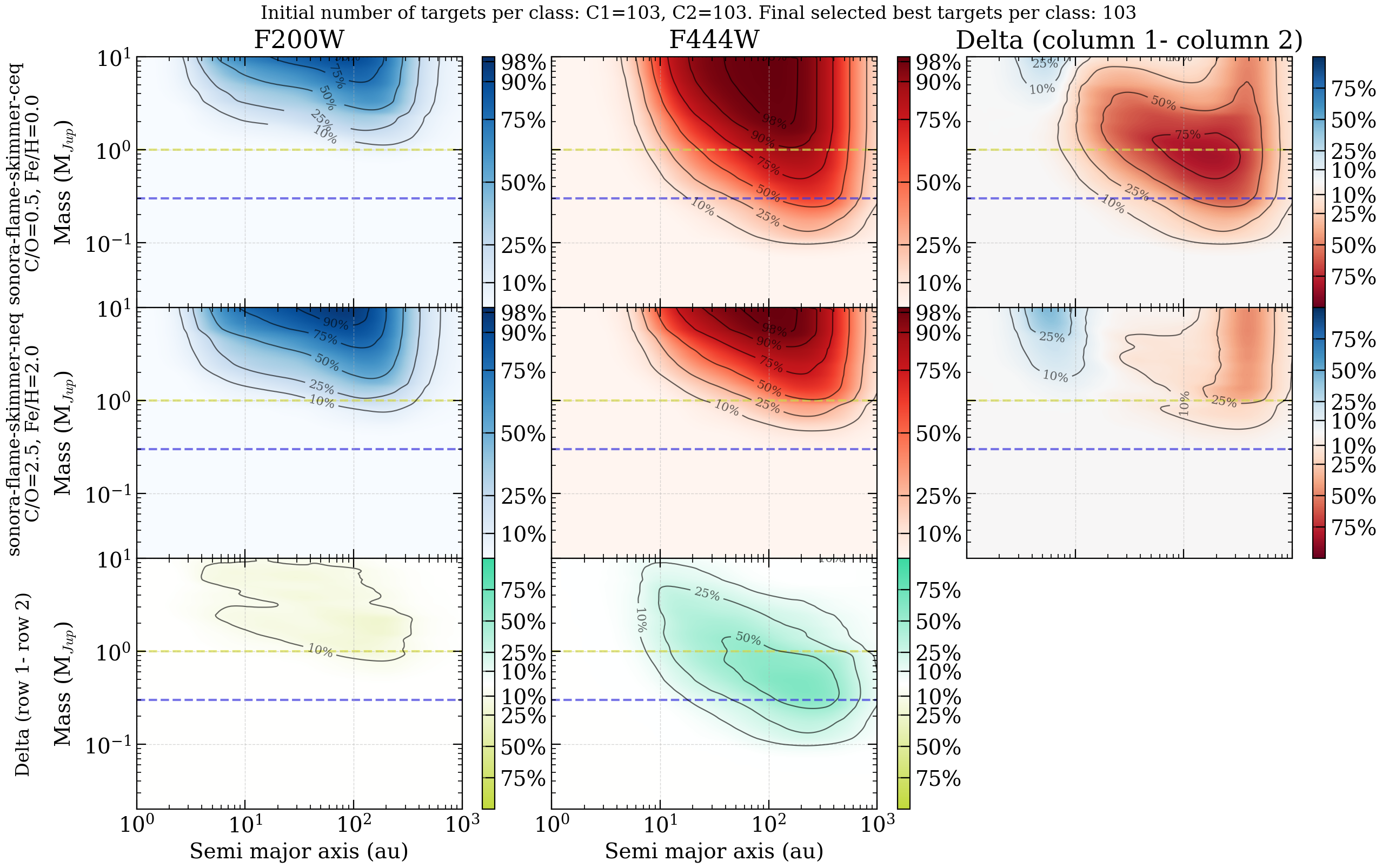}\\
    \includegraphics[width=0.95\textwidth]{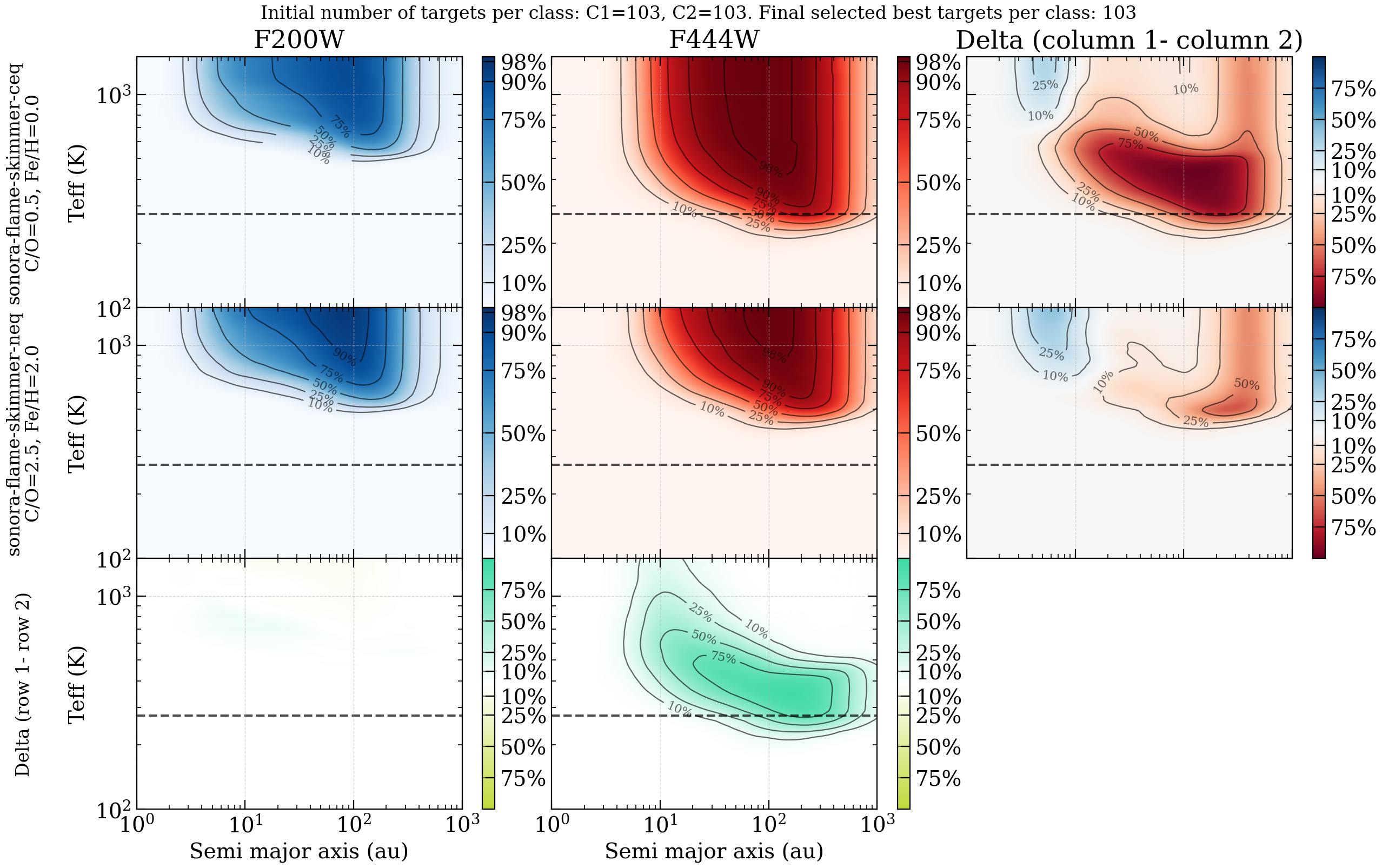}
    \caption{Mass/T$_{\mathrm{eff}}$ detection probability maps comparison for two \texttt{Sonora Flame Skimmer} \texttt{CEQ} and \texttt{NEQ} models, for filters F200W and F444W. The yellow and blue horizontal dashed lines show the Jupiter and Saturn reference masses, while the  black horizontal dashed line shows the Epsilon Ind AB reference temperature \citep{Matthews2024}. The last row  in both panels clearly show the impact of atmospheric chemistry (e.g. disequilibrium chemistry, enhanced metallicity)
    on the planet Mass/T$_{\mathrm{eff}}$ detection probabilities.}
    \label{fig: Fdmaps}
\end{figure*}

Finally, a comparison with our cloudy case (the black body line) shows that while clouds may have a suppressive effect for NIRCam coronagraphic observations \citep[as shown by the \texttt{CEQ} panel of Figure \ref{fig: mass_teff_trends}, as well as ][]{Bowens2025}, increasing metallicity might have a greater potential to limit F444W observations (see \texttt{NEQ} panel of Figure \ref{fig: mass_teff_trends}). Since both clouds fractions and metallicity are quite uncertain for this class of objects, only future atmospheric characterization of the exoplanet population uncovered by GO-4050, GO-5835, and SURVEY-6005 (i.e. the upcoming analysis from GO-10764 program) will be able to disentangle these two cases. This results, based on the analysis of 141 targets (the sum of all the UMA, ABDMG, BPMG, THA, COL and TWA target members), while mostly confirming \cite{Bowens2025} predictions, also highlight once again the tension between the atmospheric chemistry of the expected population of exoplanets and the instrument predicted sensitivity limits.

\subsection{Detection probability maps}
\label{subsec: Detection probability maps}
To be able to place population based general mass and T$_{\mathrm{eff}}$ predictions, we used {\tt{MADYS}} (see section \ref{sec: Surveys sensitivity limits}) to convert our calibrated contrast curves into detection probability maps, using the same models we introduced earlier in section \ref{subsec: Models}. 
The detection probability maps are computed over a grid in semi-major axis and companion mass spanning the range of [1, 1000] au $\times$ [0.005, 10] $M_{\rm J}$, uniformly sampled in logarithmic space. At each grid point, $N=1000$ sets of orbital elements are randomly drawn and used to convert the semi-major axis into projected separations; in particular, inclinations are sampled from a distribution uniform in $\cos i$, while eccentricities are drawn from a truncated normal distribution $\mathcal{N}(0,~0.9)$. In parallel, the stellar age constraint is sampled to convert the companion mass into a set of apparent magnitudes in the relevant filter.

For each realization, a simulated companion is considered detected if its contrast at the corresponding projected separation exceeds the 5$\sigma$ contrast limit. Averaging over all realizations yields a mean detection probability that self-consistently accounts for projection effects and age uncertainties. Detection probability maps in effective temperature are then obtained by remapping the mass–semi-major axis probability grid using the model-predicted mass–temperature relation at each sampled age.
Similar to the end of the precedent section, in this section we will also limit our analysis to two specific \texttt{Sonora Flame Skimmer} models, with a standard solar value for $\mathrm{C/O}=0.5$ and $\mathrm{Fe/H}=0.0$ for the \texttt{CEQ} case, and an enhanced extreme case of $\mathrm{C/O}=2.5$ and $\mathrm{Fe/H}=2.0$ for the \texttt{NEQ} case, as references for our discussion. 

In the following sections, we will present our detection probability maps as follows.
The first two rows show the results for the selected model, and for each selected category (left and central columns). The right column instead shows the difference between the first category (blue) and the second (red). We will refer to this direction along each row as ``model-wise". 
The left and central panels of the bottom row instead show the difference between the specific category in the \texttt{sonora-flame-skimmer-ceq} model (green) and the \texttt{sonora-flame-skimmer-neq} model (yellow). We will refer to this direction along each column as ``category-wise".

To offer a fairer comparison (since we can not show the entire distribution of performances, e.g. as in Figure \ref{fig: Mass_Teff_sens_limits}), these plots are created taking into account equal-size samples.
To do so, for each category we ranked the detection probability maps based on the sum of their detection probability below 1 $M_{J}$ (or 1000 $K$). Then we picked the best maps so that both categories have the same number of targets, and average them.
Detection probability maps for the most best-sampled groups in our programs (TWA, BPMG, COL, THA, ABDMG, and UMA) are shown in appendix \ref{apx: Group-based Detection probability maps}.

\subsubsection{Filter comparison}
\label{subsec: Filter comparison}
Figure \ref{fig: Fdmaps} shows the comparison between detection probability maps as a function of mass/T$_{\mathrm{eff}}$ versus separation for the filters F200W and F444W coronagraphic data. Note that for this comparison, we have to exclude the non-coronagraphic sources in the GO-5835 dataset, since they do not have F200W measurement, lowering the total number of targets to 103.

Model-wise, the F444W filter consistently outperforms the bluer F200W filter in mass and \(T_{\mathrm{eff}}\) sensitivity. Because the SEDs of cool (\(\lesssim 500\) K) planetary-mass companions peak in F444W while stellar flux declines, this bandpass provides the primary detection layer for identifying candidates.
Overall, the \texttt{sonora-flame-skimmer-ceq} models reach a peak of their difference at $\sim 75\%$ detection probability around $\sim 1$ M$_{\mathrm{J}}$ or $\sim 400$ T$_{\mathrm{eff}}$ at $\sim 200$ au, while the \texttt{sonora-flame-skimmer-neq} models barely reaches $\sim 25\%$ for $\sim 1$ M$_{\mathrm{J}}$, and $\sim 50\%$  for $\sim 500$ T$_{\mathrm{eff}}$. This dichotomy will have a direct impact on the final yield achieved by our three programs, and in turn, will also better inform us on the atmospheric chemistry of the exoplanet population (see also later in Section \ref{subsec: Yield estimates}).

Conversely, a filter-wise comparison reveals that F200W detection probabilities remain uniform across the models. In contrast, the F444W filter is highly sensitive to the choice of equilibrium versus disequilibrium chemistry and model parameters. For separations between $\sim 50$ and $500\text{ au}$, the model-to-model variation in F444W detection probability peaks at $\sim 50\%$ for masses between ($\lesssim 1\text{M}_{\mathrm{J}}$) and $\sim 1\text{ M}_{\mathrm{Sat}}$, and reaches $\sim 75\%$ for effective temperatures of $\sim 300\text{--}400\text{ K}$.

This result strongly suggests that atmospheric chemistry (e.g. disequilibrium chemistry, enhanced metallicity) of the planets we are considering can then heavily affect our detection thresholds (especially at the lower mass/T$_{\mathrm{eff}}$ limits). 
Then, the potential Neptune- to Saturn-mass planet inferred from cold debris disc structures at 10s of au \citep{Pearce2022},  
might be currently just inside our detectability window, depending on their own chemical composition.

\subsubsection{Spectral type comparison}
\label{subsec: Spectral type comparison}
Figure \ref{fig: Sdmaps} shows a comparison between detection probability maps as a function of mass/T$_{\mathrm{eff}}$ versus separation for different spectral types for the F444W filter.
Note that for this comparison instead, we excluded the three L-dwarfs, lowering the total number of targets to 108. We then split these 108 targets in two samples: A, F, and G spectral types stars (37 targets in total), and K and M (spectral types stars (71 targets in total). Finally, since we want to perform an equal-size comparison (as explained at the end of Section \ref{subsec: Detection probability maps}) we selected the 37 best performing targets from the K and M sample, ending up with equal number of targets for each sample.

Similar to Figure \ref{fig: Fdmaps},  we observe an increase in sensitivity towards the lowest masses for later spectral types.
Model-wise, the K and M spectral types sample outperform the A, F, and G, with a detection probability $\sim 50\%$ in favor of the later spectral types for $\lesssim 1$ M$_{\mathrm{J}}$ already $20$ au, and $\sim 75\%$ for $\lesssim 1$ M$_{\mathrm{Saturn}}$ at $60$ au for the \texttt{sonora-flame-skimmer-ceq}. The \texttt{sonora-flame-skimmer-neq} models instead shows a shift upward of the detection probability (very clear when looking at the plots spectral-type-wise, or SpT-wise for short), with a detection probability $\sim 75\%$ in favor of the K and M group for $\sim 1$ M$_{\mathrm{J}}$ at $100$ au, and just a few percent for $\sim 1$ M$_{\mathrm{Saturn}}$.
When looking at the T$_{\mathrm{eff}}$ instead, the \texttt{sonora-flame-skimmer-ceq} reach a peak of the detection probability at $\sim 50\%$ at $\sim 350$ K at $\sim30$ au, while the \texttt{sonora-flame-skimmer-neq} models reach similar values at $\sim 600$ K and $\sim 40$ au.

This dependence with the atmospheric chemistry of the exoplanets is made even clear in the last row of both panels in Figure \ref{fig: Sdmaps}, where we show the SpT-wise variation across the two models. For the mass detection probability maps (top panel), the \texttt{sonora-flame-skimmer-ceq} model provides an increase between $\sim 25-50 \%$ on the detection probability around A, F, and G stars, between one Saturn- to Jupiter-mass at $\sim 30 - 300$ au.
For the K and M stars instead, a dramatic shift appears, with the detection probability reaching up to $\sim 75 \%$ between sub Saturn ($\sim 0.1$ M$_{\mathrm{J}}$) - to Jupiter-mass at similar separations \citep[in agreement with predictions from][]{Carter2021}.
The T$_{\mathrm{eff}}$ sensitivity instead (bottom panel) appears to be more balanced between the different spectral types, with both the A, F, and G sample, and K and M sample, reaching $\sim 75 \%$ more the detection probability.

We need to notice at this point that, when building these maps, targets with different ages and distances are included in each sample, and this mixing can bias the final detection probability map for each spectral type. The mass sensitivity for example will be more sensitive to shifts in ages compared to the T$_{\mathrm{eff}}$ sensitivity.  Thus, is not easy to disentangle from this approach how much of the differences we observe is due to the different spectral types, than from the fact that the F, G, and K star sample has a higher average age versus the M stars (see Table \ref{tab: Targets summary} for reference).
To address this problem, we present in the next Section \ref{subsec: Principal Component Analysis of Detection Probability Maps} a principal component analysis of the detection probability maps, with the goal of disentangling the contribution of the spectral type, age and distance of the targets on the detection probability maps.




\subsubsection{Principal Component Analysis of Detection Probability Maps}
\label{subsec: Principal Component Analysis of Detection Probability Maps}
\begin{figure*}
    \includegraphics[width=0.9\textwidth]{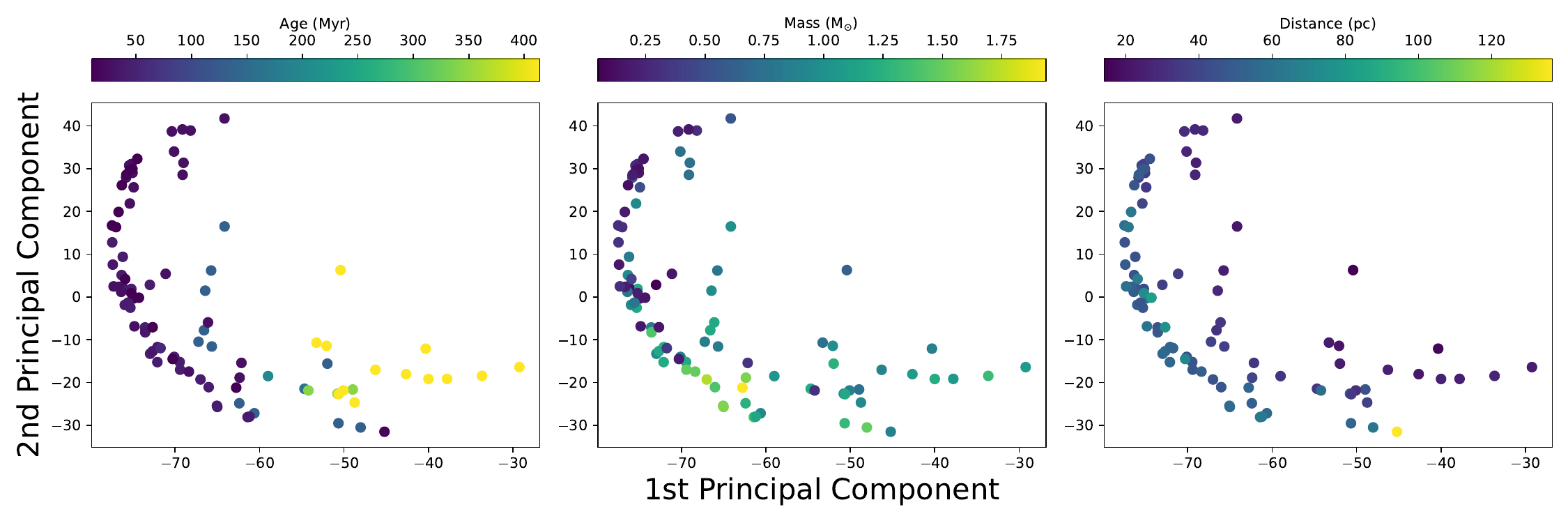}
    \caption{PC scores for the first two principal components, which account for more than 90\% of the total variance in the sub-Jupiters DPMs of single stars without disks. The PC scores appears to be clearly correlated with the system age (left), while weakly correlating with the stellar mass (center). No clear connection is observed with the distance (right).}
    \label{fig:DPM_PCA}
\end{figure*}

The analysis presented in Section~\ref{subsec: Spectral type comparison} shows how degenerate the problem of interpreting variations in the detection probability maps is with the spectral type, age, and distance of the sources  in the maps.
This inherently complex situation stems from the heterogeneity in the design of the constituent JWST programs (GO-4050, GO-5835, and SURVEY-6005).  GO-4050 and GO-5835 observed primarily lower-mass M star targets in TWA and BPMG, with ages of either $\sim10$ Myr or $\sim22$ Myr, whereas SURVEY-6005 observed a sample of solar analog FGK stars, distributed among somewhat older groups ($\sim22$ Myr for the youngest BPMG members in the sample, $\sim400$ Myr for the oldest UMA members in the sample).   Thus, given the inherent inhomogeneity of the sample, it is not entirely possible to disentangle the effects of age, distance, and spectral type.

However this configuration offers also a unique opportunity: an unprecedented large, self consistent sample of hundreds of targets with consistent observations in the same observing mode from multiple moving group and associations, covering a wide range of spectral types, ages and distances.
To more unambiguously determine the dominant variable(s) in producing the observed mass sensitivity of the sub-Jupiter survey observations, we performed a principal component analysis (PCA) on the set of individual target mass detection probability maps across the full sub-Jupiters survey sample.  We selected only single targets with no evidence of a disk in F200W imaging.  For this sample of 93 stars, we used code adapted from the \texttt{pyKLIP} pipeline \citep{Wang2015} to build a Karhunen-Lo\`eve (KL) eigenbasis from the detection probability maps of these stars.

\texttt{pyKLIP} is a PCA pipeline designed specifically for high-contrast imaging; in its more usual usage, it builds a reference library of high-contrast images (either a series of images taken over time at different position angles or a series of different reference star images), builds a KL eigenbasis from this library, and using this basis, reconstructs a target image from a selected number of modes from the KL eigenbasis \citep{Soummer2012,Amara2012}.  The reconstructed image is then subtracted from the original target image; as any astrophysical companion is either not present in the reference library (in the case of a reference library built from observations of other stars) or is moving in position angle in the reference library (in the case of a reference library built from a time series of objects taken at different position angles), this removes light from the star and speckles, while retaining light from any astrophysical object in the field.

In our adaptation of this code to the sub-Jupiter survey detection probability maps, we build a KL eigenbasis from the full set of detection probability maps, to determine how many of the constructed eigenimages are necessary to describe most of the variance in the dataset. PCA is a dimensional reduction technique; by construction, the first eigenimage / principal component (henceforth PC) explains most of the variance in the dataset, followed by each successive PC.  We find that the first PC describes 83.5\% of the total variance, while the second PC describes 7.7\% of the total variance, and the third PC describes 4.2\% of the total variance.  The remaining PCs account for only 4.7\% of the total variance.  Thus, In this case, one PC / eigenimage describes the majority of the variance in the system.

We then projected each of the detection probability map images onto the constructed KL eigenbasis, in order to associate each PC with the true physical parameters determining detection probability map sensitivity: stellar mass (used here as proxy for the spectral type), stellar age, and system distance.  In Figure~\ref{fig:DPM_PCA}, we plot the PC scores (the "weight" of the projection of each image onto each of the PCs) for the first two principal components.  In the left panel, the color of the data points reflect the age of the star, whereas the color of the data points reflects the stellar mass in the middle panel, and the distance to the system in the right panel.  These PC scores are not meaningful on their own, but can provide a sense of which of the physical parameters of the system are connected to each of the PCs.  The PC scores appears to be clearly correlated with system age, while seem to be more weakly correlated with stellar mass, and no clear connection is seen with distance.  Thus, this analysis shows that system age is the most important variable in determining the achieved detection probability map for a given star from our sample.

\begin{figure*}[!th]
    \centering
    \includegraphics[width=0.95\textwidth]{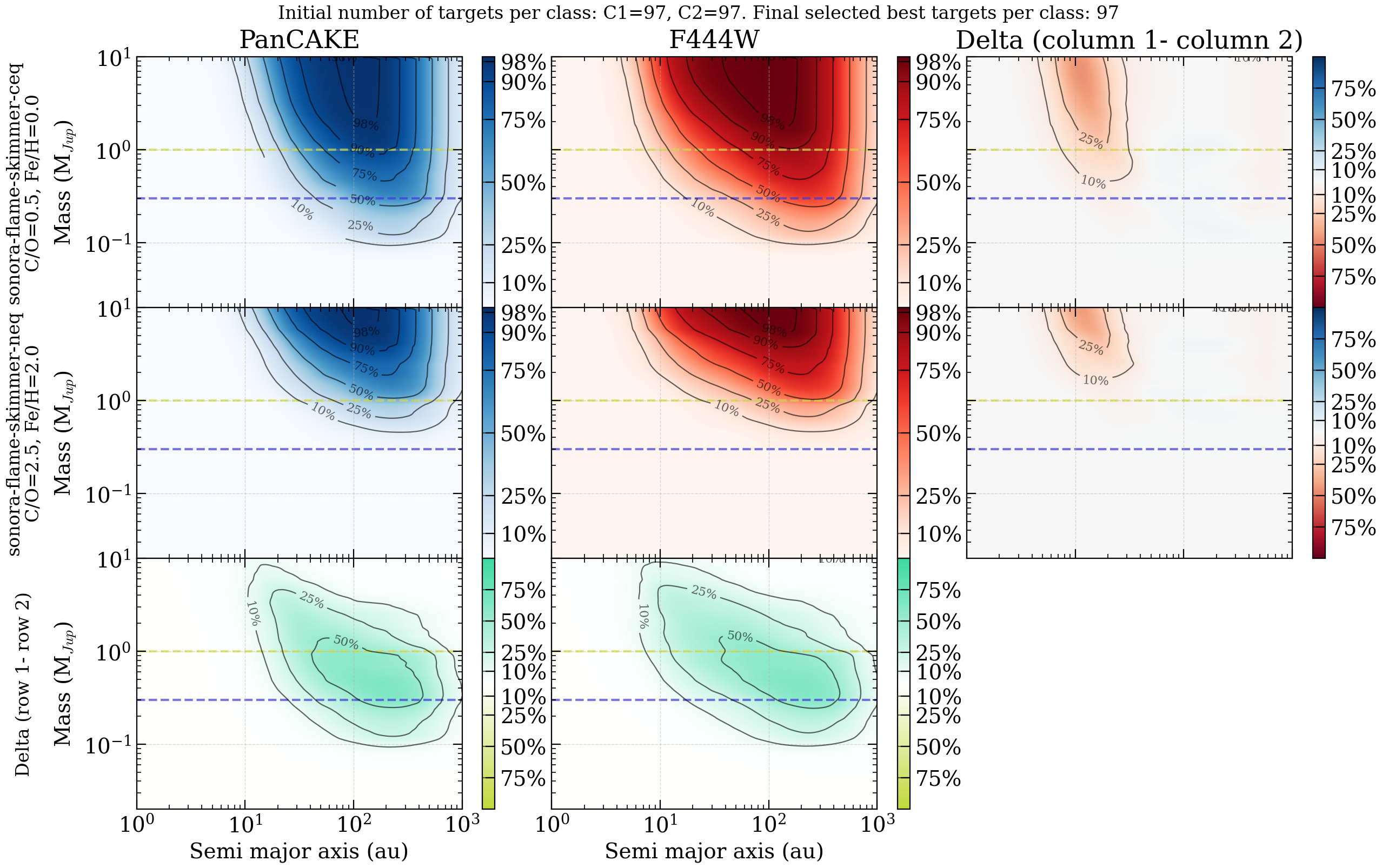}\\
    \includegraphics[width=0.95\textwidth]{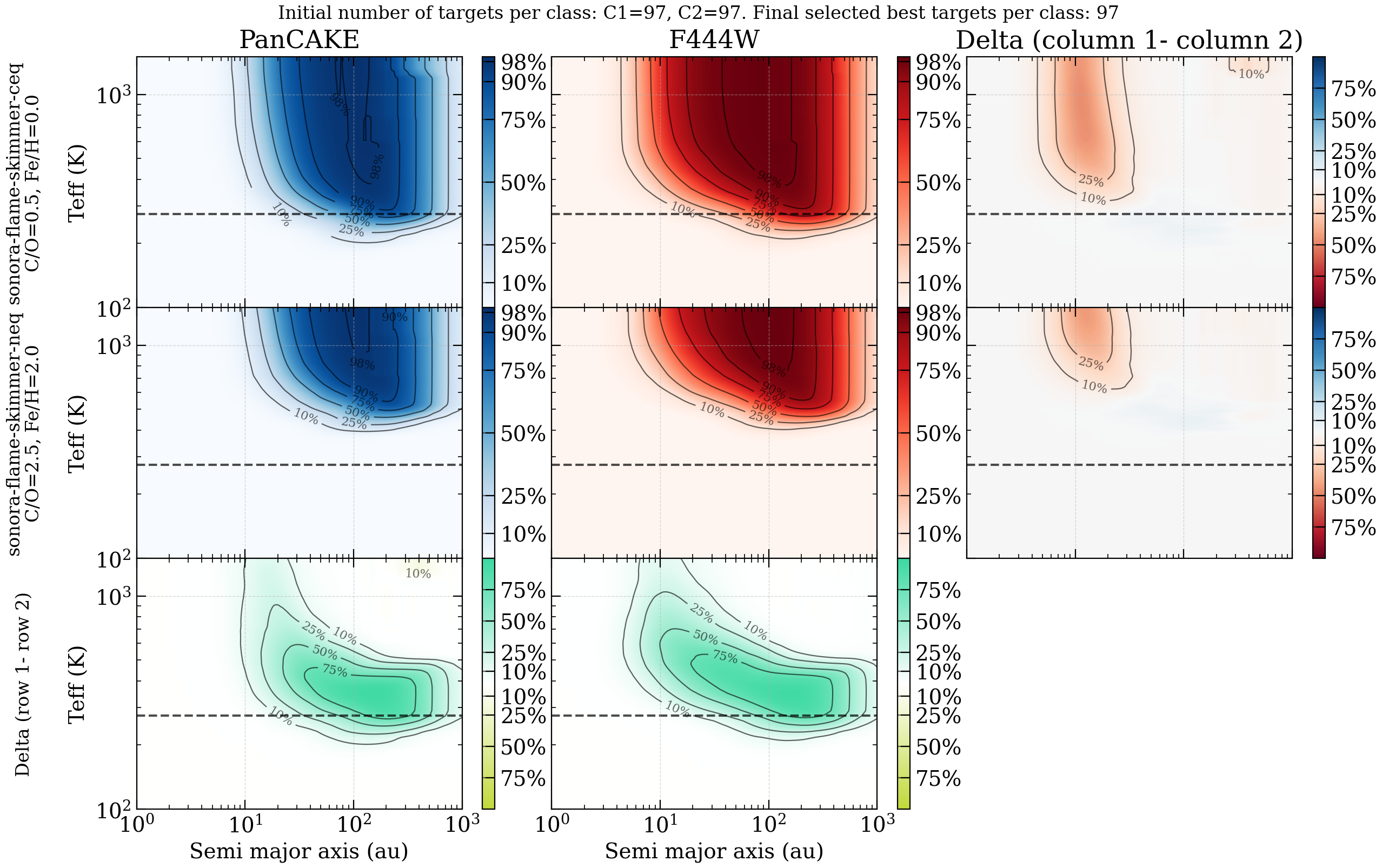}
    \caption{Mass/T$_{\mathrm{eff}}$ detection probability maps comparison for two 	\texttt{Sonora Flame Skimmer} models as explained in the text, between the \texttt{PanCAKE} simulations of our targets and our programs in F444W. 
    The yellow/blue dashed lines show the Jupiter and Saturn reference mass and T$_{\mathrm{eff}}$. All comparisons are based on equal size samples, as explained in the text.}
    \label{fig: PanCAKEdmaps}
\end{figure*}

\subsubsection{Comparison with PanCAKE simulations}
\label{subsubsec: Comparison with PanCAKE}

In our original survey design, we used \texttt{PanCAKE} \citep[an extension of the official JWST exposure time calculator;][]{Carter2021} to simulate observations of survey targets in order identify the most sensitive filters for this project. In this section, we discuss the comparison between the actual achieved detection probability performance in our programs in filter F444W and the predicted result from the  corresponding \texttt{PanCAKE}\footnote{\url{https://github.com/AarynnCarter/PanCAKE}} simulations for 97 of our coronagraphic targets. 
Figure \ref{fig: PanCAKEdmaps} illustrates the comparison between $\texttt{PanCAKE}$ simulations and our empirical data. Both panels demonstrate strong agreement across all modes, masses, and effective temperatures ($T_{\mathrm{eff}}$). The $\texttt{PanCAKE}$ simulations accurately predict NIRCam coronagraphic performance, with discrepancies remaining well below $\sim 10\%$ at separations $\gtrsim 30\text{au}$. For closer separation, our sensitivities outperform $\texttt{PanCAKE}$ prediction, with a discrepancy of  $\sim 25\%$. The likely explanation is that the employed PSF library subtraction provides improved performance over what was assumed for $\texttt{PanCAKE}$ (a 9-point single reference dither). This is supported by the fact that the improvement is limited to the closest separations ($\lesssim 30$ au), where the sensitivity limit is in a contrast-limited regime.
Moreover, these results appear to be in contrast with predictions obtained for MIRI, where $\texttt{PanCAKE}$ usually tends to under predict the instrument performance \citep{Bendahan-West2026}.

\begin{figure*}[!th]
    \centering
    \includegraphics[width=0.95\textwidth]{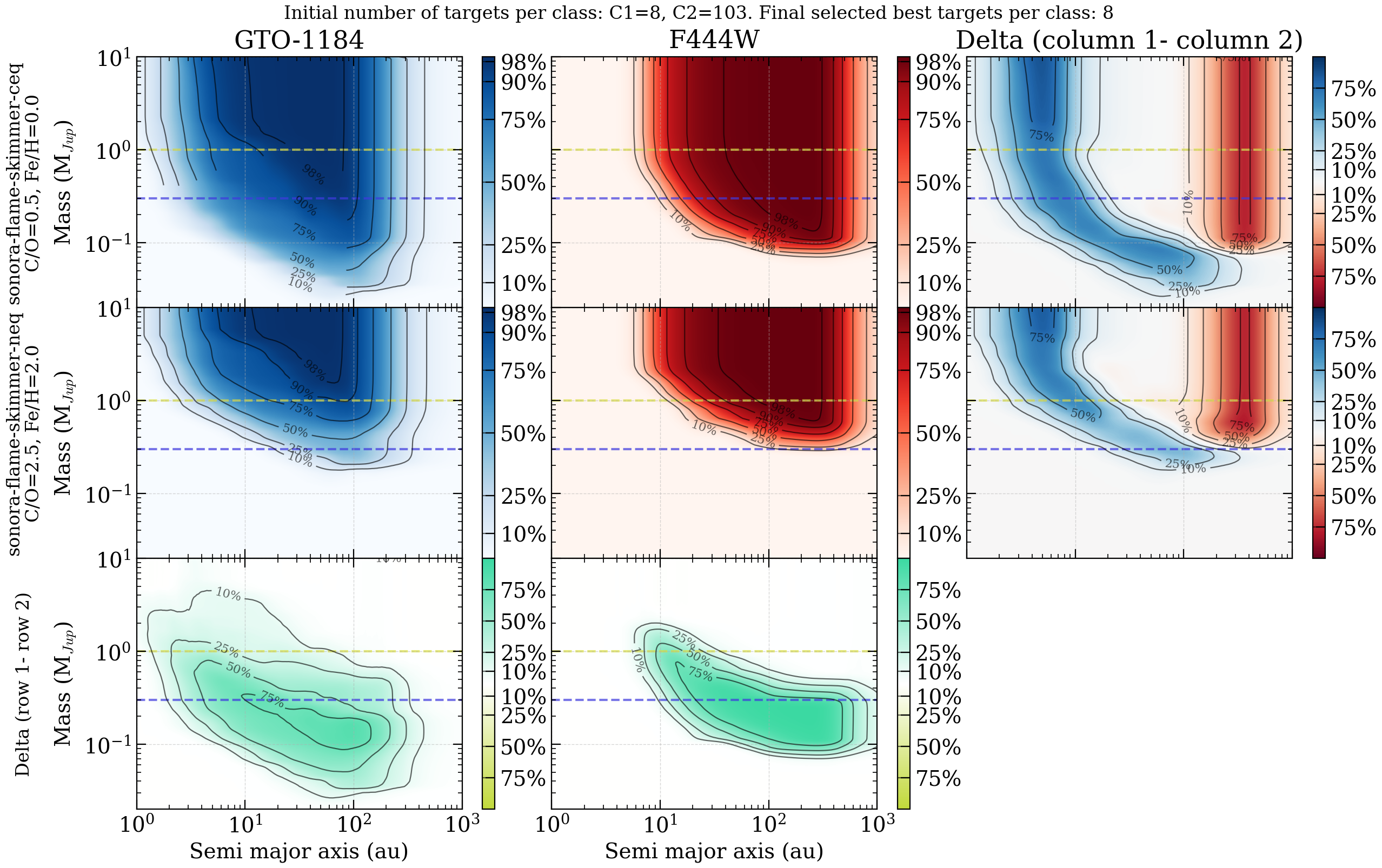}\\
    \includegraphics[width=0.95\textwidth]{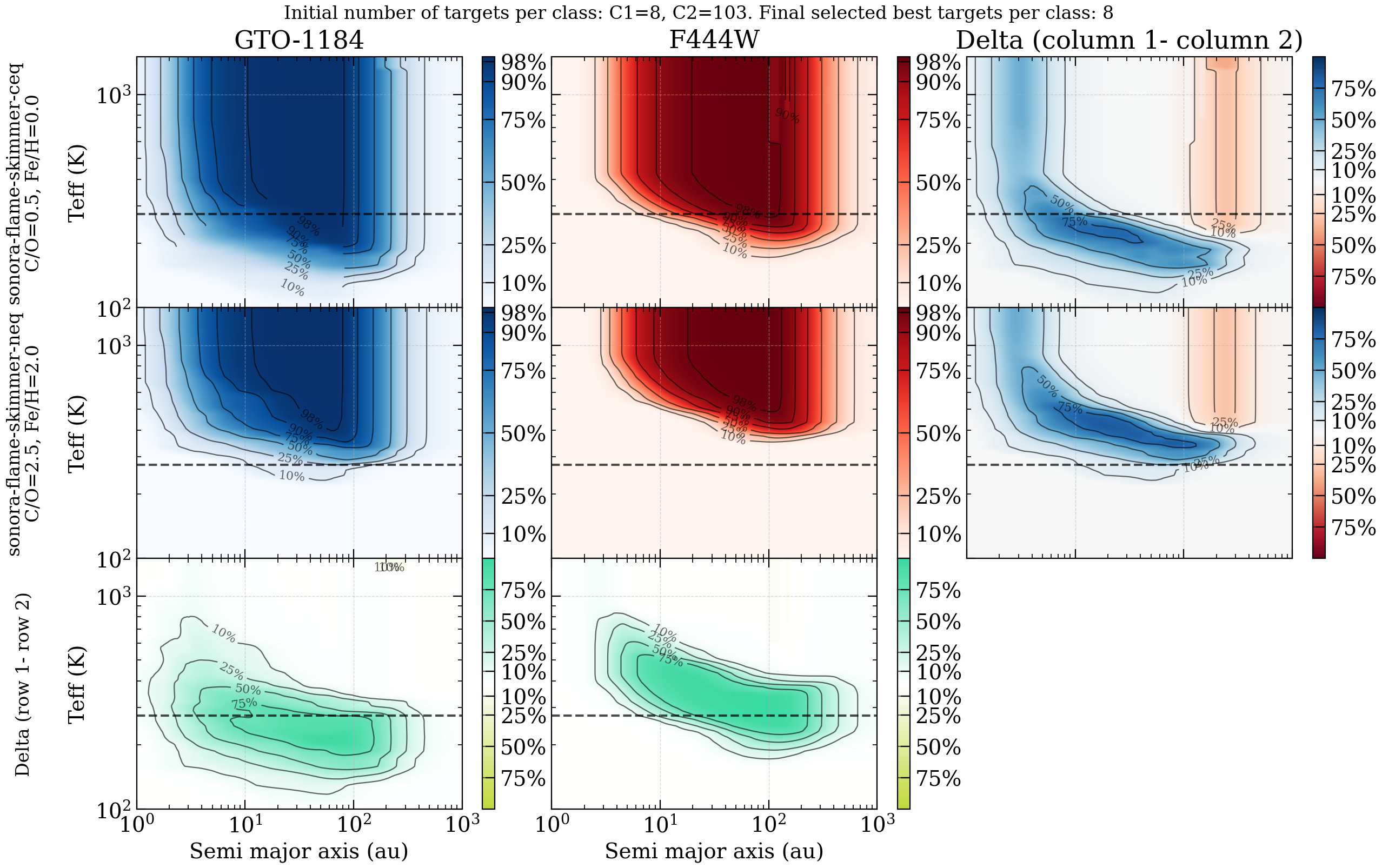}  
    \caption{Mass/T$_{\mathrm{eff}}$ detection probability maps comparison for two 	\texttt{Sonora Flame Skimmer} models as explained in the text, between GTO-1184 and our programs in F444W. 
    The yellow/blue dashed lines show the Jupiter and Saturn reference mass and T$_{\mathrm{eff}}$. All comparisons are based on equal size samples, as explained in the text.}
    \label{fig: GTOdmaps}
\end{figure*}

\subsubsection{Comparison with Bogat et al. 2025}
\label{subsubsec: Comparison with Bogat2025}
Figure \ref{fig: GTOdmaps} illustrates the comparison between our work and the GTO-1184 survey \citep{Bogat2025}. \citet{Bogat2025} obtained F356M and F444W NIRCam coronagraphic imaging of nine nearby ($\lesssim 21.1$ pc) M-dwarfs, spanning a range of ages similar to our targets' ages (i.e., $\sim24-440$ Myr). Note that to build this comparison we have to exclude non-coronagraphic sources in the GO-5835 dataset, lowering the initial number of targets in our combine programs to 103.

Since our two sample are very different in size, we decide to pick the best 9 performing targets as our reference sample for this comparison (as explained at the end of Section \ref{subsec: Detection probability maps}). This process left us with 9 M-dwarfs selected from our initial sample, with distances between $\sim 34-63$ pc, and ages $\sim 10$ Myr.

When compared to our selected sample, we see that GTO-1184 greatly overperforms at closer separations ($\lesssim 10$ au), and overall is able to potentially reach both lower detectable masses and T$_{\mathrm{eff}}$. This result clearly highlight the cardinal role that also the distance plays when trying to detect close, low-mass, cool sources.

\begin{figure*}[!th]
    \centering
    \includegraphics[width=0.95\textwidth]{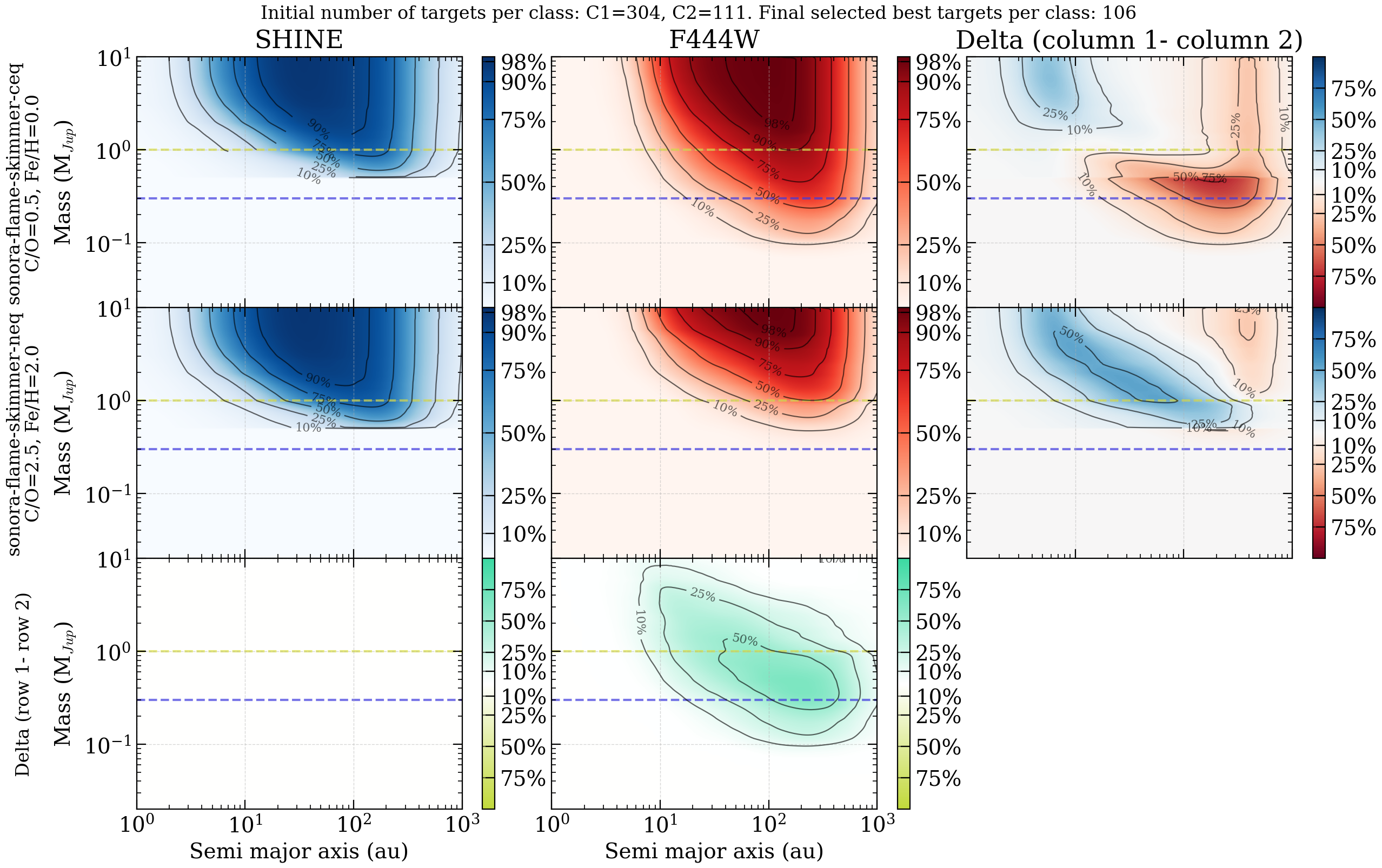}
    \caption{Mass detection probability maps comparison for the two \texttt{Sonora Flame Skimmer} models as explained in the text, between a matched sample across the SHINE survey and our programs in F444W. Note that for the SHINE survey, the \texttt{BEX-COND} models have been used instead of the one showed by the y-labels. All comparisons are based on equal size samples, as explained in the text. 
    The yellow/blue dashed lines show the Jupiter and Saturn reference mass.}
    \label{fig: SHINEdmaps}
\end{figure*}

\subsubsection{Comparison with the SPHERE SHINE survey}
\label{subsubsec: Comparison with the SPHERE SHINE survey}

The VLT SpHere INfrared survey for Exoplanets \citep[SHINE; ][]{Desidera2021,Langlois2021,Vigan2021,Chomez2025} obtained near-IR (1-2 $\mu$m) high contrast imaging of 400 nearby, young stars using the VLT/SPHERE (Spectro-Polarimetric High-contrast Exoplanet REsearch) IRDIS dual-band imager. The full SHINE sample and data reduction/processing is presented in \citet{Chomez2025,Squicciarini2026}.
Similar to our survey sample, the SHINE sample is largely drawn from nearby young moving groups and young stellar associations, but also includes other stars with indications of youth in the solar neighborhood \citep{Desidera2021}. The SPHERE-SHINE sample covers a wider range of spectral types than our own survey,
with about a third of this sample comprising of higher mass AB stars and with only 20 M stars included.  

In Figure \ref{fig: SHINEdmaps}, we present a comparison between the mass sensitivities of the SHINE survey and the JWST NIRCam survey. 
As previously, we use Sonora Flame Skimmer models to convert from contrast sensitivity to mass sensivity for JWST observations; however, the SHINE survey uses the \texttt{BEX-COND} models for this purpose \citep{Allard2001, Marleau2019}.
Given the differing coverage of spectral types between the two surveys, to ensure a robust comparison, for each spectral type in our programs (i.e. A, F, G, K, M, and L), we constructed equal-size sub-samples from the SHINE survey (see previous section \ref{subsec: Detection probability maps}) ensuring that the target count as a function of spectral type in this comparison precisely matches between the two surveys compared. At the end of this process, we ended up with a final sample of 106 targets for each category.
While this selection approach guarantees matching sample sizes and spectral type distributions, the individual targets remain distinct across the two surveys. Furthermore, because different evolutionary models were utilized to derive companion masses, direct comparisons between their mean detection probabilities cannot be interpreted as strictly one-to-one.

The SHINE survey is the most sensitive surveys for directly imaged exoplanets from the ground to date. However, the improvements in mass sensitivity provided by the extreme sensitivity of JWST compared to ground-based instruments provides access to a completely new parameter space for exoplanets, yielding sensitivity to sub-Jupiter exoplanets at wide separations from their stars.
While the SHINE survey has always used the \texttt{BEX-COND} models \citep{Allard2001,Marleau2019}, for the JWST programs in this case we alternate between the \texttt{sonora-flame-skimmer-ceq} model (our best case scenario - first row) and the \texttt{sonora-flame-skimmer-neq} model (our worst case scenario - second row). 
In the first case we are able to reach similar detection probability compared to ground observations already at $\sim 10$ au and $\sim 3$ M$_{\mathrm{J}}$, and $\sim50\%$ in mass sensitivity between $\sim50-500$ au in the region between $\sim1$ M$_{\mathrm{J}}$ to $\sim1$ M$_{\mathrm{Sat}}$. 
Since the stellar age distribution sampled by the SHINE survey is consistent with our Programs \cite[e.g. $\sim 90\%$ of the stars in the SHINE survey have an estimated age $\lesssim 450$ Myr;][]{Squicciarini2026}, the impact of individual age discrepancies on the average sensitivity is expected to be limited.

When instead we compare to our more extreme case ([$\mathrm{C/O}$, $\mathrm{Fe/H}$] = [2.5, 2.0]), we see an inversion of this trend, with the SHINE survey outperforming the JWST observation at lower masses, at almost any separations. This latest comparison though, represents more of an upper bound to our sensitivities, due to the extreme nature of the high metallicity model.  
Indeed, the detection of TWA-7b \citep{Crotts2025} in our sample already provides proof that we are actually not in this worst-case scenario presented here, and that JWST does have significant advantages over SPHERE.

\subsection{Exoplanet yield estimates}
\label{subsec: Yield estimates}

While survey observations are not yet complete and thus the true yield of our survey is not yet known, here we make predictions of the expected number of new exoplanet discoveries to come from this work, given different models of the underlying exoplanet distribution.

\subsubsection{Using core accretion synthetic population models}
\label{subsubsec: Using core accretion synthetic population models}

We restrict this analysis to single FGKM stars belonging to young moving groups and lacking disk detections, yielding a sample of 97 stars. To estimate the number of companions our JWST survey might detect, we divided the observed sample into two stellar‑mass bins, separating Sun‑like FG stars from lower‑mass KM dwarfs, which are expected to host different planetary populations. We used the ``New Generation Planetary Population Synthesis'' (NGPPS) models for the accretion of planetesimals \citep{Emsenhuber2021,Burn2021}, adopting models for 1\,M$_\odot$ and 0.7\,M$_\odot$ primaries as proxies for the underlying core accretion planet populations around FG and KM hosts. By drawing FG and KM populations separately from models that best match their respective host masses, the resulting synthetic sample better reflects the stellar mass distribution of our survey.

\begin{figure*}
    \includegraphics[width=0.48\textwidth]{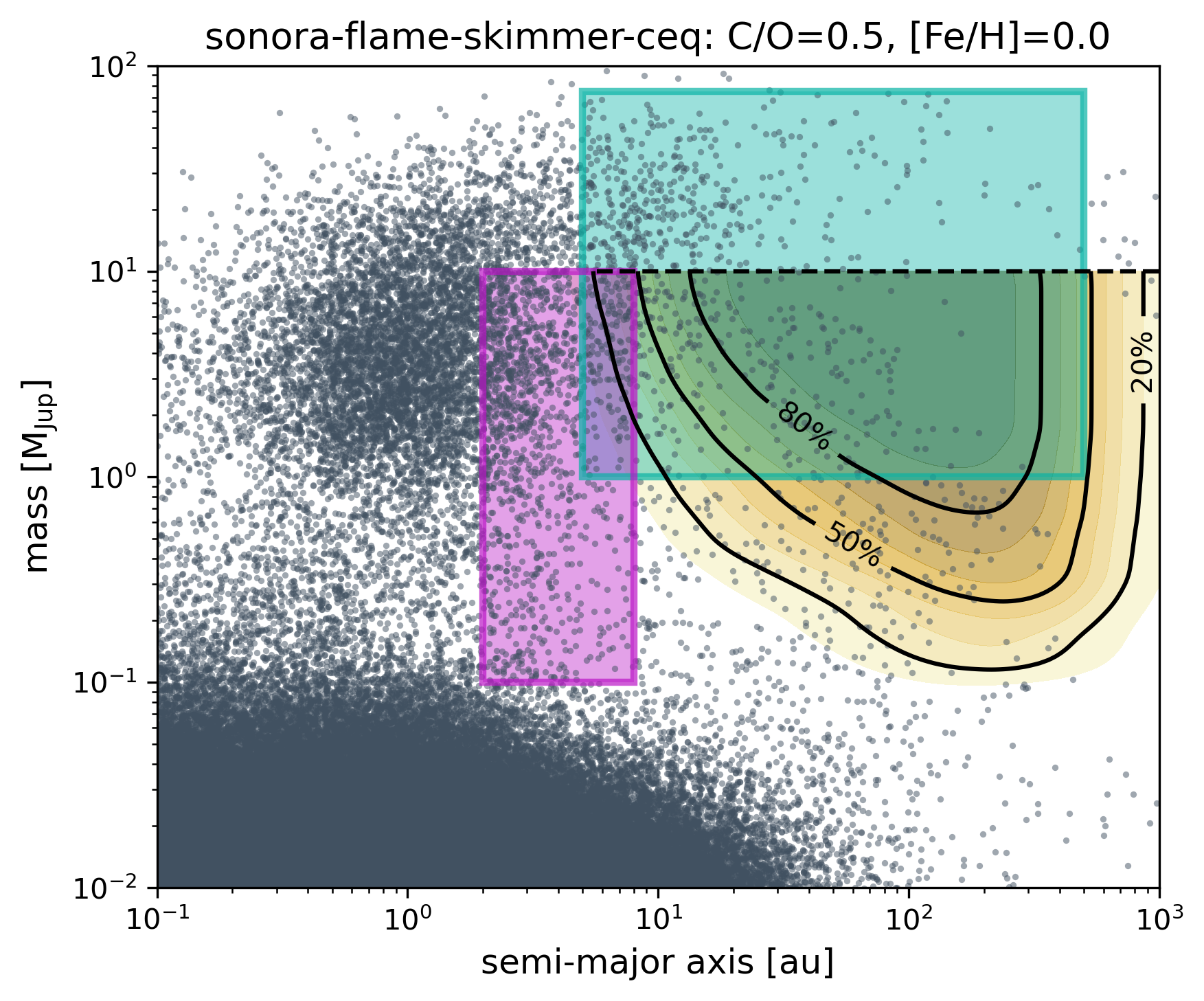}
    \includegraphics[width=0.48\textwidth]{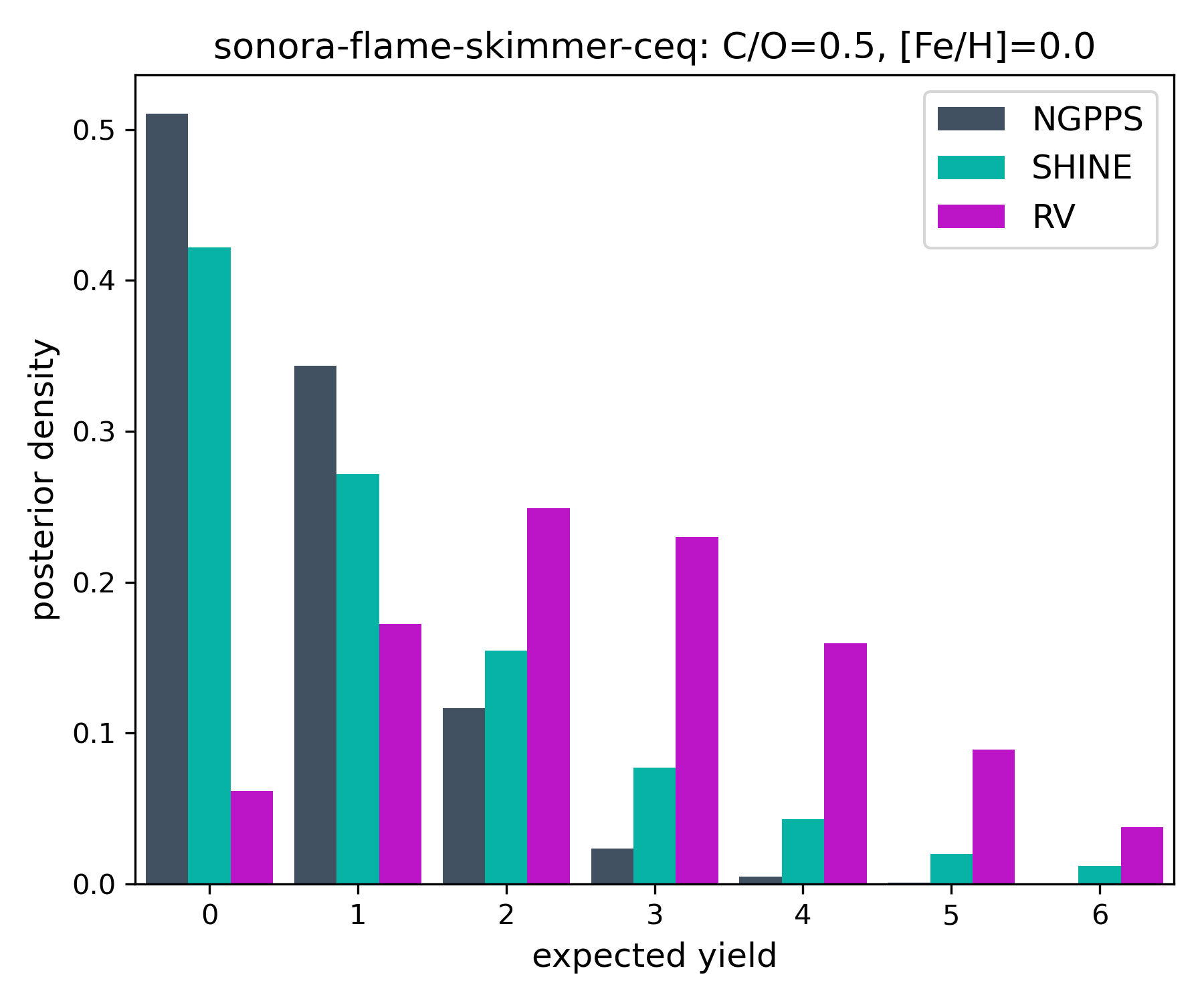}
    \caption{\textit{Left}: Combined 1\,M$_\odot$ and 0.7\,M$_\odot$ NGPPS core accretion synthetic populations, each sampled 20 times with a 0.2\,dex log-space perturbation in mass and semi-major axis to improve sampling at wide separations. The golden colormap shows the average detection completeness for the 97 stars from the survey, for companions up to 10\,M$_\mathrm{Jup}$, for the \texttt{sonora-flame-skimmer-ceq} model with C/O\,=\,0.5 and [Fe/H]\,=\,0.0, with contours marking the 20\%, 50\%, and 80\% detection probability levels. The teal and purple boxes indicate the regions of parameter space used to normalize the 1\,M$_\odot$ and 0.7\,M$_\odot$ NGPPS synthetic populations to the SHINE direct imaging and radial velocity occurrence constraints, respectively. \textit{Right}: distributions of the expected survey yield under the three normalization cases: the intrinsic NGPPS model frequencies (grey), SHINE occurrence rates (teal), and RV measurements (purple), for the same atmospheric predictions.}
    \label{fig:CA_yield}
\end{figure*}

These synthetic populations were normalized in three complementary ways: using the intrinsic occurrence predictions of the population synthesis models themselves, the SPHERE/SHINE direct imaging constraints at wide separations, radial velocity occurrence rates at tighter separations. The empirical normalizations assign a scaling factor to original synthetic populations such that the number of companions falling within a chosen calibration region reproduces the measured occurrence rates (see left panel of Figure~\ref{fig:CA_yield}). For the SHINE survey normalization, we used the occurrence constraints derived over 1--75\,M$_\mathrm{Jup}$ and 5--500\,au as part of the final SPHERE GTO statistical analysis (Chauvin et al.\ in prep.), adopting the core accretion occurrences rates measured for 0.8--1.5\,M$_\odot$ and $<$0.8\,M$_\odot$ hosts using the same synthetic population models as those employed here. For the radial velocity normalization, we used the California Legacy Survey results for giant planets between 2--8\,au around Sun-like stars \citep{Fult21}, and HARPS occurrence rates for giant planets with periods of 1,000--10,000\,days ($\sim$1.5--7\,au) around M dwarfs \citep{Mignon2025}.

To predict the survey yield under each normalization and evolutionary model, we followed a Monte‑Carlo procedure. For each realization, we drew two stellar subsamples from the normalized synthetic populations: $N_{\rm FG}$ stars from the 1\,M$_\odot$ core‑accretion model and $N_{\rm KM}$ stars from the 0.7\,M$_\odot$ model, matching the number of FG and KM stars in our survey. All planets around the drawn model stars were then evaluated against the survey sensitivity using the appropriate FG and KM detection completeness maps. Since wide-orbit giant planets are scarce in the models, we perturbed the mass and semi-major axis of each planet by a Gaussian of width 0.2\,dex in log-space to improve sampling of the population beyond 10\,au. We restricted this yield analysis to companions below 10\,M$_\mathrm{Jup}$, on the assumption that more massive substellar companions would likely already have been detected around our targets by previous surveys, and to focus the analysis on the core accretion dominated regime. Above this mass threshold, gravitational instability is expected to contribute significantly, complicating the interpretation in terms of a single formation channel. 

Detection was treated as a Bernoulli trial: for every injected planet, we drew a uniform random number in $[0,1]$ and classified the planet as detected if the draw fell below its associated detection probability $p_{\rm det}$ given by the completeness map. The total number of detections in each realization was then taken as the sum of Bernoulli successes across all injected planets. Repeating this procedure $10^4$ times yields a Monte‑Carlo distribution of the expected number of detections for the full JWST survey, under a given evolutionary model and normalization choice (right panel of Figure~\ref{fig:CA_yield}).

We show the results of this analysis in Figure~\ref{fig:CA_yield} for the \texttt{sonora-flame-skimmer-ceq} chemical equilibrium model with C/O\,=\,0.5 and [Fe/H]\,=\,0.0, which represents one of the most optimistic cases in terms of survey sensitivity. Under these atmospheric predictions, the mean expected number of detections below 10\,M$_\mathrm{Jup}$ out of our 97 FGKM stars is 0.7, 1.2, and 2.8, using the model, SHINE, and RV normalizations, respectively. The spread in predicted yields across the three normalizations reflects the significant uncertainty in the underlying demographics of gas giant planets at wide separations, with the RV normalization consistently producing the most optimistic predictions by anchoring the population to the comparatively higher occurrence rates measured at smaller separations. In contrast, the most pessimistic case corresponds to the non-equilibrium, high-metallicity atmospheres (\texttt{sonora-flame-skimmer-neq} with C/O\,=\,2.5 and [Fe/H]\,=\,2.0), where the reduced survey completeness under these atmospheric assumptions leads to 0.3, 0.5, and 1.2 mean expected number of detections under the same three normalizations.

\subsubsection{Using parametric models for planetary populations}
\label{subsec: Using parametric models for planetary populations}

As a complementary analysis, we also estimated the survey yield using an empirical parametric model for gas giant demographics. We adopted the parametric model of \cite{Meyer2025}, which represents the substellar companion population as the sum of planet and brown-dwarf components. Both components follow power-law distributions in the companion-to-star mass ratio, $q=M_{\mathrm{comp}}/M_{\star}$, and log-normal distributions in semimajor axis, $a$. The model parameters were constrained using published companion-frequency measurements from radial velocity and direct imaging surveys. The brown dwarf component is truncated below $10\,M_{\mathrm{Jup}}$ and therefore produces no companions within the mass range considered in this study, $M_{\mathrm{p}}\leq10\,M_{\mathrm{Jup}}$. The resulting yield is therefore entirely attributable to the planet component. In this model, the planet semimajor-axis distribution is truncated at 200\,au to encompass the wide-separation planet-forming region observed in extended protoplanetary disks \citep[e.g.,][]{Bae2022}

The original model predicts a rapid decline in planet occurrence at wide separations because of its log-normal semimajor-axis distribution. We also consider an alternative model that allows for a more extended population of sub-Jupiters \citep{Bogat2025}. For planets with masses below $1\,M_{\mathrm{Jup}}$, this model retains the original log-normal semimajor-axis distribution within 10\,au but transitions to a distribution that is uniform in $\log(a)$ from 10 to 200\,au, while leaving the mass-ratio distribution unchanged. 

We applied both population models to the same sample of 97 single FGKM stars used in Section~\ref{subsubsec: Using core accretion synthetic population models}, restricting the generated companions to $M_{\mathrm{p}}\leq10\,M_{\mathrm{Jup}}$ and $a\leq200$\,au. For each star, we constructed a $10\times10$ grid that is uniformly spaced in $\log(a)$ and $\log(q)$. The semi-major axis spans 1--200\,au, while the mass ratio spans the range corresponding to planet masses of $0.1$--$10\,M_{\mathrm{Jup}}$ for the given host-star mass. We calculated the expected planet occurrence in each grid cell by integrating the adopted mass ratio power-law and semimajor-axis distributions over the cell boundaries. The number of planets in each cell was then drawn from a Poisson distribution, where the expectation value is the integrated occurrence rate. Each generated planet is assigned a mass and semimajor axis by sampling uniformly within the corresponding cell. The subsequent detection procedure was identical to that described above: each simulated planet was considered detected through a Bernoulli trial with a success probability equal to its detection probability.

\begin{figure*}
    \includegraphics[width=0.95\textwidth]{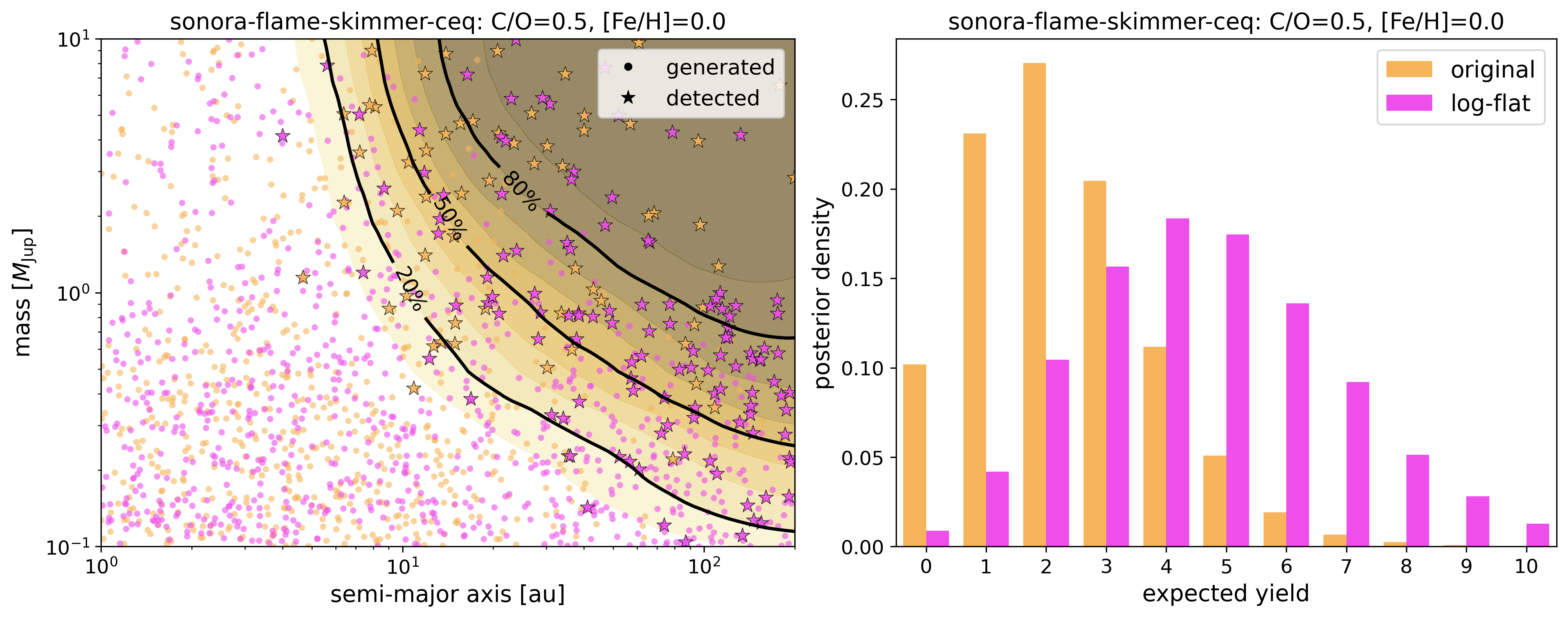}
    \caption{
    Comparison of the original parametric demographic model and the alternative log-flat model for the \texttt{sonora-flame-skimmer-ceq} atmosphere with C/O\,=\,0.5 and [Fe/H]\,=\,0.0.
    \textit{Left:} A randomly selected 0.25\% subset of the planets generated for the survey yield calculation, overlaid on the same average detection-completeness map shown in the left panel of Figure~\ref{fig:CA_yield}. Circles indicate generated planets, while star symbols mark those selected as detections through the Bernoulli trials. The original and log-flat populations are shown in orange and magenta, respectively. The log-flat prescription increases the abundance of sub-Jupiters beyond 10\,au.
    \textit{Right:} Normalized distributions of the expected survey yield, similar to the right panel of Figure~\ref{fig:CA_yield}, for the original and log-flat parametric demographic models.
    }
    \label{fig:parametric_yield}
\end{figure*}

The left panel of Figure~\ref{fig:parametric_yield} shows a randomly selected 0.25\% subset of the generated planets, with detected planets indicated by star symbols. By construction, the log-flat prescription produces an enhanced population of sub-Jupiters beyond 10\,au and consequently more detections in this region. Averaged over the 97 single FGKM stars, the integrated occurrence rate of sub-Jupiters between 10 and 100\,au is approximately 5.8\% for the original model and 12.1\% for the log-flat model.

For the same two atmospheric models discussed in Section~\ref{subsubsec: Using core accretion synthetic population models} ---the \texttt{sonora-flame-skimmer-ceq} model with C/O\,=\,0.5 and [Fe/H]\,=\,0.0, and the \texttt{sonora-flame-skimmer-neq} with C/O\,=\,2.5 and [Fe/H]\,=\,2.0--- we compare the yields predicted by the two population prescriptions. For the C/O\,=\,0.5 and [Fe/H]\,=\,0.0 atmosphere, the original parametric model predicts a mean yield of 2.3 detectable planets among the 97 FGKM stars, whereas the log-flat model increases the mean yield to 4.7 planets. The right panel of Figure~\ref{fig:parametric_yield} shows the corresponding distributions of the expected yield across the Monte Carlo simulations. The log-flat distribution is shifted toward larger yields, reflecting the larger population of wide-orbit sub-Jupiters. For the high-metallicity atmospheres, the predicted mean yield increases from 0.8 planets under the original model to 1.1 planets under the log-flat model.

\subsubsection{Limitations and prospects}

A complete picture of the expected yields across all the \texttt{BEX-ATMO} and \texttt{Sonora Flame Skimmer} family models explored in this manuscript is presented in Table~\ref{Tab: Summury yeild}. The predictions are relatively stable across atmospheric model choices, with differences driven primarily by the underlying assumptions used to normalize the synthetic populations, ranging from $<$1 expected detection under the NGPPS core accretion normalization to $\sim$4 under the parametric log-flat model. The \texttt{sonora-flame-skimmer-neq} is overall the worst case due to their significantly reduced completeness, due to the \texttt{NEQ} models more aggressively  suppressing flux than the \texttt{CEQ} models in F444W (see Section \ref{subsec: Filter comparison} for reference), that result in a lowered expected numbers of detections.

The large disparity between normalization choices reflects a fundamental limitation of any yield prediction in this regime: the region of parameter space probed by our three programs, down to sub-Jupiter companions beyond $\sim$10\,au, has not previously been constrained by any direct imaging surveys or radial velocity campaigns. As a result, all yield predictions must rely either on theoretical models that lack empirical calibration in this regime, or on extrapolations from observed demographics of higher-mass or closer-in companions. The spread in predicted yields depending on the adopted population distribution and normalization reveals that current formation models and measured occurrence rates cannot yet be reconciled into a coherent picture at wide separations and low companion masses. This inconsistency is itself a valuable result, as it underscores the need for direct observational constraints in this regime to discriminate between competing predictions and improve our global understanding of wide-orbit giant planets.

The GO-4050, GO-5835, and SURVEY-6005 campaigns were specifically designed to address these demographic questions. The upcoming GO-10764 program will shortly validate or refute the initial candidate detections from these surveys. Subsequently, matching these empirical results with model predictions (in terms of numbers and properties) 
will provide us with a wealth of new information about the underlying mechanisms behind wide-orbit Jupiter and sub-Jupiter exoplanets, and will establish a critical benchmark for future demographic frameworks.

\begin{table*}[!t]
    \centering
    \caption{Summary of the expected yields for Jupiter mass exoplanets for our programs combined, based on constrains from different synthetic population\label{Tab: Summury yeild}}
    \begin{tabular}{lccccccc}
    \hline
    Model&CO/FeH&NGPPS&SHINE&RV&Parametric (original)&Parametric (log-flat)\\
    \hline
    \hline
    bex-atmo2023-ceq&0.55/0.0&0.6&1.2&2.6&2.1&4.4\\
    \hline
    bex-atmo2023-neq-s&0.55/0.0&0.6&1.1&2.4&1.9&4.2\\
    \hline
    bex-atmo2023-neq-w&0.55/0.0&0.6&1.1&2.4&1.8&4.1\\
    \hline
    sonora-flame-skimmer-ceq&0.5/0.0&0.7&1.2&2.8&2.3&4.7\\
                            &0.5/1.0&0.7&1.2&2.7&2.2&4.6\\
                            &0.5/2.0&0.7&1.2&2.7&2.1&4.4\\
                            &1.0/0.0&0.7&1.2&2.8&2.3&4.7\\
                            &1.0/1.0&0.7&1.2&2.7&2.2&4.5\\
                            &1.0/2.0&0.6&1.1&2.6&2.1&4.4\\
                            &2.5/0.0&0.7&1.3&2.7&2.2&4.8\\
                            &2.5/1.0&0.6&1.2&2.6&2.1&4.4\\
                            &2.5/2.0&0.5&1.0&2.2&1.7&3.4\\
    \hline
    sonora-flame-skimmer-neq&0.5/0.0& 0.6& 1.0&2.4&1.8&3.8\\
                            &0.5/1.0& 0.4& 0.8&1.8&1.2&1.9\\
                            &0.5/2.0& 0.5& 0.8&1.9&1.3&2.2\\
                            &1.0/0.0& 0.5& 1.0&2.2&1.6&3.4\\
                            &1.0/1.0& 0.4& 0.7&1.6&1.1&1.7\\
                            &1.0/2.0& 0.4& 0.8&1.8&1.3&2.1\\
                            &2.5/0.0& 0.5& 0.9&2.0&1.4&2.8\\
                            &2.5/1.0& 0.3& 0.6&1.3&0.9&1.2\\
                            &2.5/2.0& 0.3& 0.5&1.2&0.8&1.1\\
    \end{tabular}
\end{table*}

\section{Conclusion}
\label{sec: Conclusion}
In this manuscript, we presented the results from the sensitivity limits analysis performed for three, recent, JWST programs GO-4050 and GO-5835 \citep{2023jwst.prop.4050C,2024jwst.prop.5835C}, and SURVEY-6005 \citep{2025jwst.prop.6005B}. These three surveys have been carefully designed to survey nearby star forming regions and moving groups to provide a statistical sample of sub-Jupiter-mass exoplanets on wide orbits for demographic study. 

We find the following key results:
\begin{enumerate}
    \item For JWST, the absolute magnitude sensitivity floor reached is governed by the instrument background noise rather than coronagraphic performance. GO-4050 and 5835 (mostly focused on M-dwarfs) exhibit poorer contrast  while converging to identical magnitudes limits as SURVEY-6005 (mostly FGK stars). Moreover, SURVEY-6005’s superior contrast did not improve its absolute magnitude sensitivity at small separations, primarily because its FGK hosts are intrinsically brighter than the M dwarfs in GO-4050/5835. 
    \item SURVEY-6005 exhibits a marginally higher limiting background compared to GO-4050/5835, as shorter readout ramps are required to prevent saturation for the brighter FGK stars observed in SURVEY-6005 compared to the fainter M stars observed in GO-4050/5835.  This causes higher read noise compared to the longer \(\texttt{DEEP8}\) ramps used in GO-4050/5835. These findings suggest that the necessary detector configuration for brighter targets limits the achievable sensitivity in the background-limited regime. \item For the GO-5835 imaging component, the absence of a coronagraph increased the overall throughput and signal-to-noise ratio achieved for this faint M-dwarf sample. This yielded improved performance at closer separation, and a background magnitude limit in F444W that is $\sim 1$ magnitude deeper than the coronagraphic limits ($\sim 22.2$ at $\gtrsim2$ arcsec).
    \item Model selection has a strong impact on the minimum mass/T$_{\mathrm{eff}}$ achievable sensitivity limits.  Nonetheless, using Sonora Flame-Skimmer models, we find that our programs are sensitive to sub-Saturn to Jupiter mass exoplanets with effective Temperature of $\lesssim 275 K$, already at $\sim 2$ arcsec; 
    \item We mapped the NIRCam/MIRI results from \citep{Bowens2025} onto our parameter space, finding consistent $T_{\mathrm{eff}}$ constraints: NIRCam demonstrates superior sensitivity relative to MIRI at distances $\gtrsim 20\text{--}30\text{ pc}$. A comparison between the cloudy cases for the \texttt{Sonora Flame Skimmer} \texttt{CEQ} and \texttt{NEQ} models shows that while clouds may have a suppressive effect for NIRCam coronagraphic observations, increasing metallicity might have a greater potential to limit F444W observations, depending on the atmospheric chemistry of the exoplanet.
    \item Comparing detection probability maps produced with  the \texttt{Sonora Flame Skimmer} \texttt{CEQ} and \texttt{NEQ} models reveals that the detection probability for sub-Jupiter exoplanets is highly sensitive to the choice of equilibrium versus disequilibrium chemistry and model parameters, as well as the spectral type of the hosting star. For separations between $\sim 50$ and $500\text{ au}$, the model-to-model variation in F444W detection probability peaks at $\sim 50\%$ for masses between $\lesssim 1\text{ M}_{\mathrm{J}}$ and $\sim 1\text{ M}_{\mathrm{Sat}}$, and reaches $\sim 75\%$ for effective temperatures of $\sim 300\text{--}400\text{ K}$.
    \item Comparing detection probability maps generated for the K and M spectral type sample with similar maps for the AFG spectral type sample, we observe an increase in sensitivity towards the lowest masses for longer wavelengths with increased detection probabilities of $\sim 50\%$ in favor of the K and M group for $\lesssim 1$ M$_{\mathrm{J}}$ at separation $\gtrsim20$ au, and $\sim 75\%$ at $\gtrsim60$ au, a stark improvement compared to ground-based adaptive optics observations, thanks to the enhanced background limit from space with JWST compared to ground-based observations;
    \item \texttt{PanCAKE} simulations accurately reproduce NIRCam coronagraphic performance, with discrepancies between expected and empirical sensitivities remaining $\lesssim 10\%$ at separations $\gtrsim 30\text{ au}$, while at closer separations our programs perform $\sim 25\%$ better compared to \texttt{PanCAKE} predictions.
    \item A comparison with the SHINE survey confirms that the improvements in mass sensitivity provided by JWST outpace ground-based observations, opening up a completely new parameter space to explore. In particular, in the region between few M$_{\mathrm{J}}$ to $\sim1$ M$_{\mathrm{Sat}}$, JWST is able to reach improved detection probabilities by up to $\sim 50\%$ compared to current ground observations at separations $\sim 50-500$ au;
    \item Exoplanet yield predictions for the survey vary significantly depending on the underlying giant exoplanet  population considered; this survey itself will be critical to refining our understanding of this population.  
\end{enumerate}

To conclude, we want to highlight the crucial role of large surveys ($>$100 stars) to build our understanding of the demographics of exoplanets. We demonstrate that the surveys presented here yield deep direct imaging sensitivity to sub-Jupiter exoplanets for the first time.  Via ongoing follow-up of the candidate exoplanet detections revealed by this novel sensitivity, these surveys will provide the first statistically significant constraints on the frequency of sub-Jupiter exoplanets at semi-major axes of 30-200 AU.

Even though these are just predictions, and the only way to truly understanding the final mass/T$_{\mathrm{eff}}$ sensitivities is to follow-up candidate detections, study their atmospheric properties, and provide a direct constraint of which models are best at describing the reality of the observations, this work still provides the framework in which these detections must be interpreted. With enough targets to investigate multiple scenarios, we were able to deliver a detailed description on how variations in ages, distances, spectral-types, and membership (i.e. the environment in which exoplanet might form) would affect the predicted sensitivities and the nature of possible candidates, while the shared set-up minimize at the same time possible sources of bias. 


\begin{acknowledgments}
BB and BJS acknowledge funding by the UK Science and Technology Facilities Council (STFC) grant nos. ST/V000594/1 and UKRI1196. This work is based on observations made with the NASA/ESA/CSA JWST. The data were obtained from the Mikulski Archive for Space Telescopes (MAST) at the Space Telescope Science Institute, which is operated by the Association of Universities for Research in Astronomy, Inc., under NASA contract NAS 5-03127 for JWST. These observations are associated with program nos. GO~4050 (PI: Carter), GO~5835 (PI: Carter), and SURVEY~6005 (PI: Biller). This research took advantage of the \texttt{spaceKLIP} \citep{Girard2022, Kammerer2022,Carter2023} pipeline, and the pyKLIP tools \citep{Wang2015}, to support the initial data reduction.
This research utilized the \texttt{MADYS} tools \citep{Squicciarini22} to generate the sensitivity limits and detection probability maps on which this work is based on.
This research made use of the Montreal Open Clusters and Associations (MOCA) database, operated at the Montr\'eal Plan\'etarium \citep{Gagne2024}. 
\end{acknowledgments}

\begin{contribution}
Dr. G. M. Strampelli led the analysis of the sensitivity limits and detection probabilities on which this paper is based, as well as the reduction of the data and the assembling of the PSF library for the imaging dataset.
Dr. A. Carter and Professor B. Biller led the three surveys GO-4050, 5835, and 6005 on which this work is based.
Dr. C. Fontanive, Professor M. Meyer, and S. Han led the yield analysis.
J. Mang led the development of the specific \texttt{Sonora Flame Skimmer} models used in this work.
R. Ferrer-Chavez led the reduction of the data and the assembling of the PSF library for the coronagraphic dataset.
Dr. V. Squicciarini developed and adapted the {\tt{MADYS}} tools that provided the framework for this study.
All the other members of the GO-4050, 5835, and 6005 collaboration equally contributed with invaluable feedback.
\end{contribution}

%
\facilities{JWST(NIRCam)}

All the {\it JWST} data used in this paper can be found in MAST:  \dataset[10.17909/3txt-jb91]{http://dx.doi.org/10.17909/3txt-jb91}.
\software{spaceKLIP \citep{Girard2022,Kammerer2022,Carter2023}, PanCAKE \citep{Carter2021}, MADYS \citep{Squicciarini22}, MOCA \citep{Gagne2024}}
This research made use of Photutils, an Astropy package for detection and photometry of astronomical sources \citep{bradley_2026_23353738}.

\FloatBarrier
\newpage
\appendix
\setcounter{table}{0}
\renewcommand{\thetable}{A\arabic{table}}

\section{The target catalog}
\label{apx: The target catalog}

\begin{longrotatetable}
\begin{deluxetable*}{cccccccccccc}
\tablecaption{Description of the 165 sources from Table \ref{tab: Targets summary}. Each source is listed with its known \texttt{Simbad} ID, 2Mass ID, distance (in parsecs), J, K, and F444W magnitudes, contrast at 2" in F444W, membership, GO number, spectral type, and extinction. All the values, with the exception of the contrast, membership, and Program number are obtained from \texttt{MOCAdb}. Ages are assumed based on their respective membership and reported in Table \ref{tab: Targets summary}. \label{tab: Catalog}}
\tablewidth{\textwidth}    
\tablehead{
    \colhead{Simbad ID} & \colhead{2Mass ID} & \colhead{Distance} & \colhead{J} & 
    \colhead{K} & \colhead{F444W} & \colhead{Contrast} & \colhead{Membership} & 
    \colhead{GO} & \colhead{SpT} & \colhead{E(B-V)}
}
\startdata
BD+22 4409 & 21310171+2320074 & 24.218 $\pm$ 0.010 & 7.075 & 6.382 & 6.20 & 2.14e-06 & ABDMG & 6005 & K8V & 0.001 $\pm$ 0.015 \\
CD-26 1643 & 04213348-2543091 & 54.566 $\pm$ 0.056 & 7.193 & 6.873 & 6.87 & 2.79e-06 & ABDMG & 6005 & F9V & 0.0 $\pm$ 0.014 \\
CD-40 1701 & 05023042-3959129 & 55.217 $\pm$ 0.031 & 8.727 & 8.05 & 8.04 & 6.97e-06 & ABDMG & 6005 & K4V & 0.0 $\pm$ 0.014 \\
HD 45270 & 06223097-6013072 & 23.873 $\pm$ 0.008 & 5.433 & 5.045 & 5.05 & 3.62e-06 & ABDMG & 6005 & G1V & 0.0 $\pm$ 0.014 \\
HD 25953 & 04064153+0141020 & 57.082 $\pm$ 0.086 & 6.892 & 6.582 & 6.58 & 1.15e-04 & ABDMG & 6005 & F5V & 0.001 $\pm$ 0.014 \\
HD 61518 & 07370278-5221570 & 61.354 $\pm$ 0.059 & 6.968 & 6.69 & 6.66 & 2.48e-06 & ABDMG & 6005 & F5V & 0.001 $\pm$ 0.014 \\
BD+41 4749 & 23193955+4215098 & 51.290 $\pm$ 0.041 & 7.605 & 7.224 & 7.17 & 3.84e-06 & ABDMG & 6005 & G0V & 0.004 $\pm$ 0.014 \\
HD 224228 & 23561067-3903086 & 21.954 $\pm$ 0.011 & 6.51 & 5.907 & 5.88 & 1.77e-06 & ABDMG & 6005 & K2V & 0.0 $\pm$ 0.015 \\
HD 217343 & 23001929-2609135 & 31.777 $\pm$ 0.022 & 6.291 & 5.943 & 5.90 & 1.76e-06 & ABDMG & 6005 & G3V & 0.001 $\pm$ 0.015 \\
BD+05 4576 & 20395460+0620118 & 36.563 $\pm$ 0.122 & 7.922 & 7.133 & 7.02 & 3.30e-06 & ABDMG & 6005 & K7Ve & 0.001 $\pm$ 0.015 \\
V* HS Psc & 01372322+2657119 & 37.755 $\pm$ 0.026 & 8.429 & 7.642 & 7.56 & 3.50e-02 & ABDMG & 6005 & K5V & 0.001 $\pm$ 0.014 \\
HD 201919 & 21130526-1729126 & 38.431 $\pm$ 0.024 & 8.348 & 7.583 & 7.53 & 4.46e-06 & ABDMG & 6005 & K6V & 0.001 $\pm$ 0.015 \\
HD 35650 & 05243016-3858106 & 17.461 $\pm$ 0.005 & 6.702 & 5.921 & 5.78 & 2.36e-06 & ABDMG & 6005 & K6V & 0.0 $\pm$ 0.014 \\
HD 152555 & 16540815-0420248 & 45.196 $\pm$ 0.047 & 6.7 & 6.363 & 6.35 & 3.12e-04 & ABDMG & 6005 & F9:V & 0.001 $\pm$ 0.016 \\
EROS-MP J0032-4405 & 00325584-4405058 & 34.551 $\pm$ 0.504 & 14.776 & 13.269 & 12.40 & 9.87e-05 & ABDMG & 5835 & L0$\gamma$ & 0.001 $\pm$ 0.015 \\
BD+20 1790 & 07234358+2024588 & 27.712 $\pm$ 0.014 & 7.643 & 6.879 & 6.84 & 2.96e-06 & ABDMG & 6005 & K3(V) & 0.0 $\pm$ 0.014 \\
BD+05 378 & 02412589+0559181 & 44.296 $\pm$ 0.042 & 7.904 & 7.069 & 6.94 & 3.50e-06 & AUSC & 5835 & M0 & 0.001 $\pm$ 0.014 \\
BD-03 5579 & 23093711-0225551 & 52.018 $\pm$ 0.052 & 8.567 & 7.818 & 7.74 & 5.54e-06 & AUSC & 6005 & K4Ve & 0.001 $\pm$ 0.015 \\
2MASS J06085283-2753583 & 06085283-2753583 & 44.209 $\pm$ 0.307 & 13.595 & 12.371 & 11.55 & 7.20e-05 & AUSC & 5835 & L0 $\delta$ & 0.0 $\pm$ 0.014 \\
RX J0019.7+1951 & 00194303+1951117 & 57.471 $\pm$ 0.072 & 10.718 & 9.864 & 9.34 & 1.63e-05 & AUSC & 5835 & M4 & 0.003 $\pm$ 0.014 \\
BD-13 6424 & 23323085-1215513 & 27.452 $\pm$ 0.016 & 7.45 & 6.569 & 6.37 & 3.84e-06 & BPMG & 5835 & M0 & 0.0 $\pm$ 0.014 \\
2MASS J19243494-3442392 & 19243494-3442392 & 51.547 $\pm$ 0.082 & 9.67 & 8.785 & 8.30 & 6.62e-06 & BPMG & 5835 & M4.6 & 0.003 $\pm$ 0.016 \\
PM J19560-3207 & 19560438-3207376 & 51.315 $\pm$ 0.047 & 8.71 & 7.846 & 7.67 & 9.78e-06 & BPMG & 5835 & M0 & 0.003 $\pm$ 0.016 \\
RX J0506.2+0439 & 05061292+0439272 & 27.630 $\pm$ 0.021 & 8.909 & 8.067 & 7.59 & 4.44e-06 & BPMG & 5835 & M4 & 0.0 $\pm$ 0.014 \\
1RXS J184206.5-555426 & 18420694-5554254 & 51.452 $\pm$ 0.061 & 9.488 & 8.584 & 8.21 & 7.28e-06 & BPMG & 5835 & M3.5 & 0.006 $\pm$ 0.016 \\
UCAC4 171-199133 & 18420483-5554126 & 51.652 $\pm$ 0.057 & 10.677 & 9.852 & 9.29 & 1.86e-05 & BPMG & 5835 & M4.6 & 0.006 $\pm$ 0.016 \\
BPS CS 22898-0065 & 21100535-1919573 & 32.365 $\pm$ 0.030 & 8.112 & 7.196 & 6.93 & 2.87e-06 & BPMG & 5835 & M2 & 0.001 $\pm$ 0.015 \\
PM J00502+0837 & 00501752+0837341 & 61.308 $\pm$ 0.196 & 9.745 & 8.897 & 8.34 & 7.43e-06 & BPMG & 5835 & M5 & 0.002 $\pm$ 0.014 \\
UCAC4 354-189365 & 21100461-1920302 & 33.619 $\pm$ 0.706 & 8.428 & 7.552 & 7.08 & 5.29e-05 & BPMG & 5835 & M5 & 0.001 $\pm$ 0.015 \\
2MASS J19355595-2846343 & 19355595-2846343 & 57.156 $\pm$ 0.925 & 13.953 & 12.712 & 11.85 & 7.43e-05 & BPMG & 5835 & M8.5 & 0.002 $\pm$ 0.017 \\
LP 467-16 & 01112542+1526214 & 21.933 $\pm$ 0.859 & 9.082 & 8.208 & 7.71 & 2.80e-03 & BPMG & 5835 & M5.5 & 0.0 $\pm$ 0.014 \\
RX J2137.6+0137 & 21374019+0137137 & 35.913 $\pm$ 0.161 & 8.802 & 7.884 & 7.46 & 7.29e-06 & BPMG & 5835 & M5 & 0.001 $\pm$ 0.015 \\
V* WW PsA & 22445794-3315015 & 20.867 $\pm$ 0.011 & 7.786 & 6.932 & 6.48 & 2.38e-06 & BPMG & 5835 & M4 & 0.0 $\pm$ 0.015 \\
1RXS J001723.4-664507 & 00172353-6645124 & 36.823 $\pm$ 0.023 & 8.563 & 7.703 & 7.39 & 3.89e-06 & BPMG & 5835 & M3 & 0.001 $\pm$ 0.015 \\
SIPS J2335-3401 & 23355015-3401477 & 37.376 $\pm$ 0.063 & 11.639 & 10.758 & 10.17 & 2.87e-05 & BPMG & 5835 & M6 & 0.001 $\pm$ 0.015 \\
PM J22088+1144 & 22085034+1144131 & 37.388 $\pm$ 0.038 & 9.901 & 9.037 & 8.54 & 8.52e-06 & BPMG & 5835 & M4.3 & 0.001 $\pm$ 0.015 \\
SIPS J1809-7613 & 18090694-7613239 & 27.277 $\pm$ 0.016 & 9.817 & 8.989 & 8.43 & 7.52e-06 & BPMG & 5835 & M5 & 0.002 $\pm$ 0.015 \\
V* AO Men & 06182824-7202416 & 39.115 $\pm$ 0.019 & 7.53 & 6.814 & 6.73 & 2.56e-06 & BPMG & 5835 & K3.5 & 0.001 $\pm$ 0.015 \\
HD 203 & 00065008-2306271 & 39.741 $\pm$ 0.054 & 5.451 & 5.24 & 5.23 & 3.73e-06 & BPMG & 6005 & F3 & 0.001 $\pm$ 0.014 \\
BD+30 397 & 02272924+3058246 & 40.945 $\pm$ 0.025 & 7.87 & 7.08 & 7.04 & 3.70e-06 & BPMG & 5835 & K8 & 0.001 $\pm$ 0.014 \\
CD-31 16041 & 18504448-3147472 & 49.457 $\pm$ 0.033 & 8.314 & 7.462 & 7.36 & 4.79e-06 & BPMG & 6005 & K8 & 0.002 $\pm$ 0.016 \\
1RXS J203336.9-255654 & 20333759-2556521 & 43.662 $\pm$ 0.077 & 9.712 & 8.877 & 8.38 & 9.09e-06 & BPMG & 5835 & M4.8 & 0.002 $\pm$ 0.016 \\
HD 207043 & 21475527-5255502 & 33.243 $\pm$ 0.019 & 6.415 & 6.05 & 6.02 & 7.99e-06 & BPMG & 5835 & G5V & 0.002 $\pm$ 0.016 \\
PM J21214-6655 & 21212873-6655063 & 31.563 $\pm$ 0.574 & 7.88 & 7.009 & 6.92 & 4.74e-06 & BPMG & 5835 & M0 & 0.002 $\pm$ 0.015 \\
2MASS J22334687-2950101 & 22334687-2950101 & 52.124 $\pm$ 0.270 & 12.941 & 11.971 & 11.39 & 7.30e-05 & BPMG & 5835 & M6 & 0.001 $\pm$ 0.015 \\
UPM J1354-7121 & 13545390-7121476 & 22.805 $\pm$ 0.010 & 8.549 & 7.672 & 7.36 & 5.06e-06 & BPMG & 5835 & M2.5 & 0.001 $\pm$ 0.015 \\
SIPS J2000-7523 & 20004841-7523070 & 29.403 $\pm$ 0.094 & 12.734 & 11.511 & 10.83 & 6.00e-05 & BPMG & 5835 & M9 & 0.002 $\pm$ 0.015 \\
2MASSW J0335020+234235 & 03350208+2342356 & 50.702 $\pm$ 0.231 & 12.25 & 11.261 & 10.68 & 4.50e-05 & BPMG & 5835 & M8.5 & 0.001 $\pm$ 0.014 \\
2MASS J19260075-5331269 & 19260075-5331269 & 47.760 $\pm$ 0.064 & 9.595 & 8.684 & 8.22 & 7.37e-06 & BPMG & 5835 & M4 & 0.006 $\pm$ 0.016 \\
Smethells 20 & 18465255-6210366 & 50.700 $\pm$ 0.048 & 8.746 & 7.854 & 7.68 & 9.04e-06 & BPMG & 5835 & M1.5 & 0.005 $\pm$ 0.016 \\
CD-57 1054 & 05004714-5715255 & 26.873 $\pm$ 0.009 & 7.095 & 6.244 & 6.07 & 2.35e-06 & BPMG & 5835 & M0 & 0.0 $\pm$ 0.014 \\
HD 173167 & 18480637-6213470 & 50.571 $\pm$ 0.056 & 6.416 & 6.136 & 6.11 & 2.56e-06 & BPMG & 6005 & F5 & 0.005 $\pm$ 0.016 \\
2MASS J19300396-2939322 & 19300396-2939322 & 60.253 $\pm$ 0.084 & 10.117 & 9.253 & 8.78 & 1.21e-05 & BPMG & 5835 & M4 & 0.003 $\pm$ 0.017 \\
Barta 161 12 & 01351393-0712517 & 37.281 $\pm$ 0.064 & 8.964 & 8.078 & 7.69 & 3.99e-06 & BPMG & 5835 & M4 & 0.0 $\pm$ 0.014 \\
2MASS J02241739+2031513 & 02241739+2031513 & 70.753 $\pm$ 0.363 & 12.48 & 11.619 & 11.00 & 4.67e-05 & BPMG & 5835 & M6 & 0.005 $\pm$ 0.014 \\
GR* 9 & 00281434-3227556 & 35.008 $\pm$ 0.029 & 10.121 & 9.278 & 8.73 & 8.90e-06 & BPMG & 5835 & M5 & 0.0 $\pm$ 0.014 \\
GJ 2006 A & 00275023-3233060 & 35.029 $\pm$ 0.048 & 8.882 & 8.012 & 7.49 & 5.24e-06 & BPMG & 5835 & M3.5 & 0.0 $\pm$ 0.014 \\
RX J0534.0-0221 & 05335981-0221325 & 34.346 $\pm$ 0.032 & 8.564 & 7.695 & 7.33 & 3.98e-06 & BPMG & 5835 & M3 & 0.0 $\pm$ 0.014 \\
V* V1005 Ori & 04593483+0147007 & 24.396 $\pm$ 0.008 & 7.117 & 6.261 & 6.08 & 2.77e-06 & BPMG & 5835 & M0 & 0.0 $\pm$ 0.014 \\
HD 358623 & 20560274-1710538 & 45.818 $\pm$ 0.207 & 7.849 & 7.039 & 6.99 & 9.53e-03 & BPMG & 6005 & K7V & 0.001 $\pm$ 0.015 \\
RX J0613.2-2742 & 06131330-2742054 & 33.772 $\pm$ 0.249 & 8.002 & 7.145 & 6.80 & 1.85e-05 & BPMG & 5835 & M3.8 & 0.0 $\pm$ 0.014 \\
RX J0529.7-3239 & 05294468-3239141 & 29.762 $\pm$ 0.017 & 9.215 & 8.316 & 7.80 & 5.22e-06 & BPMG & 5835 & M4.5 & 0.0 $\pm$ 0.014 \\
2MASS J18471351-2808558 & 18471351-2808558 & 59.915 $\pm$ 0.153 & 11.725 & 10.831 & 10.29 & 1.23e-04 & BPMG & 5835 & M5.3 & 0.003 $\pm$ 0.017 \\
HD 42270 & 05532917-8156532 & 58.822 $\pm$ 0.042 & 7.532 & 6.944 & 6.77 & 6.47e-03 & CAR & 6005 & K0 & 0.003 $\pm$ 0.015 \\
HD 44627 & 06191291-5803156 & 50.137 $\pm$ 0.031 & 7.576 & 6.981 & 6.86 & 3.92e-05 & CAR & 6005 & K1 & 0.0 $\pm$ 0.014 \\
HD 49855 & 06434625-7158356 & 55.389 $\pm$ 0.044 & 7.693 & 7.278 & 7.21 & 3.31e-06 & CAR & 6005 & G6 & 0.002 $\pm$ 0.016 \\
HD 55279 & 07003046-7941459 & 63.435 $\pm$ 0.054 & 8.265 & 7.652 & 7.66 & 1.89e-05 & CAR & 6005 & K2 & 0.003 $\pm$ 0.015 \\
S1* 329 & 08413264-6825403 & 30.972 $\pm$ 0.011 & 8.227 & 7.379 & 11.87 & 2.77e-04 & CARN & 6005 & K7 & 0.001 $\pm$ 0.015 \\
2MASSI J0421072-630602 & 04210718-6306022 & 20.248 $\pm$ 1.367 & 15.565 & 13.448 & 12.24 & 1.51e-04 & CARN & 5835 & L5 & 0.0 $\pm$ 0.014 \\
HD 62850 & 07423606-5917507 & 32.856 $\pm$ 0.017 & 6.072 & 5.701 & 5.71 & 1.58e-06 & CARN & 6005 & G2.5 & 0.001 $\pm$ 0.014 \\
HD 63608 & 07461694-5948341 & 33.764 $\pm$ 0.021 & 6.862 & 6.411 & 6.27 & 3.19e-06 & CARN & 6005 & K0V & 0.001 $\pm$ 0.014 \\
1RXS J040710.6-291823 & 04071148-2918342 & 75.962 $\pm$ 2.036 & 9.055 & 8.188 & 8.03 & 1.19e-03 & COL & 5835 & M4 & 0.001 $\pm$ 0.014 \\
HD 37484 & 05373962-2837346 & 58.838 $\pm$ 0.056 & 6.47 & 6.277 & 6.24 & 2.38e-06 & COL & 6005 & F4 & 0.0 $\pm$ 0.014 \\
V* AS Col & 05203803-3945179 & 47.340 $\pm$ 0.034 & 6.416 & 6.144 & 6.10 & 1.85e-06 & COL & 6005 & F6 & 0.0 $\pm$ 0.014 \\
HD 41071 & 06004130-4453500 & 54.113 $\pm$ 0.034 & 7.73 & 7.321 & 7.53 & 1.67e-03 & COL & 6005 & G8 & 0.0 $\pm$ 0.014 \\
HD 38397 & 05433580-3955246 & 53.594 $\pm$ 0.049 & 7.056 & 6.756 & 6.74 & 2.79e-06 & COL & 6005 & G0 & 0.0 $\pm$ 0.014 \\
HD 40216 & 05554314-3806162 & 52.767 $\pm$ 0.044 & 6.494 & 6.206 & 6.16 & 2.86e-03 & COL & 6005 & F7 & 0.0 $\pm$ 0.014 \\
2MASS J03550477-1032415 & 03550477-1032415 & 52.264 $\pm$ 0.466 & 13.08 & 11.975 & 11.32 & 1.48e-03 & COL & 5835 & M8.3 & 0.0 $\pm$ 0.014 \\
PM J05501+0511 & 05500858+0511536 & 64.060 $\pm$ 0.077 & 9.369 & 8.471 & 8.25 & 8.05e-06 & COL & 5835 & M2 & 0.001 $\pm$ 0.014 \\
UCAC4 345-006842 & 05082729-2101444 & 48.304 $\pm$ 0.139 & 9.716 & 8.833 & 8.34 & 8.99e-06 & COL & 5835 & M5 & 0.0 $\pm$ 0.014 \\
V* V1358 Ori & 06190805-0326203 & 51.672 $\pm$ 0.060 & 6.848 & 6.552 & 6.49 & 2.23e-06 & COL & 6005 & G0V & 0.0 $\pm$ 0.014 \\
CD-48 4797 & 09331427-4848331 & 46.278 $\pm$ 0.024 & 8.945 & 8.101 & 8.02 & 7.69e-06 & FCCN & 6005 & K7V & 0.001 $\pm$ 0.015 \\
* e Vir & 13164653+0925269 & 17.588 $\pm$ 0.038 & 4.392 & 4.033 & 3.75 & 2.28e-06 & FIELD & 6005 & G0IV & 0.0 $\pm$ 0.015 \\
V* LV Com & 12075091+1856558 & 41.963 $\pm$ 0.037 & 7.544 & 7.028 & 10.00 & 6.15e-05 & FIELD & 6005 & K2V & 0.001 $\pm$ 0.014 \\
2MASS J02495639-0557352 & 02495639-0557352 & 66.251 $\pm$ 0.250 & 11.963 & 11.055 & 10.43 & 3.49e-03 & FIELD & 5835 & M6 & 0.001 $\pm$ 0.014 \\
V* AF Psc & 23314492-0244395 & 34.927 $\pm$ 0.043 & 9.507 & 8.667 & 8.13 & 6.67e-06 & FIELD & 5835 & M5 & 0.001 $\pm$ 0.014 \\
HD 221239 & 23301372+3142210 & 25.086 $\pm$ 0.014 & 6.66 & 6.145 & 6.07 & 1.70e-06 & FIELD & 6005 & K2.5V & 0.001 $\pm$ 0.014 \\
PM J13219-1052 & 13215631-1052098 & 50.396 $\pm$ 0.783 & 9.524 & 8.618 & 8.26 & 1.16e-05 & FIELD & 5835 & M3.5Ve & 0.001 $\pm$ 0.016 \\
MCC 124 & 10141918+2104297 & 23.368 $\pm$ 0.103 & 7.074 & 6.261 & 6.07 & 7.10e-06 & FIELD & 5835 & M1V & 0.0 $\pm$ 0.014 \\
1RXS J112755.0-662558 & 11275535-6626046 & 109.246 $\pm$ 0.141 & 9.006 & 8.358 & 8.29 & 7.32e-05 & MEL4 & 6005 & K1V(e) & 0.016 $\pm$ 0.017 \\
TWA 46 & 12354615-4115531 & 52.931 $\pm$ 0.172 & 10.068 & 9.232 & 8.77 & 1.06e-05 & OCTN & 4050 & M3 & 0.003 $\pm$ 0.016 \\
CPD-63 1286 & 10065573-6352086 & 136.533 $\pm$ 1.272 & 9.262 & 8.58 & 8.46 & 1.25e-05 & SCOCEN & 6005 & K0V(e) & 0.014 $\pm$ 0.017 \\
CD-53 544 & 02414683-5259523 & 43.598 $\pm$ 0.087 & 7.582 & 6.763 & 6.68 & 2.74e-06 & THA & 6005 & K6 & 0.001 $\pm$ 0.014 \\
CD-78 24 & 00422033-7747397 & 49.792 $\pm$ 0.024 & 8.22 & 7.529 & 7.49 & 4.00e-06 & THA & 6005 & K3.5 & 0.002 $\pm$ 0.015 \\
HD 24636 & 03481148-7441388 & 57.112 $\pm$ 0.049 & 6.367 & 6.136 & 6.12 & 2.33e-06 & THA & 6005 & F3 & 0.002 $\pm$ 0.015 \\
HD 32195 & 04480518-8046452 & 62.474 $\pm$ 0.068 & 7.17 & 6.868 & 6.87 & 2.79e-06 & THA & 6005 & F7 & 0.002 $\pm$ 0.015 \\
Smethells 86 & 21443012-6058389 & 46.414 $\pm$ 0.030 & 8.751 & 7.874 & 7.72 & 4.81e-06 & THA & 6005 & K8 & 0.003 $\pm$ 0.015 \\
HD 207575 & 21520973-6203085 & 46.761 $\pm$ 0.032 & 6.358 & 6.027 & 6.03 & 2.93e-06 & THA & 6005 & F6 & 0.003 $\pm$ 0.015 \\
UCAC3 13-31217 & 23585674-8339423 & 55.635 $\pm$ 0.038 & 9.121 & 8.246 & 8.13 & 1.20e-05 & THA & 6005 & K5.8 & 0.003 $\pm$ 0.015 \\
CD-86 147 & 23274943-8613187 & 63.366 $\pm$ 0.046 & 7.95 & 7.5 & 7.45 & 4.19e-06 & THA & 6005 & G8 & 0.003 $\pm$ 0.015 \\
HD 202917 & 21204994-5302030 & 46.709 $\pm$ 0.034 & 7.386 & 6.908 & 6.94 & 3.42e-06 & THA & 6005 & G6 & 0.004 $\pm$ 0.015 \\
HD 987 & 00135300-7441178 & 45.864 $\pm$ 0.024 & 7.406 & 6.962 & 6.86 & 2.77e-06 & THA & 6005 & G7 & 0.002 $\pm$ 0.015 \\
HD 17250 & 02461462+0535333 & 57.043 $\pm$ 0.078 & 6.859 & 6.517 & 6.42 & 3.64e-03 & THA & 6005 & F7 & 0.001 $\pm$ 0.014 \\
2MASS J04231498-1533245 & 04231498-1533245 &  &  &  & 11.27 & 3.87e-03 & THA & 5835 & K7.5 &  \\
HD 3221 & 00345120-6154583 & 44.420 $\pm$ 0.042 & 7.337 & 6.533 & 6.40 & 3.02e-06 & THA & 6005 & K5 & 0.001 $\pm$ 0.015 \\
HD 25402 & 04003198-4144544 & 51.907 $\pm$ 0.085 & 7.203 & 6.875 & 6.64 & 1.30e-03 & THA & 6005 & G3 & 0.0 $\pm$ 0.014 \\
HD 105 & 00055255-4145109 & 38.830 $\pm$ 0.032 & 6.464 & 6.117 & 6.10 & 2.22e-06 & THA & 6005 & G0 & 0.001 $\pm$ 0.015 \\
SCR J0336-2619 & 03363144-2619578 & 46.131 $\pm$ 0.220 & 10.675 & 9.763 & 9.26 & 1.69e-05 & THA & 5835 & M5.8 & 0.0 $\pm$ 0.014 \\
V* CT Tuc & 00251465-6130483 & 44.148 $\pm$ 0.024 & 8.615 & 7.749 & 7.56 & 4.10e-06 & THA & 6005 & K7 & 0.001 $\pm$ 0.015 \\
DENIS J004135.3-562112 & 0041353-562112 & 36.789 $\pm$ 0.958 & 11.964 & 10.864 & 10.29 & 6.47e-05 & THA & 5835 & M7.8 & 0.001 $\pm$ 0.015 \\
HD 8558 & 01232126-5728507 & 45.268 $\pm$ 0.032 & 7.241 & 6.847 & 6.75 & 3.21e-06 & THA & 6005 & G6 & 0.001 $\pm$ 0.015 \\
2MASS J02045317-5346162 & 02045317-5346162 & 45.279 $\pm$ 0.035 & 10.438 & 9.563 & 9.05 & 1.28e-05 & THA & 6005 & M3.8 & 0.001 $\pm$ 0.014 \\
CD-35 1167 & 03190864-3507002 & 45.407 $\pm$ 0.034 & 8.576 & 7.723 & 7.61 & 4.86e-06 & THA & 6005 & K7 & 0.0 $\pm$ 0.014 \\
BD-15 705 & 04021648-1521297 & 54.700 $\pm$ 0.039 & 8.228 & 7.573 & 7.54 & 4.25e-06 & THA & 6005 & K3.4 & 0.0 $\pm$ 0.014 \\
V* DK Cet & 01574896-2154052 & 41.313 $\pm$ 0.042 & 6.856 & 6.472 & 6.49 & 2.10e-06 & THA & 6005 & G4 & 0.0 $\pm$ 0.014 \\
HD 10269 & 01390761-5625459 & 46.925 $\pm$ 0.036 & 6.335 & 5.989 & 5.97 & 1.77e-06 & THA & 6005 & F5 & 0.001 $\pm$ 0.015 \\
HD 10863 & 01460105-2720556 & 44.892 $\pm$ 0.062 & 5.723 & 5.456 & 5.41 & 2.09e-06 & THA & 6005 & F0IV & 0.0 $\pm$ 0.014 \\
CD-46 1064 & 03304909-4555573 & 43.895 $\pm$ 0.020 & 7.772 & 7.103 & 7.07 & 2.80e-06 & THA & 6005 & K3 & 0.0 $\pm$ 0.014 \\
PM J05334-5117 & 05332558-5117131 & 53.880 $\pm$ 0.034 & 8.986 & 8.158 & 8.01 & 8.81e-06 & THA & 6005 & K7 & 0.0 $\pm$ 0.014 \\
HD 29615 & 04384393-2702018 & 54.349 $\pm$ 0.044 & 7.273 & 6.866 & 6.87 & 3.47e-06 & THA & 6005 & G3 & 0.0 $\pm$ 0.014 \\
V* V1252 Cen & 12350424-4136385 & 57.483 $\pm$ 0.085 & 9.122 & 8.186 & 8.29 & 7.55e-06 & TWA & 4050 & M3 & 0.003 $\pm$ 0.016 \\
HD 298936 & 10131476-5230540 & 53.085 $\pm$ 0.031 & 7.87 & 7.194 & 7.16 & 5.74e-06 & TWA & 6005 & K3Ve & 0.003 $\pm$ 0.016 \\
TWA 33 & 11393382-3040002 & 48.903 $\pm$ 0.061 & 9.985 & 9.118 & 8.45 & 8.72e-06 & TWA & 4050 & M4.75 & 0.001 $\pm$ 0.016 \\
SCR J1237-4021 & 12371238-4021480 & 63.601 $\pm$ 0.101 & 9.473 & 8.515 & 8.19 & 6.23e-06 & TWA & 4050 & M2.5Ve & 0.003 $\pm$ 0.016 \\
CD-36 7429A & 11482422-3728491 & 76.489 $\pm$ 0.081 & 8.684 & 7.848 & 7.71 & 2.11e-03 & TWA & 4050 & K7IVe & 0.003 $\pm$ 0.016 \\
HD 109573 & 12360103-3952102 & 70.773 $\pm$ 0.236 & 5.784 & 5.769 & 5.82 & 7.51e-04 & TWA & 4050 & A0 & 0.004 $\pm$ 0.016 \\
* 10 Crt & 11084400-2804504 & 54.606 $\pm$ 0.224 & 5.313 & 5.235 & 5.24 & 2.58e-06 & TWA & 4050 & A2 & 0.001 $\pm$ 0.015 \\
TWA 20 & 12313807-4558593 & 80.207 $\pm$ 0.155 & 9.331 & 8.412 & 8.12 & 6.03e-06 & TWA & 4050 & M3IVe & 0.006 $\pm$ 0.016 \\
V* V1217 Cen & 11210549-3845163 & 65.404 $\pm$ 0.082 & 8.999 & 8.053 & 7.86 & 6.20e-06 & TWA & 4050 & M1Ve & 0.002 $\pm$ 0.016 \\
TWA 45 & 11592786-4510192 & 72.258 $\pm$ 0.490 & 9.934 & 9.065 & 8.64 & 9.44e-06 & TWA & 4050 & M4.5 & 0.004 $\pm$ 0.016 \\
CD-34 7390A & 11211723-3446454 & 59.847 $\pm$ 0.059 & 8.431 & 7.491 & 7.40 & 4.78e-03 & TWA & 4050 & M1Ve & 0.001 $\pm$ 0.016 \\
TWA 34 & 10284580-2830374 & 61.176 $\pm$ 0.134 & 10.953 & 10.026 & 9.03 & 1.92e-05 & TWA & 4050 & M6$\gamma$ & 0.001 $\pm$ 0.014 \\
TWA 30 & 11321831-3019518 & 47.419 $\pm$ 0.099 & 9.641 & 8.765 & 8.15 & 5.82e-06 & TWA & 4050 & M4.5Ve & 0.001 $\pm$ 0.015 \\
TWA 26 & 11395113-3159214 & 46.722 $\pm$ 0.512 & 12.686 & 11.503 & 10.78 & 6.56e-05 & TWA & 4050 & M9$\gamma$ & 0.001 $\pm$ 0.015 \\
2MASS J09462782-4457408 & 09462782-4457408 & 46.646 $\pm$ 0.068 & 10.876 & 9.992 & 9.42 & 1.85e-05 & TWA & 5835 & M5.5 & 0.001 $\pm$ 0.015 \\
CD-26 8623 & 11324124-2651559 & 45.959 $\pm$ 0.060 & 8.337 & 7.43 & 7.18 & 8.08e-06 & TWA & 4050 & M2Ve & 0.001 $\pm$ 0.015 \\
UCAC3 133-139341 & 10585054-2346206 & 43.796 $\pm$ 0.058 & 10.302 & 9.43 & 8.87 & 1.22e-05 & TWA & 4050 & M6$\gamma$e & 0.001 $\pm$ 0.015 \\
V* V1249 Cen & 12153072-3948426 & 53.604 $\pm$ 0.074 & 8.166 & 7.306 & 7.16 & 2.89e-06 & TWA & 4050 & K9IV-Ve & 0.002 $\pm$ 0.016 \\
TWA 35 & 12002750-3405371 & 70.765 $\pm$ 0.374 & 9.606 & 8.723 & 8.28 & 1.17e-05 & TWA & 4050 & M4 & 0.002 $\pm$ 0.016 \\
TWA 44 & 12175920-3734433 & 77.504 $\pm$ 0.229 & 11.572 & 10.716 & 10.19 & 3.96e-05 & TWA & 4050 & M5$\gamma$e & 0.004 $\pm$ 0.016 \\
V* CE Ant & 10423011-3340162 & 34.098 $\pm$ 0.032 & 7.792 & 6.899 & 6.60 & 2.76e-06 & TWA & 4050 & M3.25 & 0.001 $\pm$ 0.015 \\
TWA 29 & 12451416-4429077 & 82.393 $\pm$ 2.295 & 14.518 & 13.369 & 12.59 & 3.41e-04 & TWA & 4050 & M9.5$\gamma$ & 0.006 $\pm$ 0.017 \\
TWA 36 & 12023799-3328402 & 62.576 $\pm$ 0.112 & 10.686 & 9.847 & 9.36 & 1.80e-05 & TWA & 4050 & M4.8 & 0.002 $\pm$ 0.016 \\
TWA 23 & 12072738-3247002 & 56.822 $\pm$ 0.288 & 8.618 & 7.751 & 7.46 & 4.21e-06 & TWA & 4050 & M2.5Ve & 0.002 $\pm$ 0.016 \\
HD 59747 & 07330059+3701475 & 20.680 $\pm$ 0.009 & 6.09 & 5.589 & 5.61 & 1.96e-06 & UMA & 6005 & G5V & 0.0 $\pm$ 0.014 \\
HD 56168 & 07210672+6739425 & 25.811 $\pm$ 0.013 & 6.784 & 6.246 & 6.16 & 2.88e-06 & UMA & 6005 & K0V & 0.0 $\pm$ 0.014 \\
HD 238090 & 12122084+5429086 & 15.249 $\pm$ 0.003 & 6.875 & 6.059 & 5.86 & 1.95e-06 & UMA & 6005 & K7V & 0.0 $\pm$ 0.014 \\
HD 76218 & 08555568+3611464 & 25.389 $\pm$ 0.014 & 6.258 & 5.83 & 5.82 & 2.69e-06 & UMA & 6005 & G9 & 0.0 $\pm$ 0.014 \\
HD 7590 & 01162926+4256219 & 24.166 $\pm$ 0.014 & 5.515 & 5.177 & 5.19 & 7.63e-06 & UMA & 6005 & G0V & 0.0 $\pm$ 0.014 \\
HD 26923 & 04152879+0611128 & 22.075 $\pm$ 0.014 & 5.201 & 4.903 & 4.87 & 2.30e-06 & UMA & 6005 & G0IV & 0.0 $\pm$ 0.014 \\
HD 26913 & 04152578+0611588 & 22.071 $\pm$ 0.012 & 5.695 & 5.271 & 5.40 & 2.73e-06 & UMA & 6005 & G8V & 0.0 $\pm$ 0.014 \\
V* V1386 Ori & 06064047+1532317 & 15.771 $\pm$ 0.007 & 5.317 & 4.822 & 4.83 & 4.18e-06 & UMA & 6005 & K0V & 0.0 $\pm$ 0.014 \\
HD 110463 & 12414450+5543288 & 22.722 $\pm$ 0.011 & 6.569 & 6.003 & 5.96 & 1.98e-06 & UMA & 6005 & K3V & 0.001 $\pm$ 0.014 \\
HD 165185 & 18062370-3601113 & 17.115 $\pm$ 0.012 & 4.835 & 4.469 & 4.48 & 5.96e-06 & UMA & 6005 & G1V & 0.001 $\pm$ 0.015 \\
HD 217813 & 23030499+2055068 & 24.295 $\pm$ 0.014 & 5.492 & 5.148 & 5.14 & 3.93e-06 & UMA & 6005 & G1V & 0.001 $\pm$ 0.014 \\
HD 115043 & 13133699+5642298 & 25.161 $\pm$ 0.010 & 5.675 & 5.334 & 5.32 & 5.08e-04 & UMA & 6005 & G1V & 0.001 $\pm$ 0.014 \\
HD 184960 & 19341977+5114121 & 25.431 $\pm$ 0.018 & 4.7 & 4.494 & 4.50 & 1.92e-06 & UMA & 6005 & F7V & 0.001 $\pm$ 0.015 \\
* 18 Boo & 14191627+1300155 & 26.106 $\pm$ 0.052 & 4.739 & 4.394 & 4.39 & 2.33e-06 & UMA & 6005 & F5IV & 0.001 $\pm$ 0.015 \\
HD 167389 & 18130721+4128315 & 34.674 $\pm$ 0.017 & 6.224 & 5.918 & 5.89 & 1.45e-06 & UMA & 6005 & F8(V) & 0.003 $\pm$ 0.015 \\
HD 111456 & 12483942+6019115 & 26.151 $\pm$ 0.199 & 5.178 & 4.555 & 4.50 & 1.44e-05 & UMA & 6005 & F5V & 0.001 $\pm$ 0.014 \\
HD 75935 & 08534994+2654477 & 37.893 $\pm$ 0.042 & 7.102 & 6.653 & 6.69 & 2.71e-06 & UMA & 6005 & G8V & 0.001 $\pm$ 0.014 \\
V* AN CrB & 15353016+3612347 & 30.576 $\pm$ 0.013 & 7.011 & 6.496 & 9.39 & 3.82e-05 & UMA & 6005 & K0V & 0.002 $\pm$ 0.015 \\
V* DL Dra & 14420323+6115429 & 41.148 $\pm$ 0.043 & 5.477 & 5.246 & 5.21 & 1.72e-03 & UMA & 6005 & F2V & 0.004 $\pm$ 0.014 \\
BD-03 2002 & 07400289-0336130 & 14.086 $\pm$ 0.005 & 6.377 & 5.568 & 5.47 & 1.98e-06 & UMA & 6005 & K7V & 0.0 $\pm$ 0.014 \\
HD 29329 & 04460056+7636399 & 32.579 $\pm$ 0.020 & 5.535 & 5.284 & 5.27 & 1.91e-06 & VUL & 6005 & F7V & 0.001 $\pm$ 0.014 \\
\enddata 
\end{deluxetable*}
\end{longrotatetable}

\FloatBarrier

\section{Accounting for Binarity in the Imaging Data}
\label{apx: Accounting for Binarity in the Imaging Data}

\begin{figure*}[!th]
    \centering
    \includegraphics[width=0.45\textwidth]{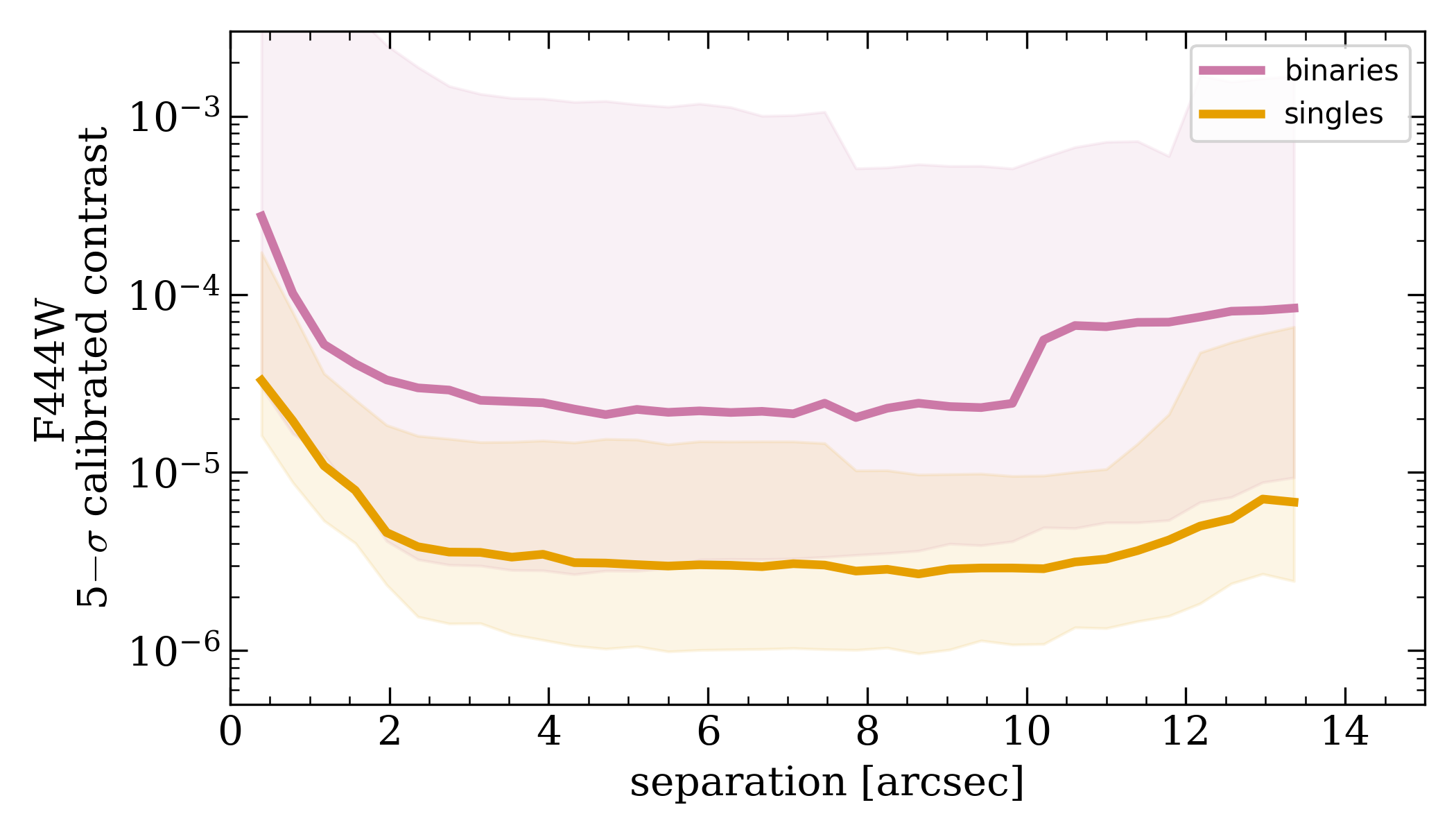}
    \includegraphics[width=0.45\textwidth]{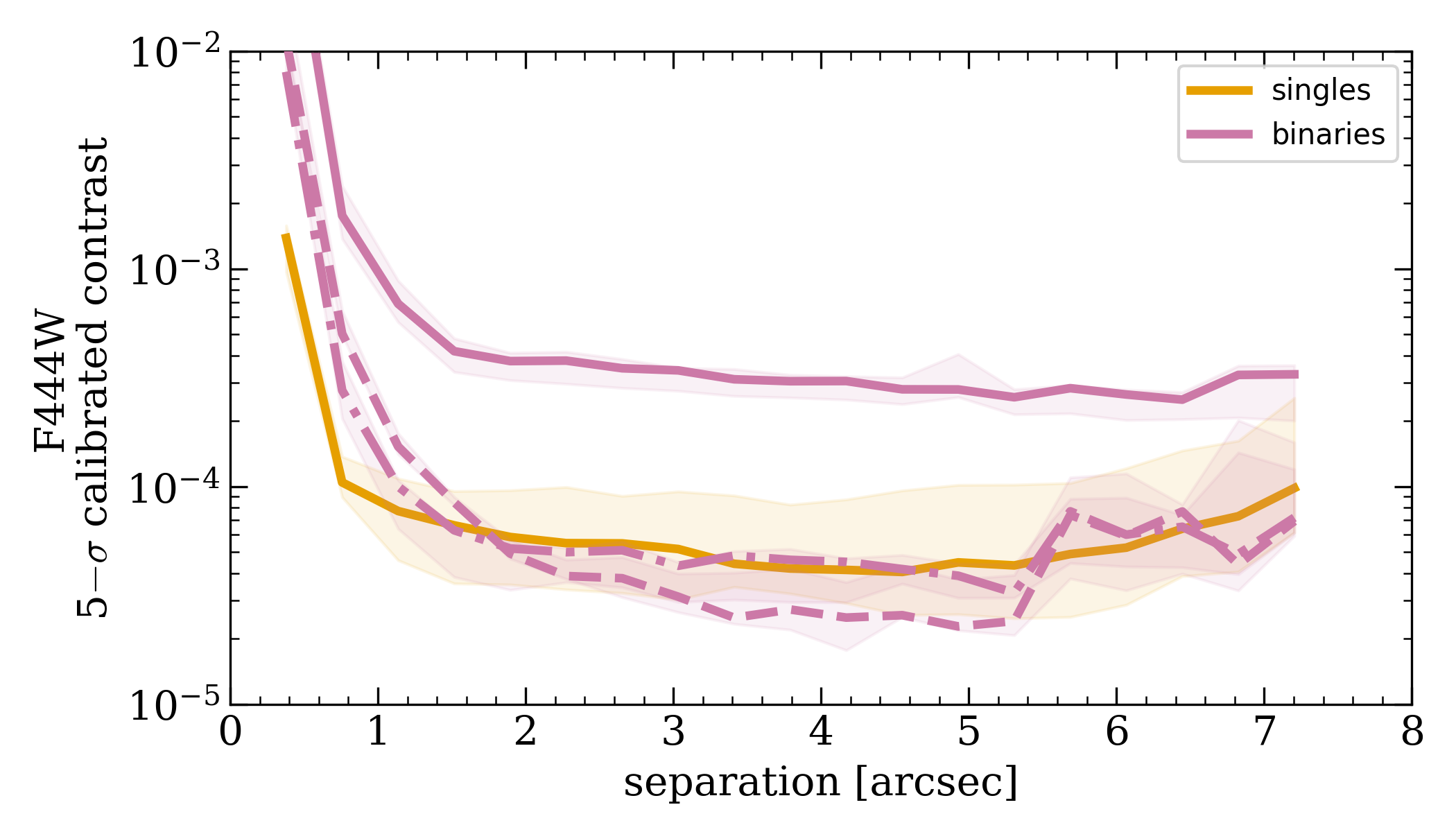}
    \caption{Left: Calibrated contrast curve comparison between isolated sources (orange) and binaries (magenta) in the GO-4050, GO-5835, and SURVEY-6005 surveys combined. Right: Comparison between median calibrated contrast curves of the three binaries in the GO-5835 imaging dataset: 2MASS J04231498-1533245, 2MASS J02495639-0557352, and 2MASS J03550477-1032415, before (magenta continuous line) and after removing the visual companion using models generated with \texttt{STPSF} (magenta dashed lines), and \texttt{Photutils EPSF} module (magenta dashed-dotted lines). For comparison is also shown the median calibrated contrast curves obtained from all the other targets in the NIRCam imaging sample that are not visual binaries (orange continuous line). In both plots, the lines marks the median of the distribution, while the shaded area cover the $16th/84th$ percentile. }
    \label{fig:binary_disk_calcon}
\end{figure*}

As shown by Table \ref{tab: Targets types}, $\sim28\%$ of our targets host a visually resolved binary companion.
Binaries require a more curated approach, since the presence of a bright close-in visually resolved companion will heavily bias any attempt at correctly recentering and aligning the primary to the center of the frame. Additionally, they can also heavily bias the residuals and the final contrast curves, if light from the companion or disk is not carefully removed. 

Thus, we must establish a new approach to handle them carefully. Even though binaries will be the focus of other papers (e.g. Bruinsma et al. in prep.), in this section we introduce a  new approach implemented in \texttt{spaceKLIP} to recenter the three visually resolved binaries in the imaging dataset (2MASS-J04231498-1533245, 2MASS-J02495639-0557352, and 2MASS-J03550477-1032415).

For this purpose, we developed a new MCMC tool for the pipeline \citep[taking advantage of \texttt{emcee}\footnote{\url{https://emcee.readthedocs.io/en/stable/}}, a Python implementation of Goodman \& Weare’s Affine Invariant Markov chain Monte Carlo - MCMC - Ensemble sampler;][]{Foreman2013} 
to fit for two different PSFs simultaneously. 
The goal is to perform a double PSF-fitting to find the best match between the image coordinates and brightness of the primary, and the separation, position angle and contrast of the companion. To do so, the tool allows the user to choose between two distinct approaches, using either \texttt{STPSF} \citep{STPSF1, STPSF2}\footnote{\url{https://stpsf.readthedocs.io/en/latest/index.html}}, or the \texttt{Photuils EPSF} module\footnote{\url{https://photutils.readthedocs.io/en/latest/user_guide/epsf.html}}. \texttt{STPSF} is a Python package that computes simulated point spread functions (PSFs) for NASA’s James Webb Space Telescope (JWST) and Nancy Grace Roman Space Telescope, taking into account detector pixel scales, rotations, filter profiles, and point source spectra. The \texttt{EPSF} module from \texttt{Photutils} instead, derives an empirical PSF, or effective PSF (ePSF), by using a set of real sources from the surveys to model what fraction of a star’s light will land in a particular pixel \citep{Anderson2016,Anderson2000}. Once the best fit is obtained, the retrieved image coordinates of the primary can be used to evaluate the shift needed to recenter the couple with the primary at the center of the frame. 

Before we can estimate contrast limits for these targets, we need to mitigate the loss of sensitivity caused by the presence of the visual binary companion in the field of view (as shown by the left panel of Figure \ref{fig:binary_disk_calcon}).

To do so, we took advantage of the \texttt{extract\_companions} functionality of \texttt{spaceKLIP} \texttt{Analysis} tool, where we extracted the bright companion from the residual of each target binary, producing a new residual where the visual companion was greatly attenuated. Similar to what we describe in section \ref{subsec: Data reduction} to perform double PSF fitting to recenter the three visual binaries, we model the PSF of the bright close-in companion using both the \texttt{STPSF} and \texttt{EPSF}.

In the right panel of Figure \ref{fig:binary_disk_calcon} we show the outcome of this process:  the \textit{calibrated} contrast curves of the three visual binaries in the GO-5835 imaging dataset pre- (continuous lines) and post-companion removal using two different methods to generate the PSF for the companion: \texttt{STPSF} (dashed lines) and \texttt{Photuils EPSF} module (dashed-dotted lines). For comparison, in the same figure we also show in orange the median \textit{calibrated} contrast curves obtained from the other targets in the GO-5835 imaging sample. As expected, when the bright companion is removed, we see a net improvement in the contrast, more in line with the median contrast curve for isolated targets. Overall, once we remove the binaries, we reach the same $\lesssim 10^{-4}$ contrast level at separations $\gtrsim 1-1.5$ arcsec of the GO-5835 isolated targets in the imaging dataset.
\FloatBarrier

\section{Group-based Detection Probability maps}
\label{apx: Group-based Detection probability maps}
\begin{figure*}
    \centering
    \includegraphics[width=0.31\textwidth]{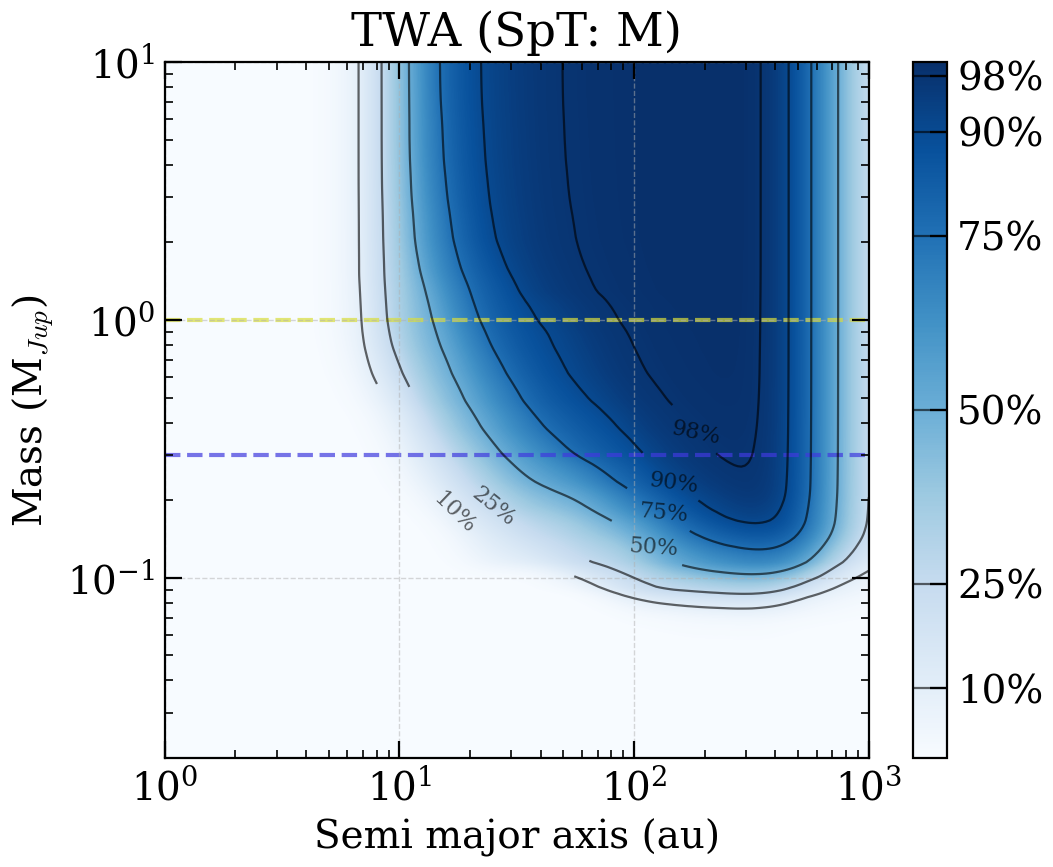}
    \includegraphics[width=0.31\textwidth]{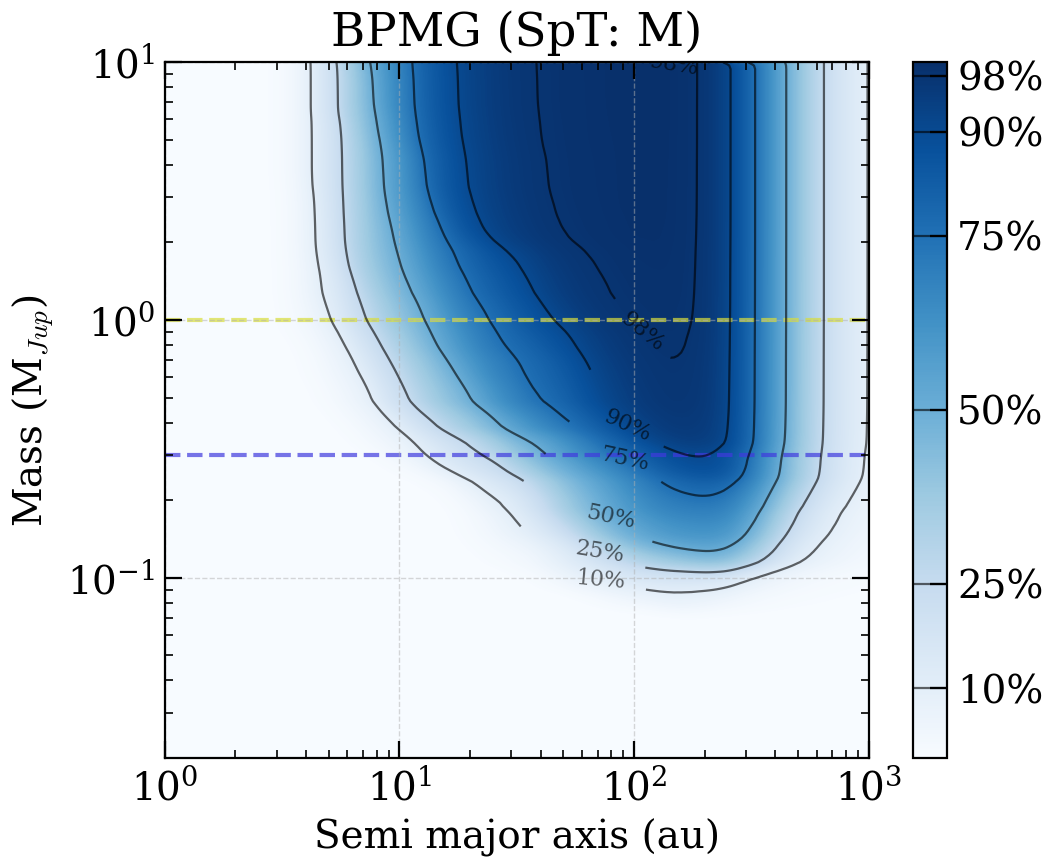}
    \includegraphics[width=0.31\textwidth]{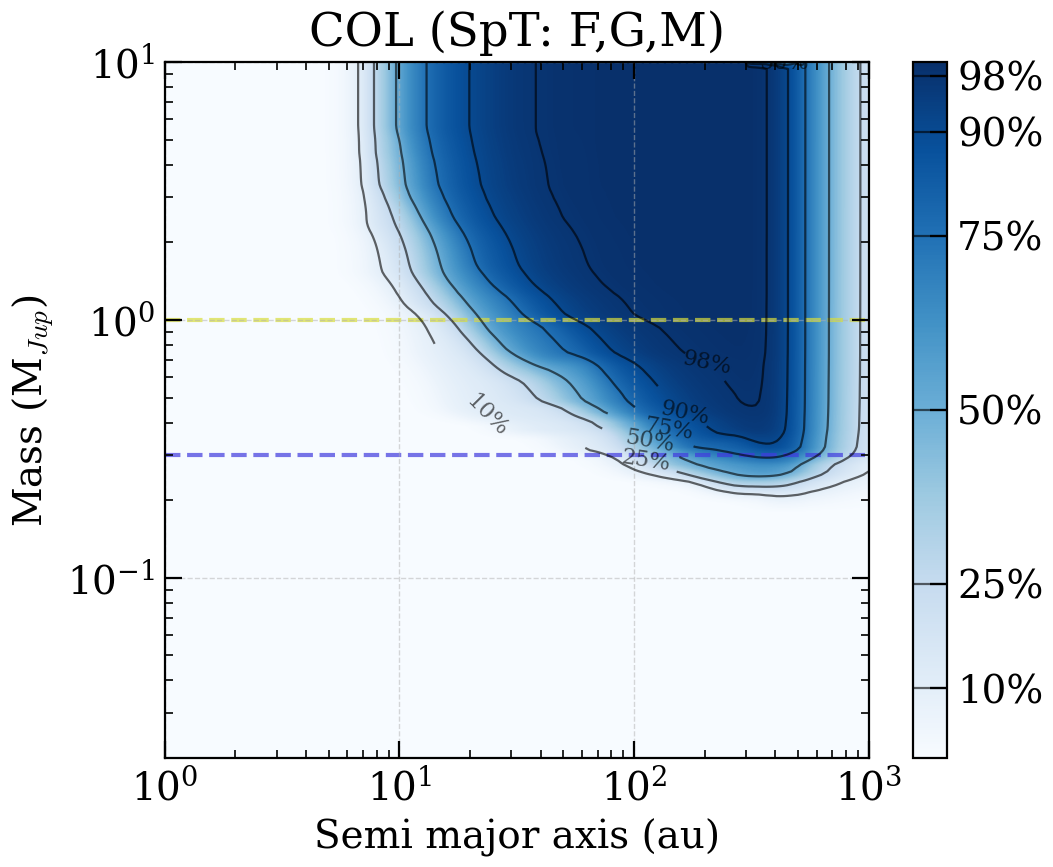}
    \includegraphics[width=0.31\textwidth]{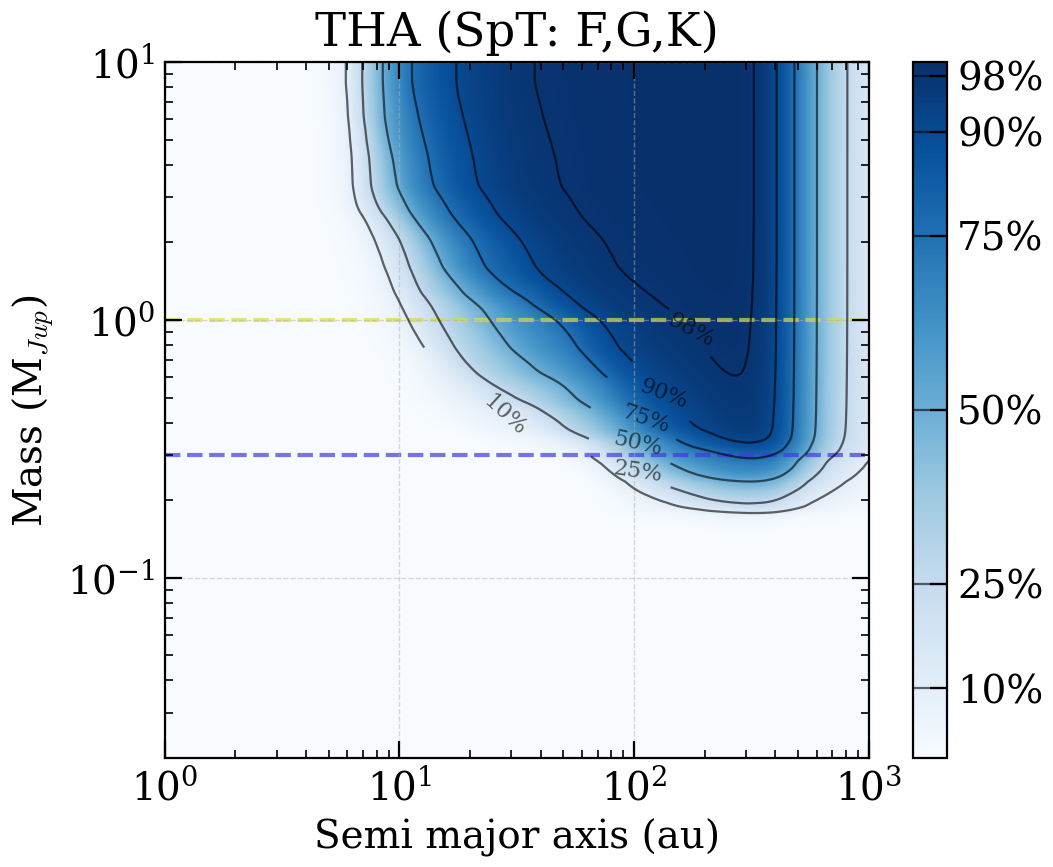}
    \includegraphics[width=0.31\textwidth]{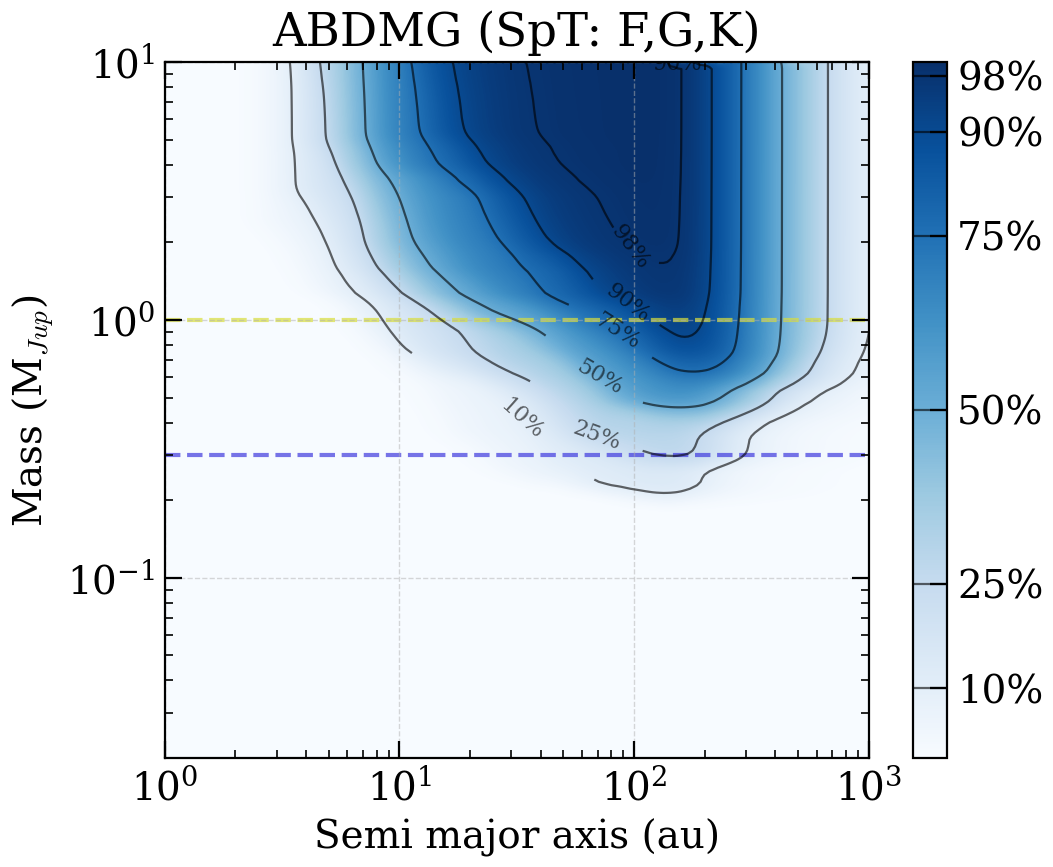}
    \includegraphics[width=0.31\textwidth]{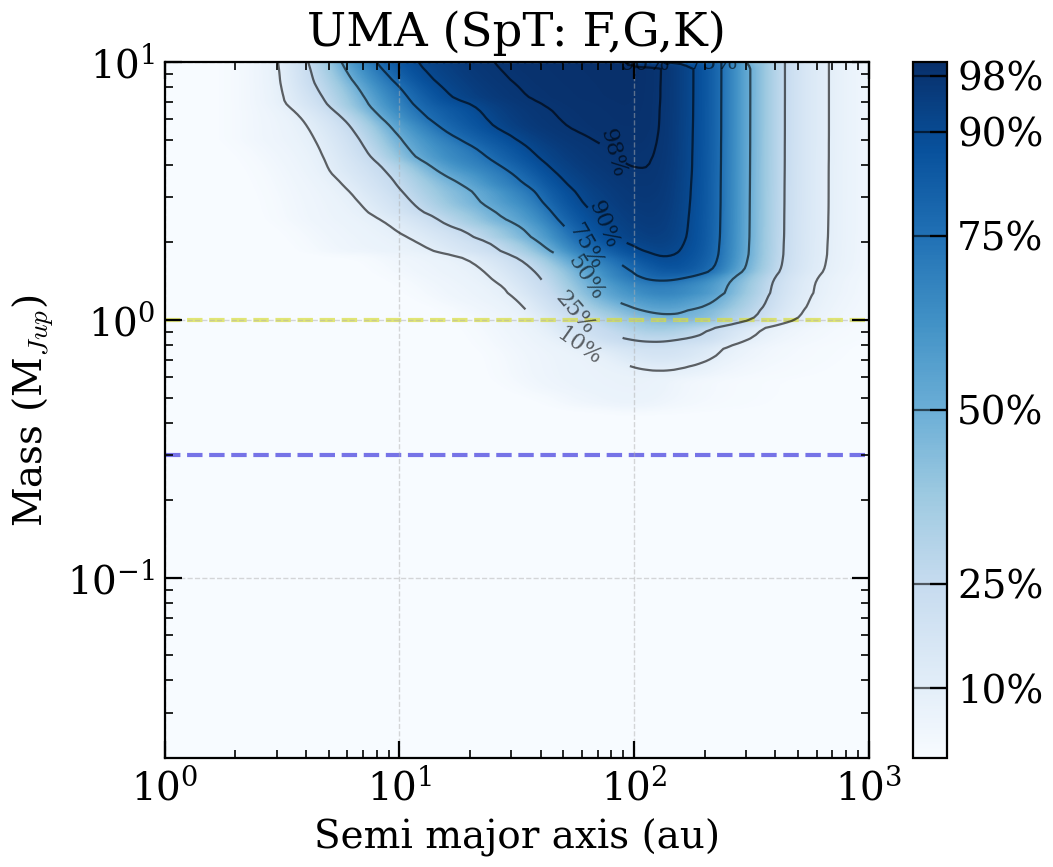}
    \caption{Mass detection probability maps comparison for two \texttt{Sonora Flame Skimmer} models, ordered by age (younger on the left).  The most common spectral types (SpT) in each sample are labeled for each group.
    All maps are derived from the F444W sensitivity limits shown in Figure \ref{fig: Contrast/Magnitude}.
    The yellow and blue horizontal dashed lines show the Jupiter and Saturn reference masses.}
    \label{fig: dmaps}
\end{figure*}

\begin{figure*}
\centering
    \includegraphics[width=0.31\textwidth]{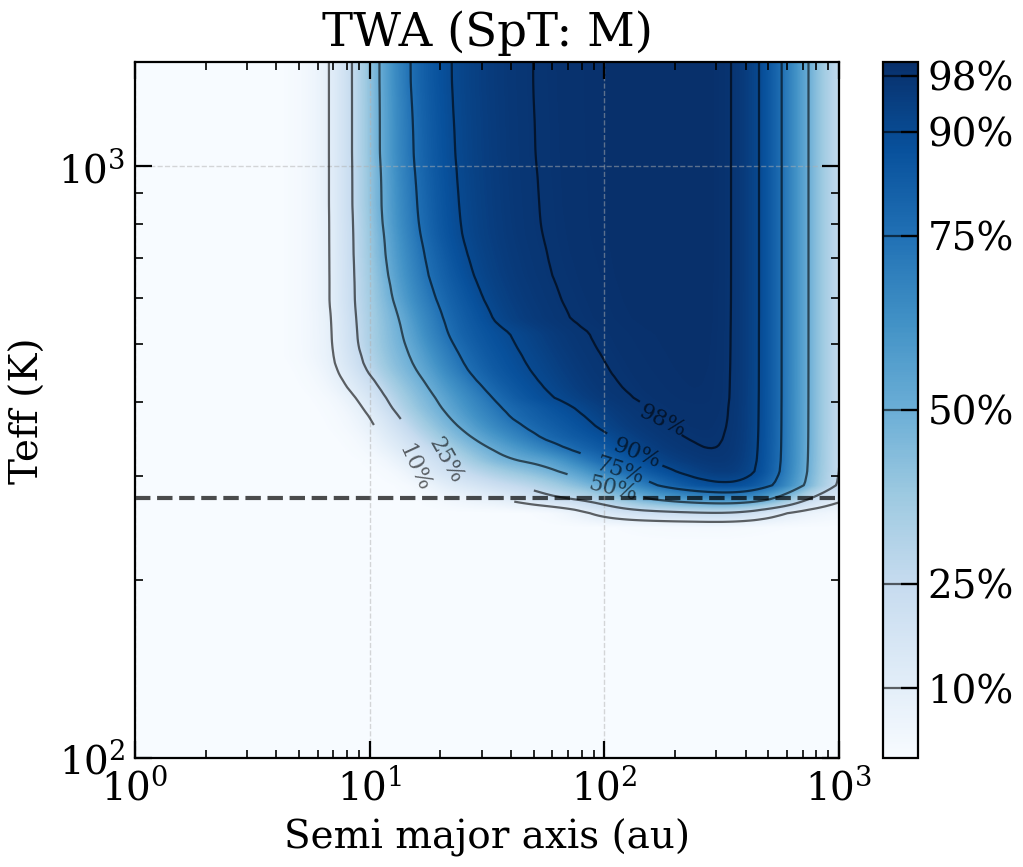}
    \includegraphics[width=0.31\textwidth]{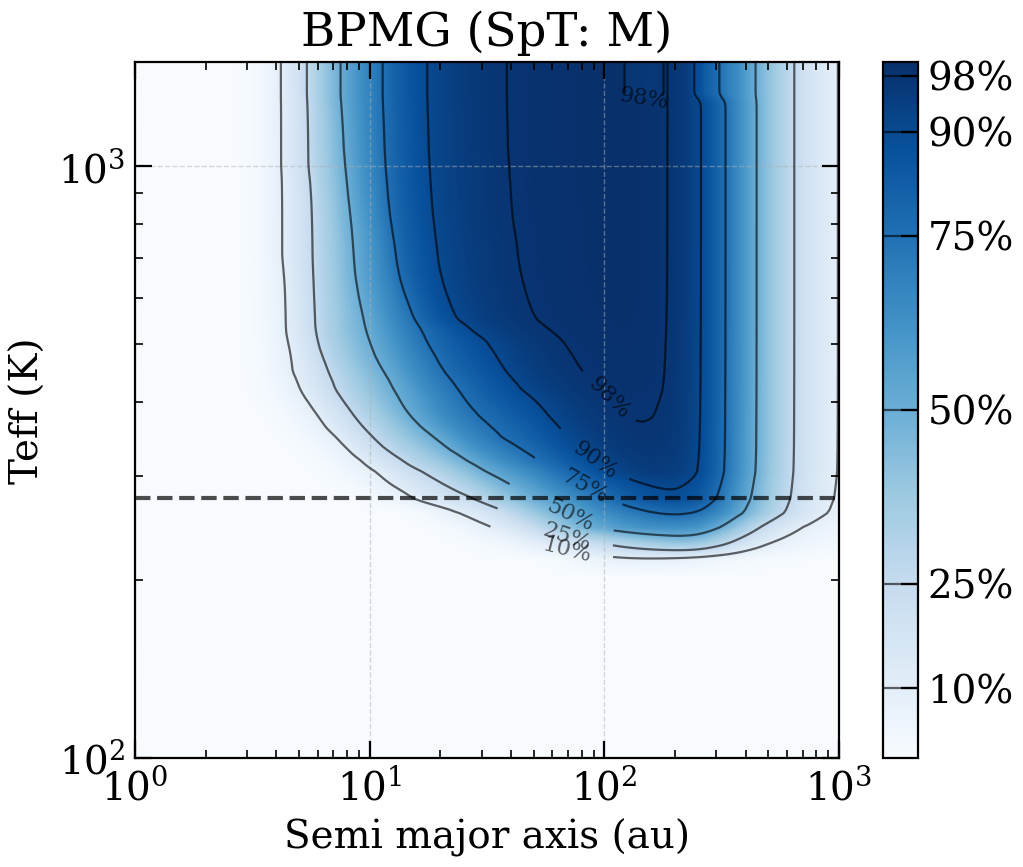}
    \includegraphics[width=0.31\textwidth]{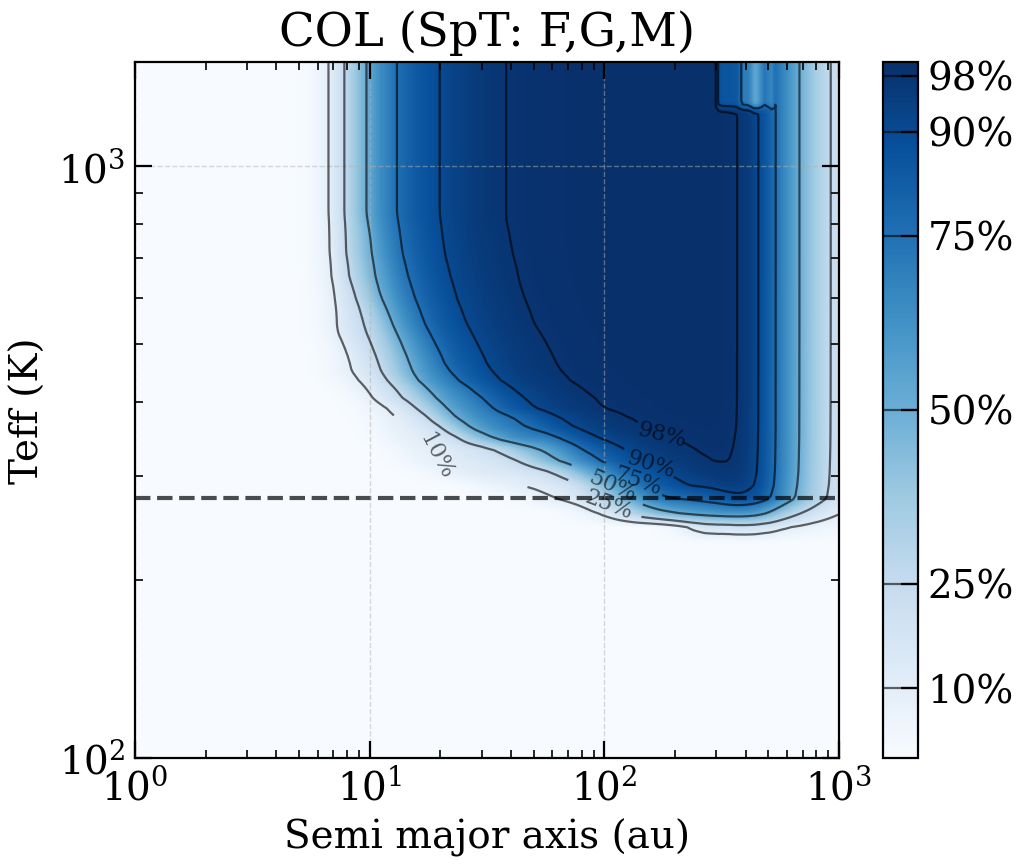}
    \includegraphics[width=0.31\textwidth]{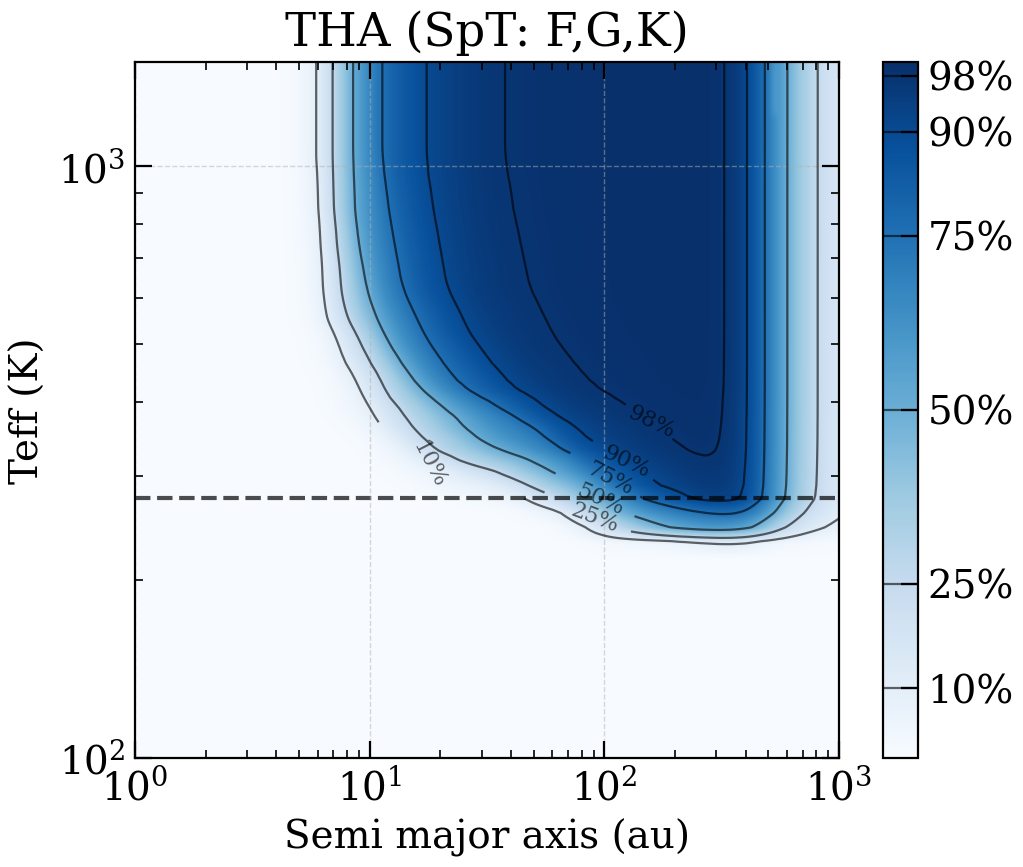}
    \includegraphics[width=0.31\textwidth]{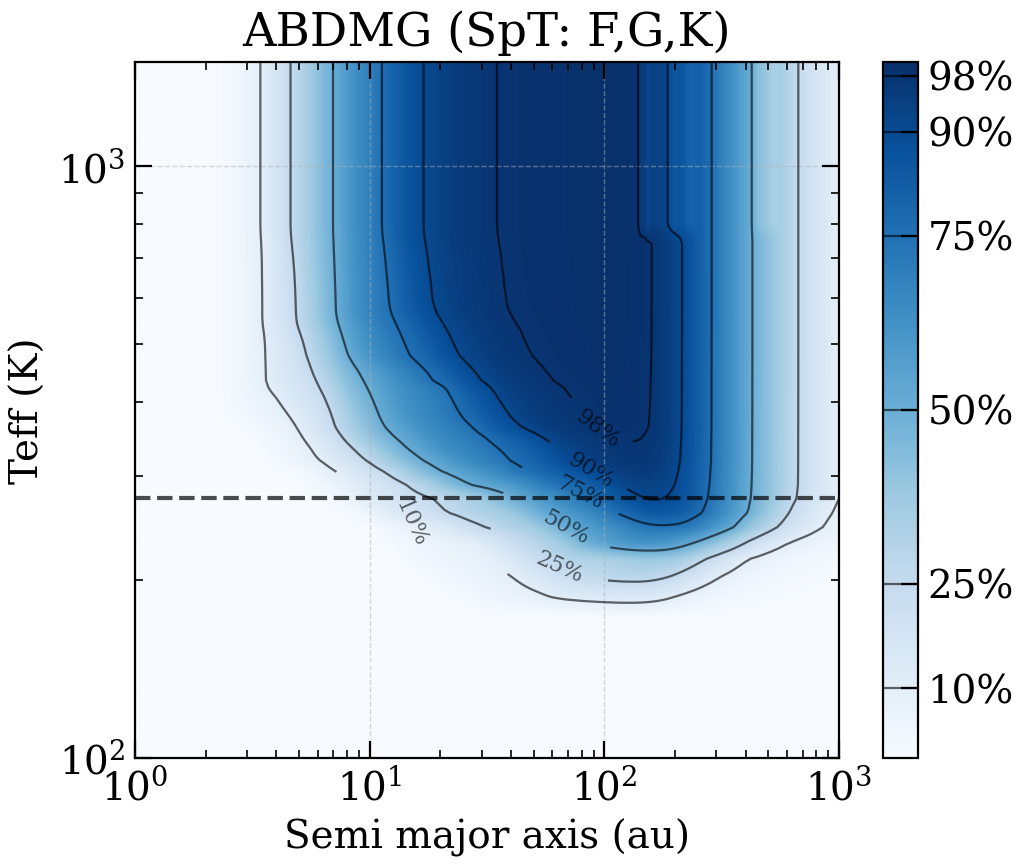}
    \includegraphics[width=0.31\textwidth]{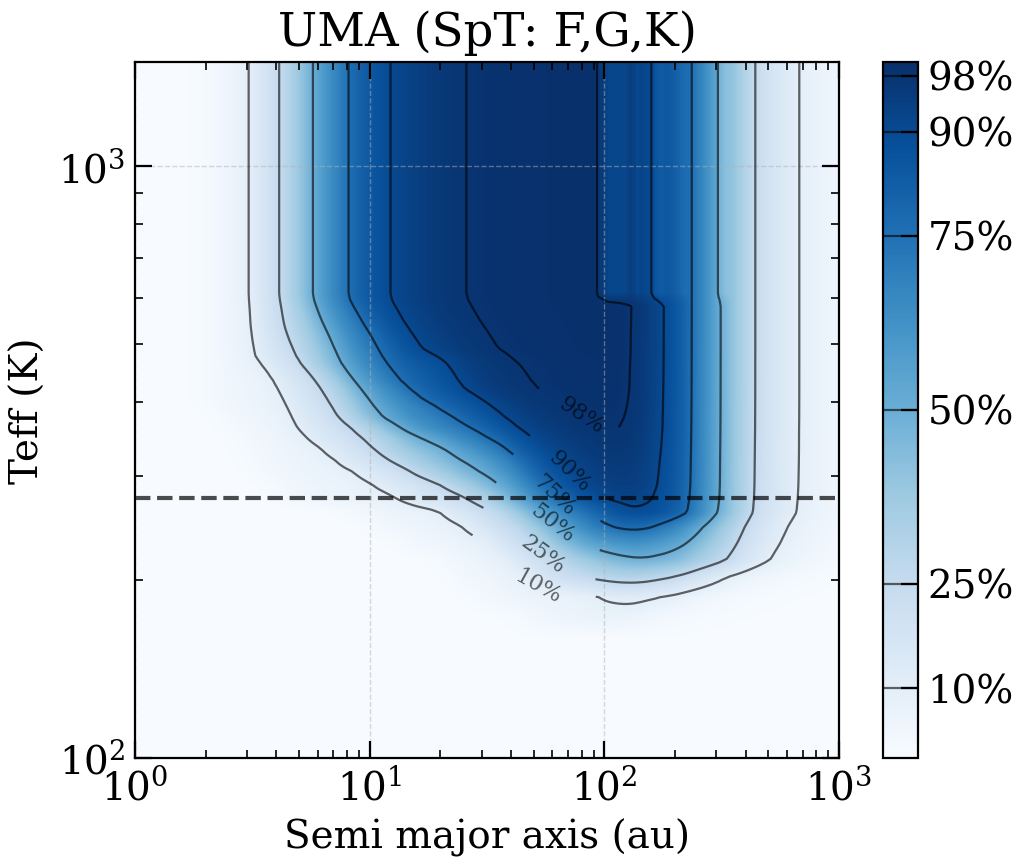}
    \caption{T$_{\mathrm{eff}}$ detection probability maps comparison for two \texttt{Sonora Flame Skimmer} models, ordered by age (younger on the left).  The most common spectral types (SpT) in each sample are labeled for each group.
    All maps are derived from the F444W sensitivity limits shown in Figure \ref{fig: Contrast/Magnitude}. The  black horizontal dashed line shows the Epsilon Ind AB reference temperature \citep{Matthews2024}. }
    \label{fig: Tdmaps}
\end{figure*}

Here we present in Figures \ref{fig: dmaps} and \ref{fig: Tdmaps} detection probability maps for the most best-sampled groups in our programs (TWA, BPMG, COL, THA, ABDMG, and UMA). All maps are derived from the F444W sensitivity limits shown in Figure \ref{fig: Contrast/Magnitude}.
Following results from section \ref{subsec: Principal Component Analysis of Detection Probability Maps}, we ordered these maps as a function of the age, from the younger group to the older.
For each group, we report the most common spectral types (SpT) within that sample, close to the group name.  
Table \ref{Tab: mass_teff_dprobs_vals} instead provide a more  quantitative representation, showing the expected mass detection probability in the region 0.3-1\,M$_{\rm Jup}$ and 50--500\,au, and the T$_{\rm eff}$ detection probability in the region 300-1000\,K and 30-200\,au, for each of the best-sampled groups.

\begin{table*}[t]  
\centering
\tabletypesize{\scriptsize}          
\setlength{\tabcolsep}{2pt}          
\renewcommand{\arraystretch}{0.95}  
\caption{Mean detection probability (\%) for mass (0.3--1.0\,M$_{\rm J}$, 50--500\,AU) and T$_{\rm eff}$ (300--1000\,K, 30--200\,AU).\label{Tab: mass_teff_dprobs_vals}}
\begin{tabular}{lc|cccccc|cccccc}
\hline
 &  & \multicolumn{6}{c|}{Mass} & \multicolumn{6}{c}{T$_{\rm eff}$} \\
Model & CO/FeH & TWA & BPMG & COL & THA & ABDMG & UMA & TWA & BPMG & COL & THA & ABDMG & UMA \\
\hline
bex-atmo2023-ceq & 0.55/0.0 & 93.6\% & 83.2\% & 77.5\% & 79.4\% & 49.4\% & 11.2\% & 93.6\% & 95.9\% & 93.4\% & 94.6\% & 96.6\% & 96.0\% \\
\hline
bex-atmo2023-neq-s & 0.55/0.0 & 93.6\% & 83.2\% & 78.1\% & 79.4\% & 49.5\% & 11.4\% & 92.6\% & 95.0\% & 92.3\% & 92.4\% & 95.8\% & 95.1\% \\
\hline
bex-atmo2023-neq-w & 0.55/0.0 & 93.7\% & 83.2\% & 78.0\% & 79.8\% & 49.4\% & 11.3\% & 92.7\% & 94.6\% & 91.9\% & 92.5\% & 94.8\% & 94.1\% \\
\hline
sonora-flame-skimmer-ceq & 0.5/0.0 & 95.2\% & 84.3\% & 81.2\% & 81.3\% & 47.3\% & 6.8\% & 94.3\% & 96.5\% & 95.1\% & 95.5\% & 97.0\% & 95.3\% \\
 & 0.5/1.0 & 94.8\% & 84.4\% & 80.0\% & 80.2\% & 49.4\% & 7.5\% & 93.3\% & 96.0\% & 93.8\% & 94.4\% & 96.8\% & 95.8\% \\
 & 0.5/2.0 & 92.9\% & 83.4\% & 79.4\% & 80.6\% & 48.9\% & 8.3\% & 81.1\% & 89.8\% & 82.2\% & 85.1\% & 92.3\% & 92.0\% \\
 & 1.0/0.0 & 95.5\% & 84.5\% & 82.7\% & 81.4\% & 48.2\% & 7.3\% & 94.8\% & 96.6\% & 95.6\% & 95.5\% & 97.2\% & 96.5\% \\
 & 1.0/1.0 & 94.7\% & 84.1\% & 78.9\% & 80.2\% & 49.9\% & 8.1\% & 93.4\% & 96.0\% & 93.6\% & 94.7\% & 96.8\% & 94.6\% \\
 & 1.0/2.0 & 92.7\% & 82.6\% & 74.7\% & 78.8\% & 51.9\% & 9.6\% & 80.9\% & 89.7\% & 82.1\% & 85.4\% & 92.1\% & 91.8\% \\
 & 2.5/0.0 & 94.6\% & 84.3\% & 80.9\% & 81.8\% & 49.5\% & 8.5\% & 94.3\% & 96.6\% & 94.7\% & 95.6\% & 97.1\% & 96.7\% \\
 & 2.5/1.0 & 93.7\% & 84.0\% & 77.1\% & 79.9\% & 49.9\% & 8.5\% & 93.0\% & 95.8\% & 92.7\% & 94.4\% & 96.6\% & 95.3\% \\
 & 2.5/2.0 & 89.9\% & 79.4\% & 64.5\% & 66.1\% & 29.6\% & 3.0\% & 79.6\% & 89.3\% & 82.1\% & 84.5\% & 92.4\% & 91.0\% \\
\hline
sonora-flame-skimmer-neq & 0.5/0.0 & 92.5\% & 80.9\% & 64.5\% & 67.9\% & 43.2\% & 4.2\% & 90.2\% & 94.7\% & 90.1\% & 91.8\% & 96.2\% & 95.0\% \\
 & 0.5/1.0 & 75.2\% & 50.5\% & 20.0\% & 22.6\% & 6.9\% & 0.0\% & 79.6\% & 87.1\% & 79.0\% & 82.2\% & 90.2\% & 89.5\% \\
 & 0.5/2.0 & 81.5\% & 56.6\% & 27.2\% & 28.5\% & 1.5\% & 0.0\% & 62.8\% & 73.4\% & 62.8\% & 67.1\% & 79.2\% & 79.0\% \\
 & 1.0/0.0 & 90.0\% & 78.6\% & 50.0\% & 53.6\% & 36.6\% & 4.9\% & 89.6\% & 94.2\% & 88.4\% & 90.3\% & 94.7\% & 94.0\% \\
 & 1.0/1.0 & 69.4\% & 45.9\% & 15.3\% & 17.8\% & 4.7\% & 0.0\% & 78.1\% & 85.8\% & 78.0\% & 81.3\% & 88.4\% & 88.5\% \\
 & 1.0/2.0 & 76.2\% & 49.8\% & 21.8\% & 22.4\% & 2.7\% & 0.0\% & 62.5\% & 74.5\% & 64.5\% & 68.0\% & 79.0\% & 77.6\% \\
 & 2.5/0.0 & 85.1\% & 76.1\% & 44.6\% & 50.5\% & 25.6\% & 1.3\% & 87.0\% & 92.9\% & 85.9\% & 88.0\% & 93.4\% & 91.4\% \\
 & 2.5/1.0 & 51.5\% & 28.5\% & 3.4\% & 7.7\% & 0.2\% & 0.0\% & 72.8\% & 80.8\% & 72.1\% & 75.1\% & 84.7\% & 80.0\% \\
 & 2.5/2.0 & 46.4\% & 20.9\% & 0.4\% & 1.3\% & 0.1\% & 0.0\% & 56.3\% & 67.6\% & 55.4\% & 59.2\% & 70.8\% & 67.1\% \\
\hline
\end{tabular}
\end{table*}

\FloatBarrier

\bibliography{sample7}{}
\bibliographystyle{aasjournalv7}



\end{document}